\documentstyle[aps,multicol]{revtex}
\font\autori = cmbx12  
\newcommand{\rvec}{{\bf r}}
\newcommand{\rrvec}{{\bf r'}}			
\newcommand{\kvec}{{\bf k}}

\newcommand{\Kinv}[1]{{\bf K}^{-1}_{#1}}	
\newcommand{\calKinv}[1]{{\bbox{\cal K}}^{-1}_{#1}}	
\newcommand{\calK}[1]{{\cal K}^{-1}_{#1}}	
\newcommand{\calG}[1]{{\cal G}^{-1}_{#1}} 	
\newcommand{\Gind}[1]{{\bf G}_{#1}}		
\newcommand{\calGind}[1]{{\bbox{\cal G}}_{#1}}	
\newcommand{\Iind}[1]{{\text{I}}^{\ast}_{#1}(k)} 
\newcommand{\Ireal}[1]{{\text{I}}_{#1}(k)}	
\newcommand{\Rind}[1]{{\bf R}_{#1}}		
\newcommand{\Sind}[1]{{\bf S}_{#1}}		
\newcommand{\Tind}[1]{{\bf T}_{#1}}		
\newcommand{\PLind}[2]{{\text{I}\!\text{#1}}_{#2}}  
\newcommand{\PLindinv}[2]{{\text{I}\!\text{#1}}^{-1}_{#2}} 
\newcommand{\sigind}[1]{\bbox{\sigma}_{#1}}	

\newcommand{\unitkk}{{\openone - \hat{\kvec}\hat{\kvec}}} 

\newcommand{\bessi}[1]{{\rm I}_{#1}(\frac{a}{\xi '})}	
\newcommand{\bessidoi}[2]{{\rm I}_{#1}(#2)}		
\newcommand{\bessk}[1]{{\rm K}_{#1}(\frac{a}{\xi '})}   
\newcommand{\besskdoi}[2]{{\rm K}_{#1}(#2)}		
\newcommand{\bessph}[1]{\text{j}_{#1}(ka)}		
\newcommand{\Ylm}[2]{\text{Y}_{#1}(#2)}			
\newcommand{\Ylmconj}[2]{\text{Y}^{\ast}_{#1}(#2)}	

\newcommand{\beq}{\begin{equation}}		
\newcommand{\eeq}{\end{equation}}
\newcommand{\beqar}{\begin{eqnarray}}
\newcommand{\eeqar}{\end{eqnarray}}
\newcommand{\fsp}{\Phi^{SP}}			
\newcommand{\fpol}{\Phi^{POL}}		
\newcommand{\la}{\left<}		
\newcommand{\ra}{\right>}		
\newcommand{\veff}{\frac{\eta^{\text{eff}}}{\eta_0} - 1}  
\newcommand{\vrelspol}{\frac{\eta^{\text{eff}}}{\eta_{\text{POL}}} - 1}	
\newcommand{\vefftext}{\eta^{\text{eff}}/\eta_0 - 1} 
\newcommand{\vrelspoltext}{\eta^{\text{eff}}/\eta_{\text{POL}} - 1}  
\newcommand{\vrelpolsp}{{\eta^{\text{eff}} \over \eta_{\text{SP}}} - 1} 
\input epsf
\begin{document}
\draft
\title{DYNAMICS OF A SUSPENSION OF SPHERES AND RIGID POLYMERS: EFFECT OF
GEOMETRICAL MISMATCH}
\author{\autori Radu P. Mondescu and  M. Muthukumar}
\address{Department of Physics  and Astronomy, and Polymer Science \&
Engineering Department and Materials Research Science and Engineering Center\\ 
University of Massachusetts, Amherst, MA 01003}
\date{\today}
\maketitle
\begin{abstract}
 An effective medium approach together with a multiple scattering formalism is
considered to study the steady-state dynamics of suspensions of spheres and
rigid stiff polymer chains without excluded volume interactions.  The polymer
chains are taken to be so long that Gaussian statistics is applicable.  We have
considered the dynamics of probe objects in a solution containing spheres and
polymers. The probe object is either a sphere or a polymer. We have studied
different conditions of the solution where some or all of the spheres are
frozen in space. The effective medium equations have been solved
self-consistently for finite volume fractions of spheres $\fsp$ and polymers
$\fpol$, respectively, and the important dimensionless variables that are
controlling the dynamical behavior have been identified.  In particular,the
role of the geometrical parameter $t={R_g\over a}$ ($a$ is the radius
of any sphere and $R_g$ the radius of gyration of a polymer chain) is
discussed.  The translational diffusion coefficients of the moving probe sphere
$D^S$ and of the probe polymer chain $D^P$, and the shear
viscosity of the suspensions have been derived.  When the polymers are present
in the solution,both the friction coefficient of the labeled mobile sphere or
that of the probe polymer chain and the shear viscosity diverge as
$\fpol\rightarrow 0.31$.  Also, when polymers are diffusing in a suspension of
fixed spheres an optimum range of $\fsp$ that maximizes the difference in the
diffusion coefficients of polymer chains characterized by distinct $t$ values
has been noticed.
\end{abstract}
\pacs{}

\begin{multicols}{2}
\section{INTRODUCTION}

	Starting with the pioneering work of Einstein \cite{Eins}, a lot of
effort was devoted to developing the hydrodynamics of suspensions at finite
concentrations, owing to both the intrinsic challenge of the problem and to the
considerable utility of such a theory in the rheology of concentrated polymer
suspensions.  Most of the work done is related to the hydrodynamics of
suspensions of hard spheres, specifically to the computation of virial-type
expansions---in the volume fraction $\Phi$ (low but non-zero) of the suspended
particles---for the change in viscosity $\delta\!\eta$ and in the friction
coefficients $\zeta$ (translational and rotational).  An incomplete review of
the development of this field could include the divergence-free theories of
Peterson and Fixman \cite{PF} and Batchelor and Green \cite{BG}, the multiple
scattering method elaborated by Freed and Muthukumar \cite{FM78,FM82} and
Muthukumar and Freed \cite{MF82,MF79}, the expansion in correlation functions
used by Beenaker \cite{BEE} and the angular momentum diagrammatic expansion
combined with the multiple scattering technique in Thomas and Muthukumar
\cite{TM}.

The viscosity and frictional properties of infinitely diluted polymer
solutions received a wide interest (see Ref.\ \cite{YAM}),but few attempts
were made to construct a detailed theory for finite concentrations
\protect\cite{FE,M81,DE}, and even less theoretical results are available for
dispersions of distinct types of elements.

In this paper we derive explicit expressions for the change in viscosity due
to suspended objects and for the translational diffusion coefficients, as
functions of the solid filling fractions of spheres and of polymer
chains, $\fsp$ and $\fpol$, respectively, for a suspension of polymers and
rigid, fixed spheres, in the stationary, time-independent limit.  Qualitative
features of the hydrodynamic behavior of the mixed suspensions are discussed
and it is shown that when the polymers are present in the solution, the
dynamics is {\sf frozen} at $\fpol$~$\simeq$~0.31.  Results in the limits of
very low \cite{KKR} and very high polymer concentrations (Rouse) and also for
the dynamics of a random array of spheres \cite{M82}, are recovered.  We have
found that the interaction between polymers, spheres and the fluid is
controlled by the dimensionless parameters ${\beta ={R_g \over \xi}}$,
$x = {a \over\xi}$ and $t={R_g \over a}$, where $\xi$~=~$\xi (\fsp,\fpol,t)$
is the hydrodynamic screening length.

Technically, our calculations are based on the multiple scattering formalism
\cite{FM78,MF82,M82,FM82} and on the assumption that the physical properties of
the suspension, averaged over the random position distribution of the
particles, are equivalent to those of an homogeneous effective fluid
characterized by the effective viscosity $\eta^{\text{eff}}$ and by the {\sf
altered Oseen propagator} of the velocity field ${\bbox{\cal G}}({\bf r},{\bf
r'})$ (to be defined later).

	The main simplifying hypothesis is the absence of excluded volume
interactions---sphere-sphere,polymer-polymer,or polymer-sphere---both the
polymer chains and the spheres being penetrable,{\sf ghost}-type objects.
Also, the solvent is assumed incompressible and described by a linearized
Navier-Stokes (N-S) equation.

	Other assumptions made and the basic features of the theory are~:

{\sl a\/}) the suspension of spheres and polymers is monodisperse, all spheres
having the radius $a$ and all polymers having the length $L$.

{\sl b\/}) except the sphere producing the flow, all others are fixed.  All
polymers are Gaussian.

{\sl c\/}) the specific calculations are valid only in the limit of
steady-state shear flow in the linear response regime,when the average force
$\la {\bf F}({\bf r}) \ra$ exerted upon the fluid is given by~: $$\left< {\bf
F}({\bf r})\right> = \left({\sf K}\, \ast \la{\bf v}\ra\right) ({\bf r})$$
where {\sf K} is a tensor-like quantity and $\ast$ is the convolution
operator: f~$\ast$~g~=~$\int f({\bf r}-{\bf r'})g({\bf r'})\,d{\bf r'}$.

{\sl d\/}) both the spheres and the polymer chains are randomly distributed,
and after the configurational average is taken the suspension becomes
isotropic.

{\sl e\/}) we employ the usual preaveraging approximation \cite{YAM}, where one
replaces the value of the effective Oseen tensor ${\bbox{\cal G}}({\bf r},{\bf
r'})$ and of its generalized inverse with their configurational averages $\la
{\bbox{\cal G}}({\bf r},{\bf r'})\ra$ and $\la\bbox{\cal
G}^{-1}(\rvec,\rrvec)\ra$.

{\sl f\/}) two models have been analyzed~:
{\small \begin{description}
\item[S-P] an ensemble of rigid spheres in a dispersion of polymers. All but
one sphere are fixed.  The tracer sphere generates the flow.
\item[P-S] a collection of polymer chains immersed in a suspension of immobile
spheres, with an external flow imposed.
\end{description}} 

The hydrodynamic interaction between fluid and particles is implemented by
enforcing stick boundary conditions.  In each example we calculate the
translational diffusion coefficient of the test object ( a polymer chain or the
tracer sphere ) and the change in the shear viscosity of the solution due to
the presence of the added particles, as functions of $\fsp$, $\fpol$ and the
{\sf coupling} parameter $t = {R_g \over a}$.

	The paper is organized as follows~: next section deals with the formal
theory of the steady-state velocity flow in a solution of $\text{N}_P$ polymer
chains and $\text{N}_{S}$ fixed spheres , in Section\ \ref{sec:eff} we apply
the derivation to the separate cases of polymers and spheres, which are needed
in Sec.\ \ref{sec:res_spec} where we solve and show the results for the four
specific examples mentioned previously.  Finally, in Section~\ref{sec:conc}, we
present, discuss, and review the main findings of our paper. For the reader
interested in reproducing and extending our results, detailed calculations are
displayed in the Appendices.

\section{FORMAL THEORY OF THE VELOCITY FIELD IN A SUSPENSION OF POLYMERS AND
SPHERES} 
\label{sec:form}
	
	To describe the dynamics of a hydrodynamic system, a knowledge of the
velocity field {\bf v}({\bf r},t) produced by the forces acting upon the fluid
is required.  Our goal is to find an expression for {\bf v}({\bf r})
(stationary limit) for the system we are interested in and to extract the
appropriate physical information (diffusion coefficient, viscosity).  This is
accomplished using the multiple scattering technique as shown below.

\subsection{Fluid and Particles}
\label{sec:fluid+part}

	Under the assumptions indicated in the previous section and taking the
fluid density equal to unity, the stationary velocity field for an
incompressible, newtonian viscous fluid is governed by the linearized N-S
equation~:
\beq
- \eta_0 \triangle {\bf v}({\bf r}) + \nabla p({\bf r}) = {\bf F}({\bf r})
\label{Navier_Stokes}
\eeq
where {\bf F}({\bf r},t) and $p(\rvec,t)$ are the
external force driving the flow ( including the random thermal
contribution ) and the pressure, and $\eta_0$ is the kinematic shear viscosity
coefficient.  
Introducing the Fourier transform of any function {\bf A}({\bf r}) as~:
\beqar
{\bf A}({\bf r})& = &\int \!{\text{d}{\bf k}\over (2 \pi)^3}\, {\bf A}({\bf k})
\exp(- i {\bf k}\cdot {\bf r}) \equiv \int_{\bbox{k}} {\bf A}({\bf k}) \exp{(-
i {\bf k}\cdot{\bf r})}\nonumber \\ {\bf A}({\bf k}) & = & \int \!\text{d}{\bf
r}\, {\bf A}({\bf r}) \exp(i {\bf k}\cdot{\bf r})
\label{Fourier}
\eeqar 
one finds the formal solution of Eq.~(\ref{Navier_Stokes}) as a convolution~:
\beq {\bf v}({\bf r}) = ({\bf G}\,\ast\,{\bf F})({\bf r}) 
\eeq 
with {\bf G}({\bf r},${\bf r'}$) being the Oseen tensor \cite{LL}\ (in dyadic
notation)~:
\beqar 
{\bf G}({\bf k}) = && \frac{\openone - \hat{\bf k}\hat{\bf k}}{\eta_0
k^2} \nonumber\\ 
{\bf G}({\bf r},{\bf r'}) = {1\over 8 \pi \eta_0 |{\bf r} -
{\bf r'}|} && \left ( \openone + {({\bf r}-{\bf r'})({\bf r}-{\bf
r'})\over{|{\bf r}- {\bf r'}|^2}} \right) 
\eeqar 
Here, $\openone$ is the unit tensor and $\hat{\bf k}$ is the unit vector
pointing in the {\bf k} direction.

Physically, ${\bf G}(\rvec,\rrvec)$---the {\sf force propagator}---is conveying
the hydrodynamic disturbances, converting a pointlike force acting at ${\bf
r'}$ to the velocity of a fluid particle located at {\bf r}.  When a fixed
object is introduced in the fluid, the velocity flow is perturbed due to the
scattering off the particle surface. One needs then to include in
Eq.~(\ref{Navier_Stokes}) a term ${\bf F}^{OBJ \to FL}$({\bf r}) corresponding
to the force exerted by the particle upon the fluid~:
\beq 
- \eta_0\triangle {\bf v}({\bf r}) + \nabla
p({\bf r})\;=\; {\bf F}({\bf r}) + {\bf F}^{OBJ\to FL}({\bf r}) 
\label{Navier_obj}
\eeq 
To couple in a simple way the dynamics of the solvent and of the object, we
choose no-slip boundary conditions~: 
\beq 
{\bf v}^{\text{FLUID}}({\bf r}_s)\;=\; {\bf v}^{\text{OBJ}}({\bf r}_s), \;\;\;
{\bf r}_s\,\in \, \text{object surface} .
\label{stick}
\eeq

	In Eq.~(\ref{Navier_obj}) {\bf F}({\bf r}) is zero inside and on the
surface of the particles and ${\bf F}^{OBJ\to FL}({\bf r})$ is non-zero only on
the surface of the suspended particles.  The equation of motion is valid
throughout the volume of the fluid.  Because we want to replace the fluid and
the suspended particles with an effective medium characterized by an average
velocity field defined everywhere, it is necessary to apply
Eq.~(\ref{Navier_obj}) inside the volume of the particles as well.  This is
true provided the full stress tensor~${\bf\Pi}({\bf r})$ \cite{BM,FM78} obeys (
neglecting the inertial term in the equation of motion of the object )~:
\beq
\begin{array}{rll}
\nabla\,\cdot\,{\bf\Pi}({\bf r}) & = 0 &,{\bf r}\in \text{inside object}. \\
\nabla\,\cdot\,{\bf\Pi}({\bf r}_s) & = {\bf F}^{OBJ\to FL}({\bf r}_s)&,{\bf
r}_s \in\text{object surface}.
\end{array}
\eeq

The formal solution of Eq.~(\ref{Navier_obj}) can be written as~:
\beq
{\bf v}({\bf r})\;=\;\left. {\bf G} \ast {\bf F}\right|_{\bf r} + \left. {\bf
G} \ast {\bf F}^{OBJ\to FL}\right|_{\bf r}		
\label{velsol}
\eeq 
Using the boundary condition (\ref{stick}), we can eliminate the unknown
quantity ${\bf F}^{\text OBJ \to FL}$({\bf r}) ---as shown explicitly in the
next section---and arrive at the exact equation that gives the microscopic
velocity field everywhere in the fluid~: 
\beq 
{\bf v}({\bf r}) = \left. {\bf G}\ast {\bf F}\right|_{\bf r} + \left. {\bf
G}\ast{\bf T}^{OBJ} \ast {\bf G}\ast {\bf F} \right|_{\bf r}
\label{obj}
\eeq
where the multiple convolution product means~:
$${\bf f}\ast {\bf g}\ast {\bf h} |_{\bf r} = \int\!
d{\bf r}\,d{\bf r'}\:{\bf f}({\bf r}-{\bf r'})\cdot{\bf g}({\bf r'}-{\bf
r''})\cdot {\bf h}({\bf r'' })$$ 
and ${\bf T}^{OBJ}$ represents the {\sf flow propagator}, a tensor operator
that transforms the velocity field incident on the object in a force field
located on its surface and acting upon the fluid. ${\bf T}^{OBJ}$ depends in
general on the position, structure and geometry of the object. The
interpretation of (\ref{obj}) is straightforward~: there are two contributions
to the velocity of the fluid {\bf v}({\bf r}) at point {\bf r}, a {\sf direct
wave} due to the external force {\bf F} acting at some point ${\bf r'}$, and an
{\sf indirect} contribution coming from the scattering off the surface of the
object of the disturbance produced by {\bf F} at ${\bf r'}$ (see Fig.~1). 

When many objects are present in the solution, one should add all the
contributions coming from all possible {\sf scattering sequences} but, even for
two bodies only, Eq.\ (\ref{obj}) becomes a series with an infinite number of
terms, the convergence of which might be problematic.  For some specific cases
(polymers, spheres) \cite{M81,M82}, these difficulties can be overcome, which
was the reason we decided to investigate solutions of mixed polymer chains and
spheres.  For example, when two objects labeled {\sf a} and {\sf b} are present
in the solution, the multiple scattering expression for the velocity field {\bf
v}(\rvec) will be~:
\beqar 
{\bf v}(\rvec) & = & \left. {\bf G}\ast {\bf F}\right|_{\rvec} +
\left. {\bf G}\ast\{ {\bf T}_a + {\bf T}_b \}\ast {\bf G}\ast {\bf
F}\right|_{\rvec}\nonumber\\ 
& & + \left. {\bf G}\ast\{ {\bf T}_{a}\ast{\bf G}
\ast{\bf T}_b + {\bf T}_{b}\ast{\bf G}\ast{\bf T}_a \}\ast {\bf G}\ast{\bf
F}\right|_{\rvec}\\ 
& & + \left. {\bf G}\ast \{ {\bf T}_{a}\ast{\bf G}\ast{\bf
T}_{b}\ast{\bf G}\ast{\bf T}_a \right. \nonumber\\
& & + \left. {\bf T}_{b}\ast{\bf G}\ast{\bf T}_{a}\ast {\bf
G}\ast{\bf T}_{b} \}\ast {\bf G}\ast{\bf F}\right|_{\rvec} + \dots\nonumber 
\eeqar

Being interested only in the average properties of the system, we will average
in Eq.\ (\ref{obj}) upon the random position of the body, thus obtaining the
average velocity field {\bf u}({\bf r})~: 
\beqar 
{\bf u}({\bf r}) = \la {\bf v}({\bf r})\ra & = & {1\over V}\int\! d{\bf
R}^0\,{\bf v}({\bf r}) = \left.  {\bf G}\ast{\bf F}\right|_{\bf r} \nonumber\\
& & + \left. {\bf G}\ast\la {\bf T}^{OBJ}\ra\ast {\bf G}\ast {\bf F}
\right|_{\bf r}
\label{usol}
\eeqar 
with ${\bf R}^0$ denoting the position of the center of mass of the object.Note
that if excluded volume interactions are considered, appropriate particle
distribution functions should be used.

To make the connection with the experimentally measurable quantities, we will
turn to Eq.~(\ref{Navier_obj}) and average it directly over the particle
distribution to get the Navier-Stokes equation describing the {\sf effective}
fluid~: 
\beq
-\eta_0 \triangle {\bf u}({\bf r}) + \nabla \la p({\bf r})\ra -
\left. {\bf\Sigma} \ast {\bf u} \right|_{\rvec} = {\bf F}(\rvec)
\label{nseff}
\eeq
In writing the equation above we introduced a new quantity, the
${\bf\Sigma}({\bf r},{\bf r'})$ operator, called the self-energy of the fluid
and defined by the relation~: \beq \left. {\bf\Sigma} \ast {\bf
u}\right|_{\rvec} = \la {\bf F}^{OBJ \to FL}(\rvec)\ra
\label{sigdef}
\eeq
Note that this relation implies a linear response regime of the shear flow.

${\bf\Sigma}(\rvec,\rvec ')$ is the essential quantity that encompasses all the
information regarding the change in the viscoelastic properties of the fluid
due to the presence of the object(s). To illustrate this, let us introduce the
Fourier transform of the self-energy, knowing that the averaged solution is
translationally invariant~: 
\beq 
{\bf\Sigma}(\kvec) = \int\!  d\rvec\,{\bf\Sigma}(\rvec - \rrvec)\exp(i
\kvec\cdot (\rvec - \rrvec)) 
\eeq 
Then Eq.\ (\ref{obj}) becomes~: 
\beq
\left[\eta_0 k^2 \openone - {\bf\Sigma}(\kvec)\right]\cdot {\bf u}(\kvec) - i
{\kvec} p(\kvec) = {\bf f}(\kvec)
\label{vfr}
\eeq 
Using the incompressibility condition $\hat{\kvec} \cdot {\bf
u}(\kvec) = 0$ and decomposing ${\bf\Sigma}(\rvec,\rrvec)$ into its transverse
and longitudinal parts applying the relations~: 
\beqar 
{\bf\Sigma}(\kvec) & = & \Sigma_{\perp}(\kvec)\,(\openone -
\hat{\kvec}\hat{\kvec}) + \Sigma_{\parallel}(\kvec)\,
\hat{\kvec}\hat{\kvec}\nonumber\\ 
(\openone - \hat{\kvec}\hat{\kvec})\cdot(\openone - \hat{\kvec}\hat{\kvec}) & =
& (\openone - \hat{\kvec}\hat{\kvec}) \label{sigtrans} \\
\hat{\kvec}\hat{\kvec}\cdot (\openone - \hat{\kvec}\hat{\kvec}) & = & 0
\nonumber 
\eeqar 
we can eliminate the pressure $p(\kvec)$ to solve Eq.\ (\ref{vfr}) for the
Fourier component {\bf u}(\kvec) of the averaged velocity field {\bf
u}(\rvec)~: 
\beqar 
{\bf u}(\kvec) & = & {\bbox{\cal G}}(\kvec)\cdot {\bf
F}(\kvec)\nonumber\\ {\bbox{\cal G}}(\kvec) & = & \frac{\openone -
\hat{\kvec}\hat{\kvec}}{\eta_0 k^2 - \Sigma_{\perp}(\kvec)} \label{oseenk}
\eeqar 
We recover then, immediately~: 
\beq 
{\bf u}(\rvec) = \left. {\bbox{\cal G}}\ast {\bf F}\right|_\rvec
\label{uk}
\eeq

The interpretation of ${\bf\Sigma}$ emerges clearly.  What we achieved is to
replace the initial system {\sf fluid $+$ object(s)} with an {\sf effective
medium} of a certain viscosity $\eta^{\text{eff}}$ where the hydrodynamic
disturbances are propagated by the modified Oseen tensor ${\bbox{\cal
G}}(\kvec)$, with all the complexities of the multiple scattering processes
involving the suspended object(s) captured by the ${\bf\Sigma}$ tensor.

In the hydrodynamic limit $|\kvec|\to 0$, the total change in viscosity is
given by~:
\beq
\frac{\eta^{\text{eff}}- \eta_0}{\eta_0} = - \lim_{\kvec\to 0}{1\over\eta_0}
{\partial\over \partial k^2} \Sigma_{\perp} (\kvec)
\label{viscchange}
\eeq
\indent If $\Sigma_{\perp}(\kvec = 0)\neq 0$ , the hydrodynamic interaction is
screened, with the screening length $\xi$ given by~:
\beq
\xi^{-2} = - {1\over\eta_0}\Sigma_{\perp}(\kvec = 0)
\eeq
Similarly, formulas for the friction coefficients $\bbox{\zeta}$
(translational, rotational, cross translational-rotational ) of the object can
be derived from ${\bf\Sigma}(\kvec)$.  We will explicitly show this, for the
translational friction coefficient, in the following sections.

More practical, from an experimental point of view, is the diffusion
coefficient $D$ of a particle moving in a solution, calculated from the
Einstein formula~:
$$D(\Phi)\;=\; {k_B T \over \zeta(\Phi)}$$ 
where T~=~temperature, $k_B$ is the Boltzmann constant and $\Phi$ is the volume
fraction of the suspended particles in the solution.  Note that in general the
friction coefficient is a tensor-like quantity, in which case the previous
expression should be adjusted correspondingly. The dimensionless variable that
we have computed and plotted in this paper is the {\sf relative translational
diffusion} coefficient~: 
\beq 
{D_t (\Phi)\over D^0_t}\;=\; {\zeta^0_t \over \zeta_t (\Phi)}
\label{diffdef}
\eeq 
with $D^0_t$ the translational diffusion coefficient of a particle in the pure
fluid (e.g. $D^0_t = {k_B T\over 6 \pi\eta_0 a}$ for a sphere and $D^0_t =
{8\sqrt{2}\over 3}\,{k_B T \over 6 \pi\sqrt{2\pi}\eta_0 R_g}$ for a single
polymer chain ).  In the remaining of this work, we will denote this relative
diffusion coefficient simply by $D$.

Having established that $\bbox{\Sigma}(\kvec)$ contains the relevant dynamical
information , we need to actually compute it by relating it to the known
quantity ${\bf T}^{OBJ}$. Here lies the essence of the multiple scattering
formalism. Averaging Eq.\ (\ref{velsol}) directly and using the definition of
${\bf\Sigma}$ from Eq.\ (\ref{sigdef}) we can rewrite the solution for the
averaged velocity field {\bf u}(\rvec) in terms of the {\bf G} propagator of
the pure fluid~: 
\beq 
{\bf u}(\rvec) = \left. {\bf G}\ast {\bf F}\right|_{\rvec} + \left. {\bf
G}\ast{\bf\Sigma}\ast{\bf u}\right|_{\rvec} 
\eeq 
Iterating indefinitely, one obtains~: 
\beqar 
{\bf u}(\rvec) & = & \left. {\bf G}\ast{\bf F}\right|_{\rvec} + \left. {\bf
G}\ast{\bf\Sigma}\ast{\bf G}\ast {\bf F}\right|_{\rvec} \nonumber\\ 
& & + \left. {\bf G}\ast
{\bf\Sigma}\ast{\bf G}\ast{\bf\Sigma}\ast{\bf G}\ast{\bf F}\right|_{\rvec} +
\dots\label{usolsig} 
\eeqar 
Representing ${\bf\Sigma}$ as a series in the number $i$ of distinct scattering
events~: 
\beq 
{\bf\Sigma}(\rvec,\rrvec) = \sum_{i=1}^{\infty}{\bf\Sigma}^{(i)}(\rvec,\rrvec)
\eeq 
substituting in Eq.\ (\ref{usolsig}) and comparing to the formal solution
from Eq.\ (\ref{usol}), it follows that~: 
\beqar 
{\bf\Sigma}^{(1)}(\rvec,\rrvec) & = & \la {\bf T}^{\text{OBJ}}(\rvec,\rrvec)\ra
\nonumber\\ 
{\bf\Sigma}^{(2)}(\rvec,\rrvec) & = & - \left. \la {\bf
T}^{OBJ}\ra\ast{\bf G}(\rvec_1 ,\rvec_2)\ast \la {\bf
T}^{OBJ}\ra\right|_{(\rvec, \rrvec)} \label{sigma} \\ 
{\bf\Sigma}^{(3)}(\rvec,\rrvec) & = & \left. \la {\bf T}^{OBJ}\ra\ast{\bf
G}\ast\la {\bf T}^{OBJ}\ra\ast{\bf G}\ast \la {\bf T}^{OBJ}\ra
\right|_{(\rvec,\rrvec)}\nonumber\\ \dots & & \nonumber 
\eeqar 
where integration over ${\bf r}_1$ and ${\bf r}_2$ in the multiple convolution
is understood.

The expressions displayed above are particular for the case of only one
particle in the solution, but they illustrate the basics of the formalism used
in this paper.  Now it is straightforward to generalize the procedure developed
before to deal with a system where more than one object, or different types of
particles are immersed in a fluid.

Finally, at any volume fraction $\Phi$ of the solute particles, we compute
self-consistently \cite{MF79} the contribution ${\bf\Sigma}(\rvec,\rrvec)$ of
the suspended particles to the divergence of the full stress tensor of the
fluid.  This is accomplished by assuming linear superposition of the
contributions of each particle to the self-energy of the fluid (assumption
which is exact when there are no thermodynamic correlations among particles, as
in the present work). Then, for a suspension of $N$ particles, one can replace
$N-1$ of them with an {\sf effective medium} described by the modified Oseen
tensor ${\bbox{\cal G}}(\rvec,\rrvec)$, which is a function of some
${\bf\Sigma}_{N-1}$ (see Eqs.\ (\ref{nseff},\ref{oseenk},\ref{uk})).  Then, it
follows that $\delta{\bf\Sigma}$, the self-energy increase due to the remaining
particle is~: 
\beq 
\delta{\bf\Sigma}\left({\bbox{\cal G}}({\bf\Sigma}_{N-1}) \right) =
{{\bf\Sigma}_{N-1}\over N-1} 
\eeq 
The left-hand side of the equation can be evaluated as outlined for the case of
one object in a fluid, but using the ${\bbox{\cal G}}$ propagator instead of
${\bf G}$ of the pure fluid.  Thus we have obtained a self-consistent equation
for calculating the full ${\bf\Sigma}_{N-1}$.

\subsection{Calculation of ${\bf\Sigma}(\rvec,\rrvec)$ for a suspension of
$\text{N}_{S}\/$ spheres and $\text{N}_P\/$ polymer chains}
\label{sec:twoB}

	We consider $\text{N}_{S}$ fixed, rigid spheres, each of radius $a$,
and $\text{N}_P$ mobile, Gaussian polymer chains, each of length $L = nl$
($l=\:$Kuhn length; $n=\:$number of beads ), immersed in an incompressible
fluid of viscosity $\eta_0$, in the regime of stationary flow.

The static velocity field in the suspension is described by the N-S equation
together with stick boundary conditions~:
\beqar
-\eta_0 \triangle {\bf v}(\rvec) + \nabla p(\rvec) & & = {\bf F}(\rvec) +
\sum_{\alpha,i\geq 1}^{\text{N}_{P},n}\,\delta (\rvec - {\bf R}_{\alpha
i})\bbox{\sigma}_{\alpha i} \nonumber\\
& + &  \sum_{b=1}^{\text{N}_{S}}\int\!  d\Omega_{b}
\delta(\rvec - {\bf R}_b)\bbox{\sigma}_{b}(\Omega_b)
\label{nspolsp}
\eeqar
\beqar
\nabla\cdot {\bf v}(\rvec) & = & 0 \nonumber\\ 
{\dot{{\bf R}}}_{\alpha i } & = & {\bf u}_{\alpha} +
\bbox{\omega}_{\alpha}\times{\bf S}_{\alpha i} = {\bf v}({\bf R}_{\alpha i})
\label{nsbound}\\ 
{\dot{{\bf R}}}_{b} & = & 0 = {\bf v}({\bf R}_{b})\nonumber
\eeqar 
Here, the greek symbols $\alpha , \beta , \dots$ are labeling the polymer
chains and are running from 1 to $\text{N}_P\/$, $i,j,k,\dots$ are indices
($\in \{1,n\}$) for the beads on any arbitrary chain and $b,c\dots$ are labels
for the spheres (range $\overline{1,\text{N}_{S}}$).  ${\bf R}_{\alpha i} =
{\bf R}^{0}_{\alpha} + {\bf S}_{\alpha i}$ is the position vector of the $i$-th
bead of the chain $\alpha$, with ${\bf R}^0_{\alpha}$ denoting the position
vector of the center of mass of the $\alpha$ chain and with ${\bf S}_{\alpha
i}$ the position vector of the {\it i\/}-th bead with respect to the center of
mass of $\alpha$.  ${\bf R}_{b} = {\bf R}^0_{b} + \rvec_{b}(\Omega_{b})$ is the
decomposition of the position vector of a point on the surface of the sphere
$b$ in the position vector of its center of mass and the relative coordinate of
the surface point in the center of mass frame ($|{\bf r}_{b}| = a,\;\text{for
any sphere}$); $\Omega$ indicates the orientation of the vector ${\bf r}_b$.
$\bbox{\sigma}_{\alpha i}$ and $\bbox{\sigma}_{b}(\Omega_b)$ are the densities
of force exerted by the $i$-th bead of chain $\alpha$ and by the sphere $b$ at
the surface point ${\bf r}_b$, respectively, upon the fluid.  When there is no
risk of confusion, we will write only $\bbox{\sigma}_b$, the dependence of
$\Omega_b$ being understood.  Finally,{\bf F}(\rvec) is some force acting at
\rvec\ that generates the flow (e.g.the effect of any sphere or polymer chain
moving with uniform velocity).

From the boundary conditions for the polymer chains and because we are
neglecting the inertial terms in the equations of motion for the polymers, the
total force and torque acting upon any chain must vanish. We can write then~:
\beqar 
- \sum_{i=1}^{n}\bbox{\sigma}_{\alpha i} & = & 0 \nonumber\\ 
- \sum_{i=1}^{n}{\bf S}_{\alpha i}\times \bbox{\sigma}_{\alpha i} & = &
0\;\;,\alpha = \overline{1,\text{N}_{P}}
\label{constr}
\eeqar

The formal solution of Eq.\ (\ref{nspolsp}) is~:
\beqar
{\bf v}(\rvec) & = & \left. {\bf G}\ast {\bf F}\right|_{\rvec} +
\sum_{\alpha,i=1}^{\text{N}_{P},n}{\bf G}(\rvec - {\bf R}_{\alpha
i})\cdot\bbox{\sigma}_{\alpha i} \nonumber\\
& & + \sum_{b=1}^{\text{N}_{S}}\int \! d\Omega_b
\,{\bf G}(\rvec - {\bf R}_b)\cdot \bbox{\sigma}_b(\Omega_b)
\label{vforcespolsp}
\eeqar 
Using the boundary conditions (\ref{nsbound}) and the constraints
(\ref{constr}) we can eliminate the unknown forces $\bbox{\sigma}_{\alpha i}$
and $\bbox{\sigma}_b$, to express the solution for the velocity field as a
multiple scattering series in terms of the single-object {\sf flow propagator}
${\bf T}_{\alpha}$ for the chain $\alpha$ and ${\bf T}_b$ for the $b$ sphere
(see Appendix A for details)~: 
\beqar 
{\bf v}(\rvec) & = & \left. {\bf G}(\rvec)\ast{\bf F}\right|_{\rvec} +
\sum_{\alpha ,b}\left. {\bf G}\ast \{{\bf T}_{\alpha} + {\bf T}_{b}\}\ast {\bf
G}\ast {\bf F}\right|_{\rvec} + \nonumber\\ 
& + & \sum_{\alpha,b} \left. {\bf G}\ast \{{\bf T}_{\alpha}\ast{\bf G}\ast{\bf
T}_b + {\bf T}_b\ast{\bf G}\ast{\bf T}_{\alpha}\}\ast{\bf G}\ast{\bf
F}\right|_{\rvec}\nonumber\\ 
& + & \sum_{\alpha\neq\beta}\left. {\bf G}\ast {\bf T}_{\alpha}\ast {\bf G}\ast
{\bf T}_{\beta}\ast {\bf G}\ast {\bf F}\right|_{\rvec} \label{vpolsp}\\ 
& + & \sum_{b \neq c} \left. {\bf G}\ast{\bf T}_b\ast{\bf G}\ast{\bf T}_c\ast
{\bf G}\ast{\bf F}\right|_{\rvec} + \dots \nonumber 
\eeqar 
where the sum should be continued over all possible scattering sequences.  In
this expression, the first factor on the right-hand side represents the {\sf
direct wave}, the second contains the single scattering processes from only one
polymer chain or one sphere, the third includes the sequences sphere-polymer
and polymer-sphere,etc... .  Any sequence of {\bf T} operators is valid except
those involving two consecutive scatterings off the same sphere or the same
polymer (the exclusion constraint).

The macroscopic equation for the effective velocity field {\bf u}(\rvec) is
retrieved by performing a configurational (position) average over the
distribution of the particles, mathematically expressed as~:
\beq
\la\,\cdot\ra = {1\over V^{\text{N}_{P}\,\text{N}_{S}}}\int\prod_{\alpha} d{\bf
R}^0_{\alpha}\int\prod_{b} d{\bf 
R}^0_b \,\langle \,\cdot\, \rangle_{\alpha i,\alpha j,\dots}  
\label{confav}
\eeq  
with the ``0'' superscript referring the center of mass and $<\,>_{\alpha
i,\alpha j,\dots}$ being an average over the distribution of the segments of
the $\alpha$ chain about its center of mass.  The probability distribution
function is taken to be Gaussian (see \cite{YAM}). Then {\bf u}(\rvec)
is calculated from~: 
\beqar 
&{\bf u}(\rvec)& =  \la {\bf v}(\rvec)\ra = {\bf G}\ast {\bf F} +
\sum_{\alpha,b}{\bf G}\ast\{\la{\bf T}_{\alpha}\ra + \la{\bf T}_b\ra\}\ast{\bf
G}\ast{\bf F}\nonumber\\ 
& + & \sum_{\alpha,b}{\bf G}\ast\{\la{\bf T}_{\alpha}\ast{\bf G}\ast{\bf
T}_b\ra + \la{\bf T}_b\ast{\bf G}\ast{\bf T}_{\alpha}\ra \}\ast{\bf G}\ast{\bf
F}\nonumber\\ 
& & + \sum_{\alpha\neq\beta}{\bf G}\ast\la{\bf
T}_{\alpha}\ast{\bf G}\ast{\bf T}_{\beta}\ra\ast{\bf G}\ast {\bf F}
\label{upolsp}\\ 
& & + \sum_{b\neq c}{\bf G}\ast\la{\bf T}_b\ast{\bf G}\ast{\bf T}_c\ra\ast{\bf
G}\ast{\bf F}+ \dots\nonumber 
\eeqar 
where all convolutions are actually functions of \rvec.

As outlined in the previous section, the meaningful physical quantity is the
self-energy tensor ${\bf\Sigma}(\rvec,\rrvec)$ defined as~:
\beqar
{\bf\Sigma}\ast{\bf u} & = & \la \sum_{\alpha,i} \delta(\rvec - {\bf R}_{\alpha
i})\bbox{\sigma}_{\alpha i} + \sum_{b} \int \! d\Omega_b \,\delta (\rvec - {\bf
R}_b)\bbox{\sigma}_b \ra \nonumber\\ & = & {\bf\Sigma^{\text{POL}}}\ast {\bf u}
+ {\bf\Sigma^{\text{SP}}}\ast{\bf u}
\label{sigdefpolsp}
\eeqar 

Next, the solution of the configurationally averaged Navier-Stokes equation is
still Eq.~(\ref{usolsig}), but now the total ${\bf\Sigma}$ includes the effects
of both the polymers and the spheres. Expanding once again the self-energy in
the number of distinct scattering events (i.e. the number of ${\bf
T}_{\alpha,b}$ operators) and comparing the terms with the same number of {\bf
T} factors in Eqs.~(\ref{usolsig},\ref{upolsp}), it follows that~:
\beqar
&{\bf\Sigma}^{(1)}&(\rvec,\rrvec) = \sum_{\alpha}\la{\bf T}_{\alpha}\ra +
\sum_{b}\la{\bf T}_b\ra\nonumber\\ 
&{\bf\Sigma}^{(2)}&(\rvec,\rrvec) = \sum_{\alpha\neq\beta}\la{\bf
T}_{\alpha}\ast{\bf G}\ast{\bf T}_{\beta} \ra + \sum_{b\neq c}\la{\bf
T}_b\ast{\bf G}\ast{\bf T}_c\ra \nonumber\\
& + & \sum_{\alpha ,b}\{\la{\bf T}_{\alpha}\ast{\bf G}\ast{\bf T}_b\ra + \la
{\bf T}_b\ast{\bf G}\ast{\bf T}_{\alpha}\ra\} \nonumber\\
& - & \sum_{\alpha ,\beta}\la{\bf T}_{\alpha}\ra\ast{\bf G}\ast\la{\bf
T}_{\beta}\ra - \sum_{b,c}\la {\bf T}_b\ra\ast{\bf G}\ast\la{\bf T}_c\ra
\label{sigpolsp}\\ 
& - & \sum_{\alpha,b}\{\la{\bf T}_{\alpha}\ra\ast{\bf
G}\ast\la{\bf T}_b\ra + \la{\bf T}_b\ra\ast{\bf G}\ast\la{\bf T}_{\alpha}\ra \}
\nonumber\\ 
& & \;\;\;\; \dots
\eeqar

Assuming that the spheres do not rotate , we can derive the following
expressions for the ${\bf T}_{\alpha,b}$ operators (see Appendix A)~:
\beqar
&{\bf T}_{\alpha} & (\rvec,\rrvec) = - \sum_{i,j}^{n}\delta(\rvec - {\bf
R}_{\alpha i})\left[{\bf K}^{-1}({\bf S}_{\alpha i},{\bf S}_{\alpha
j})\right.\nonumber\\  
& - & \sum_{l,l'=1}^{n} \left. {\bf K}^{-1}({\bf S}_{\alpha i},{\bf S}_{\alpha
l})\cdot{\bf g}^{-1}_{t}\cdot{\bf K}^{-1}({\bf S}_{\alpha l'},{\bf S}_{\alpha
j}) \right] \delta(\rvec - {\bf R}_{\alpha j}) \nonumber\\
& & \; + \; \text{rotational terms} \; + \; \dots \label{topers}\\
& & \;\;\;\;\; {\bf g}_{t} =  \sum_{i,j}^{n}{\bf K}^{-1}_{ij}\;\; ;\;\;\;{\bf
g}_t\cdot{\bf g}^{-1}_t = \openone \, . \nonumber\\
&{\bf T}_{b}&(\rvec - \rrvec) = - \int \! d\Omega_b \, d\Omega_{b}'\: \delta
(\rvec - {\bf R}_b) \,{\bf K}^{-1}_{b}(\Omega_b ,\Omega_{b}')\, \delta (\rrvec
- {\bf R}_{b}') \nonumber
\eeqar
in which ${\bf R}_b$ and ${\bf R}_{b}'$ are the position vectors of separate
points on the surface of the same sphere labeled $b$ and ${\bf
K}^{-1}_{\alpha}$ and ${\bf K}^{-1}_b$ are the generalized inverse operators
for the single polymer chain and the single sphere \cite{M81,M82}, defined by~:
\beqar
\sum_{j=1}^{n} {\bf K}^{-1}(\Sind{i},\Sind{j})\cdot {\bf G}(\Sind{j} -
\Sind{k}) & = &\openone \, \delta_{ik} \label{geninv}\\
\int\! d\Omega_{b}'' \: {\bf K}^{-1}(\Omega_b,\Omega_{b}'')\cdot {\bf
G}(\rvec_{b}(\Omega_b '') - \rvec_{b}(\Omega_b ')) & = & \openone \, \delta
(\Omega_b - \Omega_b ')
\nonumber
\eeqar
To avoid an increasingly intricate notation, we will adopt the following
short-hand notation~:
\beqar 
{\bf K}^{-1}(\Sind{\alpha i},\Sind{\alpha j}) & \to & \Kinv{\alpha
i,\alpha j} \nonumber\\ 
{\bf K}^{-1}(\Omega_b ,\Omega_b ') & \to & \Kinv{b,b'}\\ 
{\bf G}(\Rind{\alpha j} - \Rind{b}) & \to & \Gind{\alpha j,b} \nonumber
\eeqar
The Einstein summation convention over repeated indices is implied everywhere
but where explicitly not followed . We also drop the index $\alpha$ when
referring to quantities not specifically depending on a particular polymer
chain.

Some remarks are necessary before embarking upon some concrete calculations.
In the limit of our static description of the suspension of polymer chains and
spheres, Eqs.\ (\ref{sigpolsp}) are exact, for noninteracting as well as for
interacting objects, and will describe the dynamics of the suspension with any
degree of accuracy, although practically it may require a strenuous effort.  To
obtain analytical results, we are limiting ourselves to the case of
noninteracting polymers and spheres and, further, we will approximate the total
self-energy tensor ${\bf\Sigma}(\rvec ,\rrvec)$ with the first term in the
expansion (\ref{sigpolsp}) that contains the contribution from {\sf single}
scattering events~: 
\beq 
{\bf\Sigma}(\rvec ,\rrvec) \simeq {\bf\Sigma}^{(1)}(\rvec ,\rrvec)
\label{sigapprox}
\eeq 
but, for the convenience of notation we will still denote it by
${\bf\Sigma}(\rvec ,\rrvec)$.  Actually, for the noninteracting situation, this
is the leading term as can be seen from Eq.\ (\ref{sigpolsp}) by breaking the
averages, with the contribution from higher order processes vanishing due to
the lack of correlations among particles.

As already pointed out, ${\bf\Sigma}(\kvec)$ contains the sought information
about the transport properties of the suspension.  To evaluate the self-energy,
first we take the configurational average (\ref{confav}) over the ${\bf T}$
operators in Eq.\ (\ref{topers}), then we Fourier transform their expressions,
obtaining~: 
\beqar 
\la{\bf T}_{\alpha}(\kvec)\ra & = & - {1\over V}\sum_{i,j\geq
1}^{n}\la\exp{(i\kvec\cdot(\Sind{i}-\Sind{j}))}\left[\Kinv{\alpha i,\alpha
j}\right.\right. \nonumber\\ 
& - & \left.\left. \sum_{l,l'\geq 1}^{n} \Kinv{\alpha i,\alpha l}\cdot{\bf
g}^{-1}_{t} \cdot \Kinv{\alpha l',\alpha j} \right] \ra_{i,j} \\ 
\la{\bf T}_b (\kvec)\ra & = & - {1\over V}\int\! d\Omega_b
\, d\Omega_{b}'\: \Kinv{b,b'}\, \exp{[i \kvec\cdot(\rvec_{b}-\rrvec_{b})]}
\label{tavepolsp}
\eeqar
The preaverage over ${\bf K}^{-1}$ part produces terms like
$\la\Kinv{ij}\ra_{ij}\la\exp{(i \kvec\cdot (\Sind{i}-
\Sind{j}))}\ra_{ij}$---with the {\it i,j} indices signifying an average over
the distribution of the segments of any chain---so ${\bf\Sigma}(\kvec)$ can be
written as~: 
\beqar
& & {\bf\Sigma}(\kvec) \simeq  {\bf\Sigma}^{(1)}(\kvec) =
-c^{\text{POL}}\sum_{i,j\geq 1}^{n}\la \exp{(i
\kvec\cdot(\Sind{i}-\Sind{j}))}\ra_{ij} \nonumber\\
& & \times \left[\la\Kinv{ij}\ra_{ij} - \la{\bf
g}^{-1}_{t}\ra_{ij}\cdot\sum_{l,l'\geq 1}^{n}\la\Kinv{il}\ra_{il}\cdot
\la{\Kinv{l'j}}\ra_{l'j}\right] \label{sigsol}\\
& & - c^{\text{SP}}\int\!d\Omega \, d\Omega '\,{\bf K}^{-1} (\Omega,\Omega ')
\exp{\left[i \kvec\cdot(\rvec (\Omega) - \rrvec (\Omega '))\right]}\nonumber
\eeqar
where $\rvec (\Omega)$ , $\rrvec (\Omega ')$ are two generic points on the
surface of any sphere and $c^{\text{POL}} = \frac{\text{N}_{P}}{V}$ ,
$c^{\text{SP}} = \frac{\text{N}_{S}}{V}$ are the concentrations of the polymers
and of the spheres, respectively.

\section{CALCULATION OF ${\bf\Sigma}(\kvec)$ AND EFFECTIVE MEDIUM THEORY FOR
SOLUTIONS OF SPHERES OR POLYMERS ONLY}
\label{sec:eff}

In this part of the paper, the problems of constructing the effective media in
the separate cases of a suspension of spheres and a polymer dispersion are
addressed.
 
\subsection{Calculation of ${\bf W}(\kvec)$ for $\text{N}_{S}\/$ spheres
immersed in a fluid : a review} 
\label{sec:w_NS}

This problem is well documented in the literature \cite{MF79,M82} so we will
quote the main results we need.  The system consists in $\text{N}_{S}\/$,
rigid, penetrable spheres of radius $a$, immersed in an
incompressible,newtonian fluid of viscosity $\eta_0$. All spheres but one,
which is moving at constant velocity and generates the flow, are fixed. The
suspension is assumed stationary.

A remark should be made regarding the notation. Because we want to distinguish
this particular system from the others ,we will denote the self-energy of the
fluid in the presence of the spheres only by ${\bf W}^{SP}$ instead of
${\bf\Sigma}^{SP}$.

Along the lines previously exposed,we can map the suspension averaged over the
uniform distribution of the spheres to an effective fluid of viscosity
$\eta^{\text{eff}}_{SP}$ where the disturbances are propagated by the modified
Oseen tensor ${\bbox{\cal G}}^{SP}(\kvec)$ ~:
\beq
{\bbox{\cal G}}^{SP}(\kvec) = \frac{\openone - \hat{\kvec}\hat{\kvec}}{\eta_0
k^2 - W^{SP}_{\perp}(\kvec) }
\label{oseensp}
\eeq
with the transverse part of the self-energy computed from Eq.\
(\ref{sigtrans}).

 	In order to solve self-consistently for ${\bf W}^{SP}(\kvec)$ we
expand its transverse part, in the small \kvec\ limit~:
\beq
W^{SP}_{\perp}(\kvec) = \left( - \eta_0 \xi^{-2}_{SP} - \eta_0 W_{1}^{SP}(\fsp)
k^2 \right)
\label{sigexp}
\eeq 
Then the modified Oseen propagator becomes~:
\beqar
{\bbox{\cal G}}^{SP}(\kvec) & = & \frac{\openone -
\hat{\kvec}\hat{\kvec}}{\eta_0(1 + W_{1}^{SP})(k^2 + {\xi '}^{-2}_{SP})}
\nonumber\\ 
& & {\xi '}^{-2}_{SP} = \frac{\xi^{-2}_{SP}}{1 + W_{1}^{SP}}
\label{gsolsp}
\eeqar
where ${W_{1}^{SP}}$ is some function of $\fsp$. All other symbols were already
defined.

${\bf W}^{SP}(\kvec)$  has the following form (see Appendix B)~:
\beqar
& & {\bf W}^{SP}(\kvec) = -6\pi\eta_0 a c^{SP}{1 \over
\bessi{1/2}\bessk{1/2}}(1 + W_1^{SP}) \openone + \nonumber\\ 
& & +\, {3\over 2}\eta_0\fsp (1 + W_1^{SP})k^2 \left[{1 \over
\bessi{1/2}\bessk{1/2}} \right. \nonumber\\
& & - \, \left. {8 \over 9\, \bessi{3/2}\bessk{3/2}} - {2 \over 5 \,
\bessi{1/2} \bessk{5/2}} \right] \openone \label{wSP}\\ 
& & +\, {3\over 2}\eta_0\fsp (1 + W_1^{SP})\, k^2 \left[ {4\over 27\,
\bessi{3/2}\bessk{3/2}} \right. \nonumber\\
& & + \, \left. {2\over 5\, \bessi{1/2}\bessk{5/2}}\right] \hat{{\bf
z}}\hat{{\bf z}} \nonumber 
\eeqar
where I and K are the modified Bessel functions of the first kind, $\fsp =
{4\pi a^3\over 3}c^{SP}$ is the volume fraction of spheres and $\hat{{\bf z}}$
is the versor of the Oz direction.  Projecting out the transverse part of ${\bf
W}^{SP}(\kvec)$ by dotting in with $(\openone - \hat{{\bf z}}\hat{{\bf z}})$
and comparing the result with the expansion (\ref{sigexp}), one arrives at the
following set of equations for $x = {a\over\xi '}$ and $W_1^{SP}$~: 
\beqar 
& x^2 & = {9\over 2}\fsp {1\over {\rm I}_{1/2}(x){\rm K}_{1/2}(x)} \nonumber\\
& W_1^{SP}& (\fsp) =  \left\{ 1 + \frac{3}{2} \fsp\left[{1\over
\bessidoi{1/2}{x}\besskdoi{1/2}{x}} \right.\right. \label{sysSP}\\
& & - \, \left.\left. {8\over 9 \,\bessidoi{3/2}{x}\besskdoi{3/2}{x}} - {2\over
5 \,\bessidoi{1/2}{x}\besskdoi{5/2}{x}}\right]\right\}^{-1} \, - \, 1 \nonumber
\eeqar

These equations can be solved numerically and $x(\fsp)$ and $W_1^{SP}(\fsp)$
calculated for any values of $0\leq\fsp<1$,thus leading to the required ${\bf
W}^{SP}(\kvec)$.  As $\fsp\to 0$, $W_1^{SP}$ decreases to zero and also, from
the structure of Eq.\ (\ref{sysSP}), one expects a divergent behavior of
$W_1^{SP}$ as $\fsp$ increases beyond a certain limit. The effective viscosity
$\eta_{\text{SP}}$ of the medium containing the spheres and the translational
friction coefficient $\bbox{\zeta}_t$ of the moving sphere are obtained from~:
\beqar 
& & {\eta_{\text{SP}} - \eta_0\over\eta_0 } = W_1^{SP}(\fsp) \nonumber\\ 
& & \bbox{\zeta}_t = \zeta_t\, \openone = - {1\over c^{SP}}
{\bf W}^{SP}(k=0) \; = \; 6 \pi a \eta_0 \label{viscfrSP}\\
& & \; \times \,  {1\over \bessidoi{1/2}{x(\fsp)}\besskdoi{1/2}{x(\fsp)}} (1 +
W_1^{SP}(\fsp))\, \openone\nonumber 
\eeqar

One can derive the above expression for the friction coefficient by applying
the general analysis from Appendix~A to our particular case of a suspension of
$N-1$ fixed and one uniformly moving spheres . Then, the relation between
the total average force exerted upon the mobile sphere by the fluid and its
velocity will be given by Eq.\ (\ref{forces}) without the polymer term~:
\beqar
\la \int\! d\Omega\, \bbox{\sigma}(\Omega)\ra & = & \la\int\! d\Omega\,d\Omega
'\: {\bbox{\cal T}}(\Omega,\Omega ')\ra\cdot {\bf u} = \\
& & \la\int\! d\Omega\,d\Omega '\: {\bbox{\cal K}}^{-1}(\Omega,\Omega ')\ra
\cdot{\bf u} = \bbox{\zeta}_t\cdot {\bf u} \nonumber
\eeqar

Note that now ${\bbox{\cal K}}^{-1}$ is the generalized inverse of
${\bbox{\cal G}}$, to account for the presence of the fixed spheres in the
solution.  Comparing then to the sphere contribution to the self-energy from
Eq.\ (\ref{sigsol}) and making use of (\ref{Bsigzero},\ref{BKinvzero}) one can
recover the formula for $\bbox{\zeta}_t$ displayed in (\ref{viscfrSP}).  When
the filling fraction of the {\sf background } spheres approaches
zero, $W_1^{SP}\to\,0$ and one recovers the Stokes result for the friction
coefficient ${\bf\zeta_t}$ of one sphere.

As stated before, the readily measurable quantity we calculated is the relative
diffusion coefficient of the mobile sphere (relative to the bare diffusion
constant), introduced in (\ref{diffdef}).  From (\ref{viscfrSP}) we get
immediately~:
\beqar
{\bf D}^{SP} \,=\, D^{SP} \openone \,& = & \,
\bessidoi{1/2}{x(\fsp)}\besskdoi{1/2}{x(\fsp)} \\
& \times & {1\over 1 + W_1^{SP}(\fsp)}\:\openone\nonumber
\eeqar
in which $x(\fsp)$ and $W_1^{SP}(\fsp)$ are calculated from (\ref{sysSP}) for a
given fraction $\fsp$.

\subsection{Calculation of the self-energy for a solution of 
$\text{N}_P\/$ noninteracting polymer chains} 
\label{sec:w_NP}

We consider $\text{N}_P\/$ Gaussian, noninteracting, free moving polymer
chains, each of length ${L=nl}$, dispersed in an incompressible, newtonian
fluid of viscosity $\eta_0$. Some external force {\bf F}(\rvec) generates the
velocity flow.

Once again, we will use the notation ${\bf W}$ for the self-energy of the
fluid when only the polymers are present.

Following the derivation in Sec.\ \ref{sec:twoB}, we can write the
Navier-Stokes equation for the velocity field ${\bf v}(\rvec)$ similar to
(\ref{nspolsp})---but no spheres present---with the stick boundary conditions
(\ref{nsbound}). Averaging the N-S equation over the configuration of the
polymer chains and introducing the self-energy of the fluid ${\bf W}$ as in
(\ref{sigdefpolsp}), we can write the solution for the averaged velocity ${\bf
u}(\rvec) = {\bbox{\cal G}}^{POL}\ast{\bf F}$ (see Eqs.~(\ref{nseff},
\ref{oseenk},\ref{uk})), where the force propagator ${\bbox{\cal G}}^{POL}$ is
given by~:
\beqar 
{\bbox{\cal G}}^{POL} & = & {\openone - \hat{\kvec}\hat{\kvec} \over \eta_0 k^2
- W_{\perp}^{POL}(\kvec)} = \nonumber\\
& & {\openone - \hat{\kvec}\hat{\kvec} \over \eta_0 (1
+ W_1^{POL})(k^2 + {\xi_{POL} '}^{-2})} \label{oseenpol}\\ 
{\xi_{POL} '}^{-2} & = & {\xi_{POL}^{-2} \over 1 + W_1^{POL}}
\nonumber
\eeqar
where, in the limit of $k\to 0$, the following approximation was made for the
transverse part of the self-energy~:
\beq
W_{\perp}^{POL}(\kvec) = - \eta_0 \xi_{POL}^{-2} - \eta_0 W_1^{POL}(\fpol) k^2
\label{wpoltrans}
\eeq

Our goal is to find ${\bf W}^{POL}(\kvec)$ self-consistently, for any volume
fraction $\fpol$.  If ${\bf W}^{POL}(\kvec)$ were the self-energy of the fluid
containing $\text{N}_{P}-1$ chains, the contribution of one more chain added
would be (similar to (\ref{sigsol}),but with $c^{POL}= 1/V$ and no spheres)~:
\beqar
& & {{\bf W}^{POL}(\kvec) \over \text{N}_{P}-1}   = - {1\over V}\sum_{i,j\geq
1}^{n}\la \exp[i \kvec\cdot(\Sind{i}-\Sind{j})]\ra_{ij} \nonumber\\
& & \left[\la\calKinv{ij}\ra_{ij} - \la{\bf g}_t^{-1}\ra_{ij}\cdot
\sum_{l,l'\geq 1}^{n} \la\calKinv{il}\ra_{il} \cdot\la\calKinv{l'j}\ra_{l'j}
\right] \label{wdefPOL}\\  
& & {\bf g}_t = \sum_{ij}^{n}\calKinv{ij}\;;\;{\bf g}^{-1}_{t}\cdot
{\bf g}_{t} = \openone\; . \nonumber
\eeqar
Here, in contrast to (\ref{sigsol}), $\bbox{\cal K}^{-1}$ signifies the
generalized inverse of the modified Oseen tensor $\bbox{\cal G}$ because we
imagined the last chain immersed in the effective medium created by the initial
pure fluid plus $\text{N}_{P}-1$ polymer chains and the force propagator should
be changed correspondingly.

The calculations are rather laborious and are detailed in Appendix~C.  In the
hydrodynamic limit of long wavelengths, the expression of ${\bf
W}^{POL}(\kvec)$ is~:
\beqar
{\bf W}^{POL}(\kvec) & = & - {9\over\pi}\eta_0 k^2 \fpol (1 + W_1^{POL})
\text{Q}(\beta)\openone \nonumber\\ 
\text{Q}(\beta) & = & {1\over\beta}\log\left(1 + {\beta\over\sqrt{\pi}}\right)
+ {1\over \sqrt{\pi}} + {\beta \over 2 \pi} \label{sigPOL} 
\eeqar 
Symbols have the following meanings: ${\fpol= {4\pi R_g^3 \over 3}} c^{POL}$ is
the volume fraction of the polymers ; $R_g = \sqrt{{Ll\over 6}}$ is the radius
of gyration ; $\beta(\fpol) = {R_g \over \xi '(\fpol)}$ with $\xi '$ the
screening length in the effective medium, given by (\ref{oseenpol}).  In
obtaining the previous relation we used the Kirkwood-Riseman approximation that
amounts to consider only the diagonal terms in the Fourier expansion of
$\bbox{\cal G}^{-1}(s,s')$ ($s$ is the arclength along the polymer chain), as
detailed in Appendix~C.  We believe this approximation is physically
justifiable because we are working in the limit of long chains, when the
off-diagonal terms are small.

To get the self-consistent equations for the unknowns $\beta(\fpol)$ and
$W_1^{POL}(\fpol)$, we take the transverse part of ${\bf W}^{POL}(\kvec)$ by
dotting in $(\openone - \hat{\kvec}\hat{\kvec})$ and then we equate it with the
expansion (\ref{wpoltrans}) to arrive at~:
\beqar
\beta(\fpol) & = & 0 \nonumber\\
W_1^{POL}(\fpol) & = & {1 \over 1 - {9\over\pi} \fpol \text{Q}(0)} - 1
\label{sysPOL} 
\eeqar
which gives $W_1^{POL}$ as an analytic function of the volume fraction
$\fpol$. Also it is interesting to remark that $\beta (= {R_g\over a}) = 0$
implies that the hydrodynamic screening is absent in the limit of $k\to 0$
(observation that is actually manifest from the structure of
(\ref{sigPOL})). The function $\text{Q}(\beta)$ is defined in
Eq.~(\ref{sigPOL}).

We can calculate the viscosity of the effective medium $\eta_{\text{POL}}$
(using (\ref{viscchange})) and the diffusion coefficient $D^{POL}$ of a polymer
chain as follows~:
\beqar
{\eta_{\text{POL}}(\fpol) - \eta_0 \over \eta_0} & = & W_1^{POL} \nonumber\\
D^{POL}(\fpol) & = & {\zeta_t^{KR} \over \zeta_t^{POL}(\fpol)}
\label{viscdiffPOL}\\  
& = & {1 \over 1 + W_1^{POL}} = 1 \, - \, {18 \over \pi\sqrt{\pi}}\fpol
\nonumber   
\eeqar 
where we employed ${\text{Q}(0) = {2\over \sqrt{\pi}}}$ and ${\zeta_t^{KR}=(9
\pi\sqrt{\pi}/4)} \eta_0 R_g$ is the Kirkwood-Riseman friction coefficient for
the non-free-draining limit.

Limiting values for $D^{POL}$ and $\eta_{\text{POL}}$ can be obtained for small
and large polymer filling fractions $\fpol$~:
\beqar
\fpol \, & \ll &\, 1 :  \nonumber\\
& & {\eta_{\text{POL}} - \eta_0 \over\eta_0}\,\to\, {24\over \sqrt{\pi}} \,
R_g^{3} \, {1\over V} \nonumber\\
& & D^{POL}\,\to\,1 \\
\fpol (\text{large}) \,& < & \,1 : \nonumber\\
& & {\eta_{\text{POL}} - \eta_0 \over \eta_0 }\,\to\, \infty \;\; \text{as}\;\;
\fpol\,\to\,0.309\dots \nonumber\\ & & D^{POL} \,\to\,0 
\eeqar

We remark that for low $\fpol$ the result for the viscosity is twice that of
Kirkwood-Riseman \protect\cite{KKR} for one chain because we neglect the
rotational terms in calculating ${\bf W}^{POL}(\kvec)$.  It is also notable
that the constraint of the positivity of $W_1^{POL}$ in this effective medium
approach leads to the divergence of the viscosity and of the friction
coefficient,which occurs at $\fpol_{DIV} \simeq 0.309$.  This divergence is
attributed to the rigid-body dynamics assumed for the polymer.

\section{Results for Specific Examples}
\label{sec:res_spec}

In this section we apply the above theory to two types of mixed
polymers-spheres suspensions~:
\begin{description}
\item[S/P] system~: one mobile sphere among other fixed, rigid ones, in a
polymer solution.
\item[P/S] system~: free moving polymer chains in a suspension of fixed
spheres.
\end{description}

In addition, to gradually develop the theory we start by explicitly
investigating two particular cases of the general models just mentioned,
namely~:
\begin{description}
\item[1S/P] system~: one mobile sphere inside a polymer solution.
\item[1P/S] system~: one polymer chain moving in a suspension of fixed spheres.
\end{description}

\subsection{One sphere moving in a suspension of polymers}
\label{sec:res_1SP}

In this example we consider a sphere of radius $a$ moving uniformly with
the velocity ${\bf v}_0$ inside a suspension of $\text{N}_P\/$
non-interacting, Gaussian, free moving polymer chains of length $L$.  The
flow of the fluid is generated by the moving sphere.  The motion of the
particles is coupled to the motion of the solvent by stick boundary
conditions. We also mention that the total force and torque acting upon each
polymer chain are zero.

Similar to Eq.\ (\ref{nspolsp}), the equation of motion for the fluid and the
boundary conditions can be written as~:
\beq
-\eta_0 \triangle {\bf v}(\rvec) + \nabla p(\rvec) = {\bf F}_0(\rvec) +
\sum_{\alpha,i}^{NP,n}\delta(\rvec - \Rind{\alpha i} ) \bbox{\sigma}_{\alpha i}
\nonumber 
\eeq 
\beqar 
\nabla \cdot {\bf v}(\rvec) & = & 0 \nonumber\\
{\dot{{\bf R}}}_{\alpha i} & = & {\bf v}({\bf R}_{\alpha i}) \eeqar in which
${\bf F}_0(\rvec)$ is the force exerted by the mobile sphere upon the fluid
given by~: \beq {\bf F}_0(\rvec) = \int\! d\Omega_0 \; \delta(\rvec - {\bf
R}_0)\bbox{\sigma}(\Omega_0) 
\eeq
 
Averaging the Navier-Stokes equation over the distribution of the polymer
chains and introducing the self-energy of the solution~:
\beq
{\bf\Sigma}(\rvec,\rrvec) = {\bf W}^{POL}(\rvec,\rrvec) = \la\sum_{\alpha
,i\geq 1}^{NP,n}\delta(\rvec - \Rind{\alpha i})\bbox{\sigma}_{\alpha i} \ra
\eeq 
we find the formal solution (following Eqs.\ (\ref{nseff},\ref{uk}))~:
\beqar 
{\bf v}(\rvec) & = & \left. {\bbox{\cal G}}\ast {\bf F}_0 \right|_{\rvec}
\nonumber\\ 
{\bbox{\cal G}}(\kvec) & = & {\openone -\hat{\kvec}\hat{\kvec}
\over \eta_0 k^2 - W_{\perp}^{POL}(\kvec)} 
\eeqar 
and we succeeded in transforming the initial problem in the Stokes motion of a
single sphere in a fluid of viscosity $\eta_{\text{POL}}$.  Because only the
polymers are contributing to the effective viscosity of the solution,
$\eta^{\text{eff}}$ is given by (\ref{viscchange},\ref{viscdiffPOL})~: 
\beq
{\eta^{\text{eff}} - \eta_0 \over \eta_0} = W_1^{POL} = {1 \over 1 - {18 \over
\pi\sqrt{\pi}} \fpol } - 1
\label{visc1spol} 
\eeq

As it was shown in the previous section---Eq.~(\ref{sysPOL})---the hydrodynamic
screening is absent for a suspension of polymers.  In this problem,the sphere
does not contribute to the self-energy of the fluid, which implies that the
hydrodynamic interactions are still unscreened.  Then the translational
friction coefficient of the moving sphere is just the Stokes result~:
\beq
\bbox{\zeta}_t = 6 \pi a  \eta^{\text{eff}}\: \openone
\eeq
with $\eta^{\text{eff}} = \eta_0 (1 + W_1^{POL})$.  The translational diffusion
coefficient as defined in (\ref{diffdef}) reads then~:
\beq
D(\fpol) = {6 \pi a \eta_0 \over \zeta_t} = 1 - {18 \over \pi\sqrt{\pi}} \fpol
\label{diff1spol}
\eeq

It is noteworthy that in this case there is no {\sf size coupling} (no $t =
{R_g\over a}$ dependence) between the dynamics of the polymer chains and the
dynamics of the moving sphere.  The diffusion coefficient vanishes and the
viscosity diverges at $\fpol \simeq 0.31$.

\subsection{One polymer chain immersed in a suspension of fixed spheres}
\label{sec:res_1PS}

This time we consider a free polymer chain of length $L$, immersed in a
suspension of $\text{N}_{S}\/$ rigid, fixed, uniformly distributed spheres of
radius $a$.  There are no interactions other than hydrodynamic and the flow is
generated by some external force field {\bf F}(\rvec).

The velocity field is described by another variant of N-S
Eq.\ (\ref{nspolsp}), with only one chain present~:
\beqar
-\eta_0 \triangle {\bf v}(\rvec) + \nabla p(\rvec) & = & {\bf F}(\rvec) +
\sum_{i=1}^{n} \delta(\rvec - \Rind{i})\bbox{\sigma}_{i} \nonumber\\ 
& & + \sum_{b=1}^{\text{N}_{S}}\int \! d\Omega_b\, \delta(\rvec -
\Rind{b})\bbox{\sigma}_b 
\eeqar 
where all the symbols have the known meanings.

Taking the configurational average of the equation above and introducing the
self-energy tensors ${\bf\Sigma}^{POL}(\rvec,\rrvec)$ and ${\bf
W}^{SP}(\rvec,\rrvec)$ (see (\ref{sigdefpolsp})) related to the influence of
the polymer chain and of the spheres on the fluid, we can cast the
Navier-Stokes equation for the averaged velocity ${\bf u}(\rvec) = \la{\bf
v}(\rvec)\ra$ in the following form~: 
\beq 
-\eta_0 \triangle {\bf u}(\rvec) + \la\nabla p(\rvec)\ra - \left. {\bf
W}^{SP}\ast {\bf u}\right|_{\rvec} = {\bf F}(\rvec) + \left. {\bf\Sigma}^{POL}
\ast {\bf u}\right|_{\rvec} 
\eeq 
the formal solution of which being (see Section~II)~: 
\beq 
{\bf u}(\rvec) = \left. {\bbox{\cal G}}\ast {\bf F} \right|_{\rvec} + \left.
{\bbox{\cal G}}\ast {\bf\Sigma}^{POL} \ast {\bbox{\cal G}}\ast {\bf
F}\right|_{\rvec} + \dots 
\eeq 

Here $\ast$ is the usual convolution operator and ${\bbox{\cal
G}}(\rvec,\rrvec)$ is the effective Oseen tensor modified to account for the
influence of the spheres. Its expression is given by (\ref{gsolsp}) using the
expansion for small $k\/$ (\ref{sigexp}).  In this way our starting problem was
reduced to studying the stationary dynamics of a polymer chain in an effective
solution where the force propagator is ${\bbox{\cal G}}(\kvec)$.  Following the
results of Sec.~\ref{sec:twoB}---Eqs.~(\ref{sigapprox}, \ref{sigpolsp},
\ref{tavepolsp}, \ref{sigsol})---and noticing that we must make the changes
$c^{POL} = {1\over V}$, ${\bf G} \to {\bbox{\cal G}}$ and ${\bbox{K}}^{-1} \to
{\bbox{\cal K}}^{-1}$, we find the solution of ${\bf\Sigma}^{POL}(\kvec)$ in
first order in the number of {\sf scattering events}~:

\beqar 
{\bf\Sigma}^{POL}(\kvec) & = & - {1 \over V} \sum_{i,j\geq 1}^{n}\la\exp\left[
i \kvec\cdot (\Sind{i} - \Sind{j})\right]\ra_{ij}
\left[\la\calKinv{ij}\ra_{ij}\right. \nonumber\\
& & - \left. \la{\bf g}^{-1}_t\ra_{ij}\cdot \sum_{l,l'}^{n}\la
\calKinv{il}\ra_{il}\cdot \la \calKinv{l'j}\ra_{l'j} \right] 
\eeqar 
Note that ${\bbox{\cal K}}^{-1}$ is the generalized inverse of ${\bbox{\cal
G}}$~: 
\beq 
\sum_{j=1}^{n}\calKinv{ij}\cdot \calGind{jk} = \openone \delta_{ik}
\eeq

Remembering that the effective medium has replaced the spheres and the pure
fluid, we can repeat the derivation from Appendix~C to get an explicit form
for ${\bf\Sigma}^{POL}(\kvec)$, with the changes $W_1^{POL}\to W_1^{SP}$ and
$c^{POL} = {1\over V}$~:
\beqar
{\bf\Sigma}^{POL}(\kvec) = & - & {9 \over \pi}\eta_0 {4\pi R_g^3 \over
3}{1\over V} (1 + W_1^{SP}) \text{Q}(\beta_{SP}) k^2\, \openone \\
& & \beta_{SP} = {R_g \over \xi_{SP} '} \nonumber
\eeqar
where $\xi_{SP}'$ and $W_1^{SP}$ were defined in Sec.~\ref{sec:w_NS}.

To evidentiate the important dimensionless variables of this model, we 
introduce the parameters $t = {R_g \over a}$ and $x_{SP} = {a \over \xi_{SP}'}$
and rewrite ${\bf\Sigma}^{POL}(\kvec)$ as~:
\beq
{\bf\Sigma}^{POL}(\kvec) = - {9\over\pi} \eta_0 \,{\fsp\over \text{N}_{S}}\,
t^3 [1 + W_1^{SP}(\fsp)] \text{Q}(t x_{SP}) \, k^2 \,\openone 
\eeq 
Both $x_{SP}(\fsp)$ and $W_1^{SP}(\fsp)$ are characteristic to the problem of
the spheres immersed in a fluid and can be calculated from the system
(\ref{sysSP}). 
 
In a similar manner,the diffusion coefficient of the polymer chain in the
effective medium of spheres can be expressed as~:
\beq
D(\fsp) = {\zeta_t^{KR}\over \zeta_t^{POL}(\fsp)} = {3\sqrt{\pi}\over 4}
{1 \over 1 + W_1^{SP}} {\text{P}(t x_{SP})\over t x_{SP}} 
\label{diff1polsp}
\eeq
in which we used (\ref{frcoeffAppC}), with $W_1^{POL} \to W_1^{SP}$ and $\beta
\to \beta_{SP}$.  The function $\text{P}(x)$ is defined in Appendix~C
(\ref{gpAppC}).

The  effective viscosities, total and relative (which measures the
increase in the viscosity due to the polymer chain) are calculated from~:
\begin{mathletters}
\label{visc1polsp}
\beqar
{\eta^{\text{eff}} - \eta_0 \over \eta_0} & = & W_1^{SP} - \lim_{k\to 0}{1\over
\eta_0}{\partial \over \partial k^2}\Sigma_{\perp}^{POL}(\kvec) \nonumber\\
& = & W_1^{SP} + {9\over\pi} {\fsp\over \text{N}_{S}} t^3 (1 + W_1^{SP})
\text{Q}(t x_{SP}) \\ 
{\eta^{\text{eff}} - \eta_{\text{SP}}\over \eta_{\text{SP}}} & = & -\lim_{k\to
0}{1 \over \eta_{\text{SP}}}{\partial\over \partial k^2}
\Sigma_{\perp}^{POL}(\kvec) \nonumber\\
& = & {9\over\pi} {\fsp\over \text{N}_{S}} t^3 \text{Q}(t x_{SP})
\eeqar
\end{mathletters}
with $\eta_{\text{SP}} = \eta_0 (1 + W_1^{SP})$, and $x_{SP}(\fsp)$,
$W_1^{SP}(\fsp)$ obtained from (\ref{sysSP}),as already mentioned.

As will be shown below, Eqs.~(\ref{diff1polsp},\ref{visc1polsp}) are the
limiting formulas of the general situation presented in
Section~\ref{sec:res_PS}, where the dependence of $D\/$ and $\eta^{\text{eff}}$
on $\fsp$ and $t$ will be discussed also.

However, some remarks can be made. The arrest of the chain motion occurs at
$\fsp_{DIV} \simeq 0.49$---as found in Ref.~\protect\cite{M82} for a sphere
moving in a random array of fixed spheres.  The curves seem physically
plausible.  When the radius of gyration $R_g$ of the polymer chain is smaller
than or comparable to the radius of the sphere $(t \lesssim 1)$ the viscosities
(both total and relative) are relatively insensitive to variations in the $t$
parameter. Still, although being related only to the presence of the polymer
chain, the relative viscosity depends on $\fsp$ and $t$ variables through
$\text{Q}(t~ x_{SP})$ (see \ref{visc1polsp}), so the contribution of the chain
to the viscosity of the solution is a function of the effective solvent.

\subsection{One mobile sphere in a suspension of fixed spheres and polymers}
\label{sec:res_SP}

We now investigate the stationary dynamics of a suspension containing
$\text{N}_{S}\/$ uniformly distributed, rigid spheres of radius $a$ and
$\text{N}_P\/$ Gaussian, free moving polymer chains of length $L\/$. All
spheres but one that is moving with the constant velocity {\bf u$_0$},thus
creating the flow, are fixed, and no interactions beside the hydrodynamic
coupling occur.  We calculate the diffusion coefficient $D\/$ of the moving
sphere and the total $\eta^{\text{eff}}$ and the relative $\vrelspol$ effective
viscosities as functions of the volume fractions $\fsp$ for spheres,$\fpol$ for
polymers and of the coupling parameter $t = {R_g\over a}$.

We proceed closely to the derivations exposed in the previous sections.  First,
by properly identifying the ${\bbox{\cal G}}(\kvec)$ tensor, we {\sf construct}
an effective medium replacing the polymers and the pure fluid,in this way
reducing the problem to that of a self-consistent computation of the viscosity
and diffusion coefficient of an ensemble of spheres, as discussed in
Refs.~\cite{M82,FM82,MF79} and in Sec.~\ref{sec:w_NS}.  Solving the resulting
equations, we obtain the desired quantities,namely the diffusion coefficient of
the moving sphere and the effective viscosities.
 
The velocity field obeys the N-S Eq.\ (\ref{nspolsp}), with ${\bf F}(\rvec)\to
{\bf F}_0(\rvec)$ now being the force exerted by the moving sphere upon the
fluid. Averaging first over the distribution of the polymers and then over the
position of the fixed spheres and introducing the self-energies ${\bf
W}^{POL}(\rvec,\rrvec)$ and ${\bf\Sigma}^{SP}(\rvec,\rrvec)$ as defined in
(\ref{sigdefpolsp}), we deduce the N-S equation for the averaged velocity ${\bf
u}(\rvec) = \la{\bf v}(\rvec)\ra$~: 
\beq 
-\eta_0 \triangle {\bf u}(\rvec) + \la \nabla p(\rvec)\ra - {\bf W}^{POL}\ast
{\bf u} - {\bf\Sigma}^{SP}\ast {\bf u} = {\bf F}_0(\rvec) 
\eeq 
As usual, we denote the self-energy of the fluid resulting from the presence of
the polymers with the ${\bf W}^{POL}$ symbol and not with ${\bf\Sigma}^{POL}$
to stress that this is the background effective medium we have related our
calculations to.  This also means that ${\bf W}^{POL}$ depends only upon the
polymer properties, but ${\bf\Sigma}^{SP}$---the contribution of the spheres
to the self-energy of the fluid---could depend on $\fpol$ and on some coupling
parameter,like $t$ and,implicitly,on other characteristics of the background
effective fluid.  For ${\bf W}^{POL}$ we will use the explicit expressions
found in Sec.~\ref{sec:w_NP} (Eqs.~\ref{sigPOL},\ref{sysPOL}).

The formal solution of the N-S equation is then~:
\beq
{\bf u}(\rvec) = \left. {\bbox{\cal G}}\ast {\bf F}\right|_{\rvec}
\eeq 
with the modified Oseen tensor having the following form~:
\beq
{\bbox{\cal G}}(\kvec) = {\openone - \hat{\kvec}\hat{\kvec} \over \eta_0 k^2 -
W_{\perp}^{POL}(\kvec) - \Sigma_{\perp}^{SP}(\kvec)} 
\eeq 
Working in the $k\to 0$ limit, we approximate ${\bf\Sigma}^{SP}(\kvec)$ with~:
\beq
\label{sigspol}
{\bf\Sigma}^{SP}(\kvec) = (-\eta_0 \xi_{SP}^{-2} - \eta_0 \Sigma_1^{SP}
k^2)\,\openone 
\eeq 
where $\xi_{SP}$ is the screening length felt by the spheres in the effective
medium that supplants for the polymers and the pure fluid.  Substituting back
in the Oseen tensor formula and replacing $W_{\perp}^{POL}$ with its expression
from (\ref{wpoltrans},\ref{sysPOL}) we get~: 
\beq 
{\bbox{\cal G}}(\kvec) = {\openone - \hat{\kvec}\hat{\kvec} \over \eta_0 (1 +
\Sigma_1)(k^2 + {\xi '}^{-2})} = {\openone - \hat{\kvec}\hat{\kvec} \over
\eta^{\text{eff}} (k^2 + {\xi '}^{-2})} 
\eeq 
in which~: 
\beqar 
\Sigma_1 & = & W_1^{POL} + \Sigma_1^{SP} \nonumber\\ {\xi '}^{-2} & = & {
\xi^{-2}_{SP} \over 1 + \Sigma_1}
\label{sigdefspol} 
\eeqar 
where $\xi '$ is the total screening length in the resulting effective
medium that replaced the initial pure fluid, the polymers and the spheres. Note
that the polymers do not contribute to the screening, as found in
Sec.~\ref{sec:w_NP}.

It follows then that the total and the relative (due to the presence of spheres
only) effective viscosities of the solution and the translational friction
coefficient of the moving sphere (see Sec.~\ref{sec:w_NS}) can be computed
from~:
\beqar
\veff & = & W_1^{POL} + \Sigma_1^{SP} \nonumber\\ 
{\eta^{\text{eff}}\over \eta_{\text{POL}}} - 1 & = & - \lim_{k\to
0}{1\over\eta_{\text{POL}}} {\partial \over \partial
k^2}\Sigma_{\perp}^{SP}(\kvec) = {1\over 1 + W_1^{POL}}\Sigma_1^{SP}
\nonumber\\
\zeta_t & = & -{1\over c^{SP}} \Sigma_{\perp}^{SP}(k = 0) = \eta_0 {1\over
c^{SP}} \xi_{SP}^{-2} \label{viscfrspol}
\eeqar

Next step is to find ${\bf\Sigma}^{SP}(\kvec)$ self-consistently. This can be
done exactly as shown in Appendix~B,but taking care of the effective medium
that replaces the polymers, by using the new ${\bbox{\cal G}}$, substituting
$\Sigma_1$ for $W_1^{SP}$ and changing the meaning of $\xi '$, which will now
represent the {\sf total} screening length. Thus~:
\beqar
& & {\bf\Sigma}^{SP}(\kvec) = - 6 \pi a \eta_0 \, c^{SP}\, (1 + \Sigma_1)
{1\over \bessi{1/2}\bessk{1/2}}\,\openone \nonumber\\ 
& & \;\; + \; {3\over 2}\eta_0 (1 + \Sigma_1) \, \fsp k^2 \, \text{Z}({a
\over\xi '})\,\openone + \; (\text{longit. part})\, \hat{\bf z}\hat{\bf z} \\  
& & \text{Z}(y) =  {1\over \bessidoi{1/2}{y}\besskdoi{1/2}{y}} - {8\over
9\,\bessidoi{3/2}{y}\besskdoi{3/2}{y}} - {2\over
5\,\bessidoi{1/2}{y}\besskdoi{5/2}{y}} \nonumber
\eeqar

Comparing this expression with (\ref{sigspol}) and using the definitions of
$\Sigma_1$ and $\xi '$ from (\ref{sigdefspol}) and of $W^{POL}_1$ from
(\ref{sysPOL}), we finally obtain the system of equations for
${\bf\Sigma}^{SP}(\kvec)$~:
\beqar
\Sigma_1^{SP}(\fsp,\fpol) & = & {1\over 1 - {18\over
\pi\sqrt{\pi}}\fpol}\left[{1 \over 1 + \frac{3}{2} \fsp \text{Z}(x)} - 1
\right] \nonumber\\ 
x^2 & = & {9\over 2}\fsp {1 \over \bessidoi{1/2}{x}\besskdoi{1/2}{x}}
\label{sysspol}  
\eeqar 
with $ {x = x(\fsp) = {a \over\xi '}}$ and the function $\text{Z}(x)$ being
defined in the previous equation.  One could remark that there is no polymer
dependence in the total screening length $\xi '$.

Solving this system for $x(\fsp)$ and $\Sigma_1^{SP}(\fsp,\fpol)$ allows us to
compute the diffusion coefficient of the moving sphere
from~(\ref{viscfrspol})~:
\beqar
D(\fsp,\fpol) & = &  {6 \pi a \eta_0 \over \zeta_t } =
\bessidoi{1/2}{x}\besskdoi{1/2}{x} \label{diffspol}\\
& & \times \, (1 - {18 \over\pi\sqrt{\pi}}\fpol) (1 + {3
\over 2}\fsp \text{Z}(x)) \nonumber 
\eeqar 
and the total and relative effective viscosities as~:
\beqar
\veff \, & = & \, {1 \over (1 - {18\over\pi\sqrt{\pi}} \fpol)[1 + {3\over
2}\fsp \text{Z}(x)]} - 1 \label{viscspol}\\ 
{\eta^{\text{eff}}\over \eta_{\text{POL}}} - 1 \, & = & \, {1\over 1 +
W_1^{POL}} \Sigma_1^{SP} = {1 \over 1 + {3\over 2}\fsp \text{Z}(x)} - 1
\nonumber 
\eeqar

In Figs.~[\ref{dSP1SP_P},\ref{dSP_S}]; [\ref{vT_SP1SP_P},\ref{vT_SP_S}];
\ref{six} the results for $D\/$, the total $\vefftext$ and the relative
${\vrelspoltext}$ viscosities have been plotted against $\fpol$ (first plot of
each pair and Fig.~\ref{six}) and against $\fsp$ (in the second plot).

We have already seen that the total screening length $\xi '$ as given by
(\ref{sysspol}) is function only on $\fsp$ (as the polymers were absent).  It
is also notable that there is no dependence upon the coupling parameter $t =
{R_g \over a}$.  Both observations are related to the absence of screening in a
dispersion of non-interacting polymers (as found in Sec.~\ref{sec:w_NP}), which
prevents the spheres to {\sf sense} the existence of the polymer chains.

Two important points, at $\fsp_{DIV} \simeq 0.49$ and at $\fpol_{DIV}\simeq
0.31$, where the viscosity of the solution and the friction coefficient
diverge, are evident in Figs.~\ref{dSP1SP_P}--\ref{six}.  Note that $\fsp_{DIV}
+ \fpol_{DIV} < 1$, even if the polymer chains and the spheres are penetrable,
non-interacting objects.

At low $\fsp$ and $\fpol\to 0$ one recovers the Einstein result for the
viscosity of spheres in a pure fluid and the Stokes friction coefficient for a
sphere ($D(\fsp,0)\to 1$).

When $\fsp\to 0$ (recall $\fsp = {\text{N}_{S} - 1\over V}$), we retrieve the
results from the last section for one mobile sphere immersed in a polymer
solution (compare Eqs.\ (\ref{diffspol},\ref{viscspol}) to
Eqs.~(\ref{diff1spol},\ref{visc1spol})). The results of this limit are marked
with the $\bbox{+}$ symbol in Figs.~\ref{dSP1SP_P} and \ref{vT_SP1SP_P}.

\subsection{ $\text{N}_P\/$ polymer chains moving in a suspension of 
$\text{N}_{S}\/$ randomly distributed fixed spheres }
\label{sec:res_PS}

In this section we explore the reverse of the previous model.  There are
$\text{N}_P$ non-interacting, free-moving polymer chains of length $L$
dispersed in a suspension containing a random array of $\text{N}_{S}$
rigid, penetrable, fixed spheres.  We wish to compute---using the effective
medium approach---the translational diffusion coefficient $D$ for a polymer
chain and the total and relative (the spheres contribution) effective
viscosities as function of $\fpol$, $\fsp$ and of the coupling parameter $t =
R_g/a$.  The velocity flow is externally generated (one could consider it to be
produced by the motion of one polymer chain without any influence on the final
results).

Once again, we develop the necessary theory in two stages. First, we substitute
the spheres and the pure fluid with the corresponding effective medium
presented in Sec.~\ref{sec:w_NS}, then we calculate self-consistenty the
transport properties of the polymers dispersed in the {\sf effective} solution,
for any volume fraction $\fpol$, following the procedure detailed in
Sec.~\ref{sec:w_NP}.  Conceptually, we are modeling this system with a
superposition of two effective media, a {\sf fixed} background related to the
influence of the spheres and a {\sf foreground} related to the polymers,
coupled through hydrodynamic interactions.

Apparently, there should be a symmetry between the present system and the
reversed one (spheres in a solution of polymers), but this assumption 
proves to be incorrect, as we will further show.  The main difference is that
in the present case the {\sf size coupling } is active and $t = {R_g \over
a}$ controls both the mobility $D$ of the chains and the viscosity of
the dispersion.

We start with the Navier-Stokes (N-S) equation (\ref{nspolsp}) for the velocity
field {\bf v}(\rvec) in the suspension,with {\bf F}(\rvec) some external force
density.  After we perform the configurational average (\ref{confav}) and we
replace the averages over the forces with the ${\bf W}^{SP}(\rvec,\rrvec)$ and
${\bf\Sigma}^{POL}(\rvec,\rrvec)$ as defined in (\ref{sigdefpolsp}) the N-S
equation for the average velocity field {\bf u}(\rvec) is~: 
\beq 
-\eta_0 \triangle {\bf u}(\rvec) + \la p(\rvec) \ra - {\bf W}^{SP}\ast {\bf u}
- {\bf\Sigma}^{POL}\ast {\bf u} = {\bf F}(\rvec) 
\eeq 
where, as discussed in the previous sections, we substitute ${\bf W}^{SP}$ for
${\bf\Sigma}^{SP}$ to distinguish between the background effective fluid
associated with the spheres and the contribution to the self-energy of the
added polymer chains.  ${\bf W}^{SP}(\rvec,\rrvec)$ tensor is function only of
$\fsp$ volume fraction and it was calculated in Sec.~\ref{sec:w_NS},
Eqs.~(\ref{wSP},\ref{sysSP}).

To obtain the self-consistent equations for ${\bf\Sigma}^{POL}$ we proceed as
in the last section.  First, we write the formal solution of the N-S equation
as a convolution~: 
\beq
{\bf u}(\rvec) = {\bbox{\cal G}} \ast {\bf F}
\eeq
where the modified Oseen tensor has its Fourier transform given by~:
\beqar
{\bbox{\cal G}}(\kvec) & = & {\openone - \hat{\kvec}\hat{\kvec} \over \eta_0
k^2 - W_{\perp}^{SP}(\kvec) - \Sigma_{\perp}^{POL}(\kvec)} \nonumber\\
& = &  {\unitkk \over \eta_0 (1 + \Sigma_1)(k^2 + {\xi '}^{-2})}
\label{oseenpolsp}\nonumber\\  
\Sigma_1 & = & W_1^{SP} + \Sigma_1^{POL}\; ; \; {\xi '}^{-2} = {\xi^{-2}_{SP} +
\xi^{-2}_{POL} \over 1 + \Sigma_1}\nonumber 
\eeqar
in which we inserted $W_{\perp}^{SP}$ from (\ref{sigexp}) and we used the
following approximation , in the $k\to 0$ region~:
\beq
{\bf\Sigma}^{POL}(\kvec) = (-\eta_0 \xi^{-2}_{POL} - \eta_0 \Sigma_1^{POL}
k^2)\,\openone 			
\label{sigpolpolsp}
\eeq 
Also, $\xi '$ is the total screening length and $\xi_{SP}$ is function only
of the concentration of spheres.  We can readily write the solution for
${\bf\Sigma}^{POL}(\kvec)$ by observing the structures of the Oseen tensors
found in (\ref{oseenpol}) and (\ref{oseenpolsp}).  In comparison with the
effective medium theory for polymers (as in Sec.~\ref{sec:w_NP} and
Appendix~C), we need to transform (\ref{sigPOL}) by working with $\xi '$
instead of $\xi_{SP}'$ and with the total self-energy $\Sigma_1$ instead of
$W_1^{POL}$.  We immediately arrive to~: 
\beq 
{\bf\Sigma}^{POL}(\kvec) = - {9\over\pi}\eta_0 \fpol (1 + \Sigma_1)
\text{Q}(\beta) \,k^2\,\openone 
\eeq 
with $\beta = {R_g\over \xi '}$ and the function Q(x) defined by
(\ref{sigPOL}). Comparing to (\ref{sigpolpolsp}), we obtain the system~: 
\beqar
\xi^{-2}_{POL} & = & 0 \nonumber\\ 
\Sigma_1^{POL} & = & {9\over \pi} \, \fpol \, (1 + W_1^{SP} + \Sigma_1^{POL})
\, \text{Q}(\beta) 
\eeqar 
Introducing the natural dimensionless variables ${x(\fsp) =} {a\over\xi
'_{SP}}$ and ${t={R_g\over a}}$, we finally determine
${\bf\Sigma}^{POL}(\kvec)$ through~:
\beqar 
\Sigma_1^{POL}(\fpol,\fsp,t) & = & [1 + W_1^{SP}(\fsp)]\, \nonumber\\
& & \times \, {1 \over 1 - {9\over\pi} \fpol \text{Q}(\beta)}
\label{syspolsp}\\  
\beta^2 & = & (t\, x)^2 \, (1 - {9\over\pi} \, \fpol \, \text{Q}(\beta))
\nonumber  
\eeqar 
where $W_1^{SP}$ and $x(\fsp)$ are computed from the system (\ref{sysSP}) as
functions of $\fsp$ only.  Note that now $\beta (=R_g/\xi ')$ is in general a
non-zero function of $\fpol, \fsp$ and $t$, which means that the polymers are
subjected to a screened hydrodynamic interaction, in contrast with the previous
system.

The effective viscosities---total and relative, due to added polymers---are
determined from~: 
\beqar
\veff \; & = & \; W_1^{SP} + \Sigma_1^{POL} \nonumber\\
& = & \; (1+ W_1^{SP})\, {1 \over 1 - {9\over\pi} \fpol \text{Q}(\beta)} \, -
\, 1 \label{viscpolsp}\\ 
\vrelpolsp \; & = & \; {1 \over 1 + W_1^{SP}}\, \Sigma_1^{POL} \,=
\, {1 \over 1 -{9\over\pi} \fpol \text{Q}(\beta)} \,-\, 1
\nonumber
\eeqar 

For a mobile polymer chain the translational diffusion coefficient (actually
defined as relative to the Kirkwood-Riseman result) is shown to be (see
Appendix~C)~:
\beq
D = {\zeta_t^{KR} \over {\cal G}_{00}^{-1}}
\eeq
with ${\cal G}_{00}^{-1}$ being the zero-th coefficient of a Fourier series
expansion of $\la{\bbox{\cal G}}^{-1}\ra$ (see Appendix~C). Because in our
approximations ${\cal G}_{qq'}$ is diagonal, ${\cal G}_{00}^{-1}$ becomes
(\ref{gpAppC})~:
\beq 
{\cal G}_{00}^{-1} = ({\cal G}_{0})^{-1} = 3 \pi \eta_0 R_g {\beta \over
\text{P}(\beta)} (1 + \Sigma_1) 
\eeq 
The diffusion coefficient of any moving polymer chain is given by~: 
\beq 
D = {3\sqrt{\pi} \over 4} {1 \over 1 + W_1^{SP}} {\text{P}(\beta) \over \beta}
(1 - {9 \over \pi}\fpol \text{Q}(\beta))
\label{diffpolsp}
\eeq
The functions $\text{P}(x)$ and $\text{Q}(x)$ have been defined in
Appendix~C, Eqs.\ (\ref{gpAppC},\ref{sigpolAppC}).
 
The prominent feature of the final results for the solution viscosity and the
diffusion coefficient of a polymer chain (\ref{viscpolsp},\ref{diffpolsp}) is
their dependence on the $t$ parameter---the stationary dynamics of the spheres
and the polymer chains is {\sf size coupled}---as advertised in the beginning.
Also note that in contrast to the problem of spheres in a polymer suspension,
the variable $\beta = R_g/{\xi '}$ depends upon $\fsp$ too, which
implies---Eq.~(\ref{viscpolsp})---that the relative contribution of the
polymers to the viscosity is enhanced, function of $\fsp$ and $t$, by the
presence of the {\sf background effective fluid} related to the spheres.  It is
sensible to believe that these new phenomena have been generated by the spheres
acting as fixed scattering centers of the velocity field and had we analyzed a
system of polymer chains and Brownian spheres, we might have observed that the
dynamics of the two types of particles would have been decoupled, in the
mentioned sense (no $t$ dependence, $\beta$ function only on $\fpol$).

Checking the limit $\fsp \to 0$ we recover the results for the suspension of
polymers only ($W_1^{SP}\to 0$, $\beta\to 0$, compare to (\ref{viscdiffPOL})),
and in the limit $\fpol\to 0$ we retrieve the results for the {\bf 1P/S}
problem~: $\beta \to tx_{SP}$; $\fpol\to {1\over V} 4\pi R_g^3/3 =
{\fsp\over\text{N}_{S}} t^3$ (compare to Eqs.\ (\ref{diff1polsp},
\ref{visc1polsp})).

The values of the total and relative effective viscosities and of the diffusion
coefficient given by Eqs.\ (\ref{viscpolsp},\ref{diffpolsp}) have been plotted
as functions of $\fpol$ and $\fsp$ at different $t$. Four sets of data
corresponding to ${t = \{0.01;0.1;1;10\}}$ have been calculated and displayed.

In Figs.~[\ref{dPS1PS_S_t001}--\ref{dPS1PS_S_t10}] the diffusion coefficient
$D$ of a moving polymer chain is plotted against $\fsp$ with $\fpol$ parameter
and for each of the $t$ values mentioned, and in
Figs.~\ref{dPS_P_t01}--\ref{dPS_P_t10} $D$ is plotted as a function of $\fpol$
with $\fsp$ parameter, with $t=\{0.1,1,10\}$ for convenience.

 $D(\fpol)$ depends linearly on $\fpol$ and for fixed $\fsp$ and $\fpol$,the
diffusion coefficient $D$ decreases as $t$ increases---longer chains are less
diffusive---so the $t$ parameter acts like a mobility selector.  The
sensitivity of $D$ to $t$ variations is higher when the size ($R_g$) of the
chains becomes comparable to the radius of the spheres (compare the relative
drop in $D$ at the same $\fsp$ in Fig.~\ref{dPS_P_t01} ($t = 0.1$) to
Fig.~\ref{dPS_P_t1} ($t~=~1$), and Fig.~\ref{dPS_P_t1} to Fig.~\ref{dPS_P_t10}
($t~=~10$). 

The same observation can be made when looking at the $\fsp$ dependence of $D$,
in Figs.~[\ref{dPS1PS_S_t001}--\ref{dPS1PS_S_t10}], for the same values of
$t\/$.  Additionally, for $t = \{0.01;0.1\}$, $D$ exhibits a change in the
curvature and an almost {\sf plateau region} in the middle of the $\fsp$ range
for higher $\fpol$ (Figs.~\ref{dPS1PS_S_t001},\ref{dPS1PS_S_t01}).

We now consider the behavior of the limiting case corresponding to the {\bf
1P/S} system as shown by the $\bbox{+}$ symbols in
Figs.~[\ref{dPS1PS_S_t001}--\ref{dPS1PS_S_t10}]. One can notice a change in the
curvature of the diffusion coefficient of one polymer chain when going from $t
= 0.01$ to $t = 10$. It appears that for relatively small (in comparison to the
spheres) polymer chains $(t = 0.01)$ or relatively large $(t = 10)$ a {\sf
plateau region} appears, where $D$ is slowly varying with respect to $\fsp$.
Physically, this could signify that when the polymer is small (large), a
certain built up in $\fsp$ is required before the diffusion coefficient $D$ is
affected more markedly. Additionally, for a given radius $a$ of the spheres,the
difference in the mobilities of two polymer chains of distinct lengths ($t_1
\neq t_2;t_i = {R_g^{(i)}\over a}$) is maximized for a certain range of $\fsp$
values, fact depicted in Fig.~\ref{eleven} , where the difference in the
diffusion coefficient of one polymer chain for four pairs of $t$ values is
plotted against the filling fraction $\fsp$ (e.g. $\fsp \in [.15;.3]$ roughly
for two types of polymer chains with $t_1=0.01$ and $t_2=1$).  This may have
applications in the design of polymer sieves.

Comparing now the behavior of the polymer moving in the suspension of spheres
(Figs.~[\ref{dPS1PS_S_t001}--\ref{dPS1PS_S_t10}]) to that of the mobile sphere
inside a polymer dispersion (Fig.~\ref{dSP1SP_P}) at equivalent relative sizes
of the tracer particle ($t = 10$ in the first case is similar to $t = 0.1$ in
the second one), we notice that the $D(\fsp)$ for a small mobile polymer chain
($t = 0.01;0.1$) is higher than that of a small sphere ($t = 10$) moving in a
polymer suspension, but a larger sphere ($t = 0.1$) exhibits an increased
mobility ($D\/$) comparative to a large polymer chain ($ t=1;10$).
Eventually, because $(\fsp_{DIV}\simeq 0.49) > \fpol_{DIV} \simeq 0.31$, the
polymer immersed among spheres will retain a non-zero mobility at higher volume
fractions than the sphere moving in a polymer suspension.

 The same discussion applies in the general {\bf P/S} case. The mobility of a
small $t = 0.01$ or large $t = 10$ polymer chain is only slightly influenced by
the volume fraction of the spheres (see slopes of graphs in
Figs.~\ref{dPS1PS_S_t001},\ref{dPS1PS_S_t10} in the middle region).  This
feature indicates that, for a given radius $a$ of the background spheres and a
pair of polymer chains characterized by distinct $t$ values (e.g. $t_1 = 0.01$
and $t_2 = 1$), the difference in the diffusion coefficients of the two chains
will reach its maximum for a certain range of the spheres volume fraction
$\fsp$.

The two points of divergence $\fpol_{DIV} \simeq 0.31$ and $\fsp_{DIV}\simeq
0.49$ are clearly marked. Because $\fpol_{DIV} < \fsp_{DIV}$ the suppressing
of the mobility of the moving particles (either sphere or a mobile polymer
chain) is stronger when polymers are added than when the sphere concentration
is increased.
   
Comparing the data for $D(\fsp)$ in Fig.~\ref{dSP_S}, for the {\bf S/P} example
(note that there is no $t$ dependence, to those for $D(\fpol)$ at $t = 0.1$ in
Fig.~\ref{dPS_P_t01} and $t = 10$ in Fig.~\ref{dPS_P_t10}, the same observation
as in the {\bf 1P/S} case can be made~:\ at the same $t$ ratio between the
sizes of the moving object and of the background component and for the same
background concentration, a small ($t = 0.1$) polymer chain in the {\bf P/S}
system is more mobile than a small ($t = 10$) sphere moving in the {\bf S/P}
system, but as $t$ gets larger, a big sphere becomes more diffusive than a long
polymer chain ($t = 10$).  In this situation, the control parameter is the
concentration of the added (e.g.,in the {\bf P/S} example, polymers are the
added elements) components, but the same conclusion is reached if comparing the
data for $D(\fpol)$ from Fig.~\ref{dSP1SP_P} in the {\bf S/P} case to those
from Fig.~\ref{dPS1PS_S_t01} ($t = 0.1$) and Fig.~\ref{dPS1PS_S_t10} ($t =
10$) for the {\bf P/S} system, the control parameter now being the
concentration of the background elements.

The limiting case of one polymer chain ($\text{N}_{P}\to 1$) is very well
approached in the {\bf P/S} example, as one can notice in
Figs.~\ref{dPS1PS_S_t001}--\ref{dPS1PS_S_t10} by comparing the uppermost
curves ({\bf 1P/S} case) to those corresponding to the parameter value $\fpol =
0.01$, representing $D(\fsp)$ when $t = \{0.01; 0.1; 1; 10\}$, respectively.

Viscosities calculated from (\ref{viscpolsp}) have been plotted in
Figs.~\ref{vT_PS_P_t001}--\ref{vR_PS_S_1t10}~: the total effective viscosity
change in Figs.~\ref{vT_PS_P_t001},\ref{vT_PS_P_t10} as a function of $\fpol$
and against $\fsp$ in Figs.~\ref{vT_PS_S_t001},\ref{vT_PS_S_t10} with $t =
\{0.01; 10\}$; the relative effective viscosity in Fig.~\ref{vR_PS_P} as a
function of $\fpol$ for $t = \{1; 10\}$ and as a function of $\fsp$ in
Figs.~\ref{vR_PS_S_t001}, \ref{vR_PS_S_1t10} with $t = \{0.01; 1; 10\}$.  Two
observations could be made regarding the viscosities~:

{\sl a)} both for $\eta^{\text{eff}}(\fpol)$ with $\fsp$ parameter
(Figs.~\ref{vT_PS_P_t001},\ref{vT_PS_P_t10}) and for $\eta^{\text{eff}}(\fsp)$
with $\fpol$ parameter (Figs.~\ref{vT_PS_S_t001},\ref{vT_PS_S_t10}), as $t$
increases, the total viscosity change is more sensitive to variations in
$\fpol$ or $\fsp$ (compare Fig.~\ref{vT_PS_P_t001} when $t = 0.01$ to
Fig.~\ref{vT_PS_P_t10} when ${t = 10}$ and Fig.~\ref{vT_PS_S_t001} ($t = 0.01$)
to Fig.~\ref{vT_PS_S_t10} (${t = 10}$)).  We note that a larger polymer chain (
in comparison to the radius of the background sphere) is more sensitive to the
environment.  This behavior clearly contrasts that encountered for the total
effective viscosity in the {\bf S/P} case (see Fig.~\ref{vT_SP1SP_P}).

{\sl b)} the relative effective viscosity due to the presence of the polymers
(Figs.~\ref{vR_PS_P}--\ref{vR_PS_S_1t10}) depends not only on $\fpol$ but on
$\fsp$ and on $t$ as well, once again in contrast to the {\bf S/P} system (see
Fig.~\ref{six}).  As suggested before, $t$ acts like a control parameter that
when increased (compare ${t = 1}$ to ${t = 10}$ calculations in
Fig.~\ref{vR_PS_P} or Fig.~\ref{vR_PS_S_t001} to Fig.~\ref{vR_PS_S_1t10}) makes
the contribution of the polymers to the viscosity more markedly dependent on
both $\fpol$ and $\fsp$.  Actually, when the radius of gyration $R_g$ is
comparable to the sphere diameter $a$ ($t \leq 1$) , the relative viscosity
$\vrelpolsp$ is relatively insensitive to variations in $\fsp$ or $t$ (compare
Fig.~\ref{vR_PS_S_t001} to $t = 1$ data in Fig.~\ref{vR_PS_S_1t10}). 

Obviously, these observations may have some significance from an experimental
point of view.

\section{Conclusions}
\label{sec:conc}

In the present paper we have investigated the steady-state transport properties
of a heterogeneous suspension of rigid polymers and spheres without excluded
volume interactions, applying the multiple scattering formalism
\cite{FM78,FM82,MF82,FM78st} and the effective medium
theory~\cite{MF79,M82,FM78st}.

The main approximations and features of our approach were~:
\begin{itemize}
\item[a.] the absence of any interactions among polymers and spheres. All
particles are penetrable.
\item[b.] the solvent is assumed incompressible and it is described by a
linearized Navier-Stokes equation.
\item[c.] the stationary, time-independent limit is assumed. All results are
valid in the long range hydrodynamic limit ($k \to 0$).
\item[d.] the dynamics of the solvent, polymers and of the spheres are coupled
by stick boundary conditions.
\item[e.] the spheres are monodisperse, having radius $a$.
\item[f.] the polymers are monodisperse with chain length $L\/$, and
they translate freely as rigid bodies. We do not consider the rotation of the
polymer as a rigid body. For convenience, the distribution function for the
inter-segment distance of any chain was taken to be Gaussian.
\item[g.] ${\bf\Sigma}(\kvec)$, the self-energy of the fluid in the presence of
the immersed spheres and polymer chains was approximated with the first-order
component ${\bf\Sigma}^{(1)}(\kvec)$, in an expansion in the number of distinct
scattering events (one event means one interaction of the velocity field with
one particle).  When no correlations (interactions) among particles occur,this
is actually the leading term of the exact ${\bf\Sigma}(\kvec)$.
\end{itemize}
Under these approximations, two general systems have been analyzed~:
\begin{enumerate}
\item Moving sphere in a solution of fixed spheres and freely translating
polymer chains.
\item Moving polymer chains dispersed in a suspension of fixed spheres.
\end{enumerate}
Also, two significant limiting cases of the previous models have been
investigated~: 
\begin{itemize}
\item[a.] Moving sphere in a solution of polymers.
\item[b.] Moving polymer chain immersed in an array of fixed spheres.
\end{itemize}
For each example the underlying theory has been exposed and developed and
equations were explicitly displayed, from which the translational diffusion
coefficient $D$ of the probe particle and the effective viscosities of the
solution, total $\vefftext$ and relative, have been computed.

We have identified the proper dimensionless variables characterizing the
problem as follows~: $\beta = {R_g\over \xi '}$, $x = {a\over \xi '}$ and $t =
{R_g\over a}$, where $R_g$ and $a$ are the radius of gyration of a polymer
chain and the radius of a sphere, and $\xi '$ is the total screening length in
the solution.

It has been shown that the polymers do not contribute to the screening of the
hydrodynamic interactions at large length scales and, as functions of polymer
filling fraction $\fpol$, both the effective viscosity $\eta^{\text{eff}}$ and
the translational friction coefficient $\zeta_t$ diverge, in all examples, at a
volume fraction $\fpol_{DIV} \simeq 0.31$ much less than the corresponding
divergence point for the $\fsp$ dependence ($\fsp_{DIV}\simeq 0.49$).  While
for suspensions of uniform particles showing no long-range order---in
particular spheres---there is wide experimental support for the divergence of
viscosity and of the friction coefficient for $\fsp_{DIV} \in [0.5,0.6]$
(e.g. Ref.~\cite{MZ}), in the case of solutions of rigid polymers and
specifically for suspensions of polymers and spheres less data are available,
to our knowledge.

Perhaps the most interesting parameter is $t$, which is controlling the
response of $\eta^{\text{eff}}$ and $D$, both for spheres and polymers, to
changes in the environment (variations in $\fsp$ and $\fpol$).  The
understanding of the role of the $t$ variable might be of interest in
experimental applications, mainly related to gel permeation chromatography and
macromolecular sieving. A rather noticeable prediction of our analysis related
to the process of diffusion of the polymers in a suspension of fixed spheres is
the existence of an optimum range of the volume fraction $\fsp$ of the spheres
that maximizes the difference in the diffusion coefficients for two types of
chains characterized by distinct $t$ values (see Fig.~\ref{eleven} for the {\bf
1P/S} system and Figs.~\ref{dPS1PS_S_t001}--\ref{dPS1PS_S_t10} in the {\bf P/S}
case).

It has been also noticed that at the same $t$ ratio between the size of the
tracer (moving) particle and the size of the background element and the same
background concentration, the spheres and the polymers are behaving
differently.  When the polymer is small in comparison to the spheres, it is
more diffusive than a small sphere moving among polymer chains, but a large
polymer chain in a suspension of fixed spheres is less mobile than a large
sphere moving in a polymer dispersion. We must note that this observation is
valid before reaching $\fpol_{DIV}\simeq 0.31$.

Despite the approximations made to yield the calculations analytically
tractable, the formalism presented is essentially general and also applicable
when excluded-volume (or other types of) interactions are turned on, although
in practice the effort could be quite laborious. Yet, in this straightforward
approach we have managed to draw a physically reasonable picture of the
dynamical behavior of such heterogeneous solutions of polymers and spheres and
some conceivable findings emerged. Further extensions of this work are
possible, for example the study of rigid rodlike polymers and spheres,
microemulsions (deformable particles) or investigating polydisperse suspensions
of spheres and polymers.  This last aspect is particularly meaningful in
industrial applications (e.g.  ceramic fabrication, where processing at volume
fractions higher than the limiting values $\fsp_{DIV},\fpol_{DIV}$ is necessary
but very difficult due to the divergent viscosity),because experimentally it
has been shown that a proper adjustment of the particle size distribution could
induce substantial increases in the limiting volume fractions \cite{MZ}, but
theoretically this observations is still not fully understood.

We seek to address this question and the physically more appealing problem of a
heterogeneous suspension of particles with potential interactions in a future
paper.

\acknowledgments

Acknowledgment is made to the Materials Research Science and Engineering Center
at the University of Massachusetts, and the NSF Grant DMR 9221146001.

\appendix

\section{Solution of the Navier-Stokes equation for a suspension of
$\text{N}_P\/$ polymer chains and $\text{N}_{S}\/$ spheres}

In Sec.~\ref{sec:twoB} we have found that a formal solution of the
Navier-Stokes equation (\ref{nspolsp}) is given by Eq.\ (\ref{vforcespolsp}),
where the force densities $\bbox{\sigma}_{\alpha i}$ and $\bbox{\sigma}_b$
exerted upon the fluid by the {\it i\/}-th bead of the polymer chain $\alpha$
and by the point ${\bf R}_b$ on the surface of the sphere {\it b\/}, are
unknowns.  Using the stick boundary conditions (\ref{nsbound}), we 
eliminate the forces to express the velocity field in terms of the
single-object {\sf flow propagators} ${\bf T}_\alpha$ of any chain $\alpha$ and
${\bf T}_b$ of any sphere.

To simplify the notation, we replace the variables ${\bf R}_b$,${\bf R}_{\alpha
i}$ with the indices $b$ and $\alpha i$,respectively, when appearing in the
arguments of the Oseen tensor ${\bf G}$ and we apply the Einstein convention of
summation over repeating indices.  In the case of the spheres this will also
imply a surface integral over the solid angle $\text{d}\Omega$.  We drop the
chain index $\alpha$ for relations generally valid independent of the chain
variables. The convolution operation is denoted by $\ast$---${\bf f}\ast{\bf g}
= \int\! d\rrvec\,{\bf f}(\rvec - \rrvec)\cdot {\bf g}(\rrvec)$ with {\bf f}
and {\bf g} tensors or vectors---and it is understood that its result is a
function of some position vector \rvec, when not otherwise specified by a
subscript.  Due to the translational invariance in the suspension, ${\bf
G}(\rvec,\rrvec)$ is actually a function of the relative distance~: ${{\bf
G}(\rvec -\rrvec)}$. Then, Eq.\ (\ref{vforcespolsp}) can be rewritten as~: 
\beq
{\bf v}(\rvec) = {\bf G}\ast{\bf F} + \Gind{\rvec,\alpha i}\cdot\sigind{\alpha
i} + \Gind{\rvec,b}\cdot \sigind{b}
\label{vAppA}
\eeq
The sums over $i,j$,... run from 1 to $n$,the sums over $\alpha,\beta,\ldots$
from 1 to $\text{N}_P\/$and the sums over $b,c,\dots$ types of indices run from
1 to $\text{N}_{S}\/$.  Inserting the previous expression in the boundary
conditions (\ref{nsbound}) we get~:
\beqar
{\bf u}_{\alpha} + \bbox{\omega}\times\Sind{\alpha k} & = & \left. {\bf G}\ast
{\bf F}\right|_{\alpha } + \Gind{\alpha k,\beta j}\cdot \sigind{\beta j} +
\Gind{\alpha k,b}\cdot \sigind{b} \nonumber\\ 
0\;\; & = & \left. {\bf G}\ast {\bf F}\right|_{b} + \Gind{b,\alpha i}\cdot
\sigind{\alpha i} + \Gind{b,c}\cdot \sigind{c}
\label{nsboundAppA}
\eeqar
Here ${\bf u}_\alpha$ and $\bbox{\omega}_{\alpha}$ are the velocity of the
center of mass and the angular velocity of chain $\alpha$.  To eliminate the
unknown quantities $\sigind{\alpha i}$ and $\sigind{b}$ we need the generalized
inverse operators $\Kinv{\alpha}$ and $\Kinv{b}$ \cite{M81,M82} defined by~:
\beqar 
& & \sum_{j=1}^{n}\Kinv{\alpha}(\Sind{i},\Sind{j})\cdot {\bf G}(\Sind{\alpha j}
- \Sind{\alpha k}) = \openone \, \delta_{\alpha i,\alpha k} \nonumber\\ 
& & \int\! d\Omega_{b}''\,\Kinv{b}(\rvec ,{\bf r}'')\cdot {\bf G}({\bf r}'' -
\rrvec) = \openone \, \delta (\Omega_b- \Omega_b ') 
\eeqar 
where $\rvec(\Omega_b)$,${\bf r}''(\Omega_b '')$ and $\rvec (\Omega_b ')$ are
position vectors (about the center of mass frame) of distinct points on the
surface of the sphere $b$ and $\Sind{\alpha i}$ is the position vector of the
$i$-th bead of the chain $\alpha$ with respect to its center of mass.  Applying
the inverse operators in (\ref{nsboundAppA}) we obtain~:
\begin{mathletters}
\label{forcesAppA}
\beqar
& & \sigind{\alpha i} = \sum_{k=1}^{n}\Kinv{\alpha i,\alpha k}\cdot \left( {\bf
u}_{\alpha} - \left.{\bf G}\ast {\bf F} \right|_{\alpha k} -  \Gind{\alpha
k,b}\cdot \sigind{b} \right. \nonumber\\
& & - \left. \left. \Gind{ \alpha k,\beta j}\cdot \sigind{\beta j}
\right|_{\beta\neq\alpha}  \right)  + \, \sum_{k=1}^{n} \Kinv{\alpha i,\alpha
k}\cdot (\bbox{\omega}_{\alpha}\times \Sind{\alpha k}) \\
& & \sigind{b}(\Omega_b) = - \Kinv{b b'}\cdot \left. {\bf G}\ast {\bf
F}\right|_{b'} \,- \, \Kinv{bb'}\cdot \Gind{b',\alpha i}\cdot \sigind{\alpha i}
\,\nonumber\\
& & \;\;\;\; - \, \left.  \Kinv{bb'}\cdot \Gind{b'c}\cdot
\sigind{c}\right|_{b\neq c} 
\eeqar
\end{mathletters}
Note that there is no summation over $\alpha$ in the first equation and over
$b$ in the second one.

Analyzing the structure of these force factors, it appears manifest that the
force upon the fluid due to one object, is proportional to the relative
increase in the velocity of the fluid about the local velocity owing to the
action of all other bodies and external forces, which is a sensible physical
expectation. Also, the inverse operators $\Kinv{\alpha i,\alpha k}$ and
$\Kinv{bb'}$ will produce the friction coefficients of the polymer chains and
of the mobile sphere, after one performs the configurational average
(\ref{confav}) and sums over the indices $i,k$ in the polymer case or
integrates over $\Omega_b$ and $\Omega_b '$ for the sphere.

As shown in Ref.~\cite{M81}, the angular velocity of the chain
$\bbox{\omega}_{\alpha}$ in the force equations (\ref{forcesAppA})
leads, eventually, to the appearance of the rotational and cross
translational-rotational/rotational-translational friction coefficients of the
polymer chains.  The former has been calculated in Ref.~\cite{M81} and the
latter vanish upon preaveraging.  In the present work we will ignore the
rotational term, which does not alter the screening properties of a polymer
solution.  Then, using the conditions (\ref{constr}) that the total external
force and torque acting on any polymer chain are zero, we get the velocity
${\bf u}_{\alpha}$~: 
\beqar
{\bf u}_{\alpha} \,& = &\, {\bf g}_{t}^{-1}\cdot \sum_{i,k}^{n} \Kinv{ik}\cdot
\left. {\bf G}\ast{\bf F}\right|_{ik} + {\bf g}^{-1}_{t}\cdot
\sum_{i,k}^{n}\Kinv{\alpha i,\alpha k} \nonumber\\
& & \, \cdot \Gind{\alpha k,b}\cdot \sigind{b} + {\bf g}^{-1}_{t}\cdot
\sum_{i,k}^{n}\left. \Kinv{\alpha i,\alpha k}\cdot \Gind{\alpha k,\beta j}\cdot
\sigind{\beta j}\right|_{\beta\neq\alpha} \nonumber\\
& & {\bf g}_t = \sum_{i,j}^{n}\Kinv{ij}\;\; ; \;\; {\bf g}_t \cdot {\bf
g}_t^{-1} = \openone 
\eeqar
Inserting back in (\ref{forcesAppA}) and relabeling the dummy indices, we
finally calculate~:
\beqar
\sigind{\alpha i} & = & \Tind{\alpha i,\alpha k}\cdot \left. {\bf G}\ast {\bf
F}\right|_{\alpha k} + \left. \Tind{\alpha i,\alpha k}\cdot \Gind{\alpha
k,\beta j}\cdot \sigind{\beta j}\right|_{\beta \neq \alpha} \nonumber\\
& & \, + \, \Tind{\alpha i,\alpha k}\cdot \Gind{\alpha k,b}\cdot \sigind{b}
\label{forces}\\
\sigind{b} & = & \left. \Tind{bb'}\cdot {\bf G}\ast {\bf F}\right|_{b'} +
\Tind{bb'}\cdot \Gind{b',\alpha i}\cdot \sigind{\alpha i} \nonumber\\ 
& & \, + \left. \Tind{bb'}\cdot \Gind{b'c}\sigind{c}\right|_{b\neq c} \nonumber
\eeqar 
where there is no summation over $\alpha$ in the first equation and over $b$ in
the second and we have introduced the notation~: \beqar \Tind{\alpha i,\alpha
k} & = & -\left[ \Kinv{\alpha i,\alpha k} - \sum_{l,l'=1}^{n} \Kinv{\alpha
i,\alpha l}\cdot {\bf g}_t^{-1}\cdot \Kinv{\alpha l',\alpha k} \right]
\nonumber\\ \Tind{bb'} & = & - \Kinv{bb'} \eeqar

Following indefinite iterations in (\ref{forces}) and substituting the results
for $\sigind{\alpha i}$ and $\sigind{b}$ back in (\ref{vAppA}), we obtain the
velocity field as~:
\beqar
{\bf v}(\rvec) & \, = \, & {\bf G}\ast {\bf F} + \left. \Gind{\rvec \alpha
i}\cdot \Tind{\alpha i,\alpha k}\cdot {\bf G}\ast {\bf F}\right|_{\alpha k}
\nonumber\\ 
& + & \left. \Gind{\rvec\,b}\cdot \Tind{bb'}\cdot {\bf G}\ast {\bf F}
\right|_{b'} \nonumber\\
& + & \left. \Gind{\rvec\,\alpha i}\cdot \Tind{\alpha i,\alpha k}\cdot
\Gind{\alpha k,\beta j}\cdot \Tind{\beta j,\beta l}
\right|_{\beta\neq\alpha}\cdot \left. {\bf G}\ast{\bf F}\right|_{\beta l}
\nonumber\\
& + & \left. \Gind{\rvec\,b}\cdot\Tind{bb'}\cdot \Gind{b'c}\cdot\Tind{cc'}\cdot
{\bf G}\ast{\bf F} \right|_{b\neq c} \label{viterAppA}\\ 
& + & \left. \Gind{\rvec,\alpha i}\cdot \Tind{\alpha i,\alpha k}\cdot
\Gind{\alpha k,b}\cdot\Tind{bb'}\cdot {\bf G}\ast {\bf F}\right|_{b'}
\nonumber\\
& + & \left. \Gind{\rvec\,b}\cdot \Tind{bb'}\cdot \Gind{b',\alpha i}\cdot
\Tind{\alpha i,\alpha k} \cdot {\bf G}\ast {\bf F}\right|_{\alpha k} \, + \,
\ldots \nonumber
\eeqar
Next, the single chain and single sphere {\sf flow propagators} ${\bf
T}_{\alpha}$ and ${\bf T}_b$ will be defined as~:
\begin{mathletters}
\beqar
\Tind{\alpha}(\rvec - \rrvec) & = & \sum_{i,k=1}^{n}\delta(\rvec -
\Rind{\alpha i})\, \Tind{\alpha i,\alpha k}\, \delta(\rrvec - \Rind{\alpha k})
\\ 
\Tind{b}(\rvec - \rrvec) & = & \int\!\!\!\int\! d\Omega_b \, d\Omega_b' \:
\delta(\rvec - \Rind{b})\, \Tind{bb'}\, \delta(\rrvec - \Rind{b'}) \eeqar
\end{mathletters} 
It is worth noticing that both operators still depend upon the centers of mass
through $\Rind{\alpha i} = \Rind{\alpha}^{0} + \Sind{\alpha i}$ and $\Rind{b}
= \Rind{b}^{0} + \rvec_b(\Omega_b)$.

Eventually, the above definitions allow us to write (\ref{viterAppA}) in a
physically more suggestive manner, as a sequence of multiple convolutions that
are representing the scattering events~:
\beqar
{\bf v} &(\rvec)& = {\bf G}\ast{\bf F} + \sum_{\alpha}{\bf G}\ast
\{{\bf T}_{\alpha}\ast {\bf G}\ast {\bf F} + \sum_{b} {\bf G}\ast \Tind{b}\ast
{\bf G}\ast {\bf F} \nonumber\\ 
& + & \, \sum_{\alpha}\!\sum_{b}{\bf G}\ast
\left\{ {\bf T}_{\alpha}\ast{\bf G}\ast{\bf T}_b + {\bf T}_b\ast{\bf
G}\ast{\bf T}_{\alpha} \right\} \ast{\bf G}\ast{\bf F} \nonumber\\ 
& + & \, \sum_{\alpha}\!\sum_{\beta(\neq\alpha)}{\bf G}\ast {\bf
T}_{\alpha}\ast {\bf G}\ast {\bf T}_{\beta}\ast {\bf G}\ast {\bf F} \\
& + & \sum_{b}\!\sum_{c(\neq b)} {\bf G}\ast{\bf T}_b\ast{\bf G}\ast{\bf
T}_c\ast {\bf G}\ast{\bf F} + \; \dots \nonumber 
\eeqar
where the dependence on \rvec\ of the multiple convolutions is implied and the
sum is continued over all permissible {\sf scattering sequences}.

This is Eq.\ (\ref{vpolsp}) we wished to derive. 

\section{Self-consistent calculation of the ${\bf W}(\kvec)$ tensor for
 $\text{N}_{S}\/$ spheres immersed in a fluid}

The main tool we need is the following expansion in spherical
harmonics~: 
\beqar 
{\cal A}(\rvec(\Omega),\rrvec(\Omega ')) & = &
\sum_{l,l'\geq 0}^{\infty}\sum_{m= -l}^{l}\sum_{m' = -l'}^{l'}\!
\tilde{{\cal A}}_{lm;l'm'} Y_{lm}(\Omega) Y_{l'm'}^{\ast}(\Omega')
\nonumber\\ 
\tilde{{\cal A}}_{lm;l'm'} & = & \int\!\!\!\int\!  d\Omega\, d\Omega '\, {\cal
A}(\Omega,\Omega') Y_{lm}^{\ast}(\Omega) Y_{l'm'}(\Omega ')
\eeqar 
where $\rvec$ and $\rrvec$ are the position
vectors for points on the surface of a sphere ($|\rvec| = |\rrvec| =
a$) with angular directions $\Omega$ and $\Omega '$, $Y_{lm}(\Omega)$
are the spherical harmonics and ${\cal A}$ is any tensor (vector)-like
quantity.

Consider now $\text{N}_{P}-1$ fixed, non-interacting spheres of radius $a$
immersed in a pure fluid of viscosity $\eta_0$.  One can calculate the exact
self-energy ${\bf\Sigma}^{SP}(\kvec)$ of the suspension as in
Sec.~\ref{sec:twoB} (dropping the factors related to the polymer chains) but a
more direct approach is to seek a self-consistent approximation , ${\bf
W}^{SP}(\kvec)$ , of the exact self-energy. As discussed in
Sec.~\ref{sec:fluid+part}, we can replace our system of spheres and pure fluid
averaged over the distribution of the spheres, with an effective medium in
which the force propagator is ${\bbox{\cal G}}^{SP}(\kvec)$ given by
(\ref{oseensp}).  Adding one more sphere, its contribution to the approximate
self-energy of the fluid is found from the second factor in (\ref{sigsol}) with
the replacements $c^{\text{POL}}\to 1/V$ and ${\bf K}^{-1}\to {\bbox{\cal
K}}^{-1}$~:
\beqar
{{\bf W}^{SP}(\kvec) \over \text{N}_{S} - 1} = & - & {1\over V} \int\!
d\Omega\,d\Omega '\: {\bbox{\cal K}}^{-1}(\Omega,\Omega ') \, \nonumber\\
& & \times\, \exp[i \kvec\cdot(\rvec(\Omega) - \rrvec(\Omega '))]
\label{vdefAppB}
\eeqar
where ${\bbox{\cal K}}^{-1}$ is the single sphere generalized inverse of the
modified Oseen tensor ${\bbox{\cal G}}$~:
\beq
\int\! d\Omega\, d\Omega '' \: {\bbox{\cal K}}^{-1}(\Omega,\Omega '')\cdot
{\bf\cal G}(\rvec(\Omega ''),\rrvec(\Omega ')) \, = \, \openone \,\delta(\Omega
- \Omega ')
\label{kinvB}
\eeq
Expanding in (\ref{vdefAppB}) in spherical harmonics we get~:
\beqar
{\bf W}^{SP}(\kvec) & = & - c^{SP} \int\! d\Omega\, d\Omega '\:
\sum_{lm}\!\sum_{l'm'} \, \calKinv{lm;l'm'}\, \nonumber\\
& & \times\, \exp[i \kvec\cdot(\rvec - \rrvec)]
\, \text{Y}_{lm}(\Omega)\text{Y}^{\ast}_{l'm'}(\Omega ') \label{wsphexpB}
\eeqar 
But the quantities of interest (shear viscosity, diffusion coefficient)
are related to the transverse part of ${\bf W}^{SP}(\kvec)$, which should be
isotropic in the \kvec\ space (in our linearized model). Thus we can perform
the angular integral in the equation above by choosing $\kvec \parallel
\hat{{\bf z}}$ direction and employing the plane-wave expansion~: 
\beqar
\exp[i k (z &-& z')] = \sum_{l,l' = 0}^{\infty} (-1)^{l'}\,
i^{l + l'}(2l + 1)(2l' +1) \, \nonumber\\
& \times & \, j_{l}(ka)\, j_{l'}(ka) P_{l}(\cos\theta) P_{l'}(\cos\theta ') \\ 
& & \;\;\; \cos\theta = \hat{{\bf k}}\cdot\hat{\rvec}\; ;\;\;\;  \cos\theta ' =
\hat{\kvec}\cdot \hat{\rrvec} . \nonumber
\eeqar
we arrive at~:
\beqar
{\bf W}^{SP}(\kvec) & = & - 4\pi c^{SP}
\sum_{l=0}^{\infty}\!\sum_{l'=0}^{\infty} (-1)^{l'}  i^{l +
l'} \sqrt{(2 l+ 1)} \nonumber\\
& & \times \, \sqrt{(2l' + 1)} \, j_{l}(ka) j_{l'}(ka)\,\calKinv{l0;l'0}
\label{wbesselB}
\eeqar
Note that this equation is valid for any $k$ values. A particular value of the
self-energy ${\bf W}^{SP}(\kvec)$ that is meaningful and readily obtainable
from (\ref{wsphexpB}), is~:
\beq
{\bf W}^{SP}(k=0) \,=\, - 4\pi c^{SP} \calKinv{00;00}
\label{Bsigzero}
\eeq 

 For evaluating $\calKinv{lm;l'm'}$, we start by expanding the
definition (\ref{kinvB}) in spherical harmonics to transform it in the
equivalent form~: 
\beq 
\sum_{l''=0}^{\infty}\sum_{m'' = -l''}^{l''} \calKinv{lm;l''m''} \cdot
\calGind{l''m'';l'm'} \,= \, \openone\: \delta_{ll'}\,\delta_{mm'} 
\eeq
where the coefficients of the Oseen tensor expansion in spherical harmonics
are~:
\beq 
\calGind{lm;l'm'} = \int\!\!\!\int\! d\Omega\,d\Omega '\: {\bf\cal
G}[\rvec(\Omega),\rrvec(\Omega ')]\:\Ylmconj{lm}{\Omega} \Ylm{l'm'}{\Omega '}
\eeq
Substituting ${\bbox{\cal G}}(\rvec,\rrvec) = {\bbox{\cal G}}(\rvec - \rrvec)$
with its Fourier transform (calculated as in (\ref{Fourier})) yields~:
\beq 
\calGind{lm;l'm'} = {2\over\pi}\, (-1)^{l}\, i^{l + l'}\int_0^{\infty}\! dk\:
k^2 \,{\bessph{l}\bessph{l'}\over \eta_0 k^2 - W_{\perp}^{SP}(k)}\,
\PLind{P}{lm;l'm'} 
\eeq 
with $\bessph{l}$ the spherical Bessel functions of order $l$ and
$\PLind{P}{lm;l'm'}$ the tensor~:
\beq 
\PLind{P}{lm;l'm'} = \int\! d\Omega_k\: (\openone - \hat{\kvec}\hat{\kvec})\:
\Ylmconj{lm}{\Omega_k}\Ylm{l'm'}{\Omega_k} 
\eeq
Now inserting $W_{\perp}^{SP}(k)$ from (\ref{sigexp}) and integrating over the
$k$ variable produces~\cite{M82}~: 
\beqar 
\calGind{lm;l'm'} & = & {1 \over \eta_0 a (1 + W_1^{SP})} \:
\bessi{l+\case{1}{2}}\, \bessk{l' + \case{1}{2}} \, \nonumber\\ 
& & \times \, (\delta_{l\,l'} + \delta_{l\,l'+2}) \, \PLind{P}{lm;l'm'}
\label{GlmB}
\eeqar 
Here, I and K are the modified Bessel functions of the first kind and $\xi '$
is the effective screening length in the suspension of spheres from
(\ref{gsolsp}).

The construction of the inverse $\calKinv{lm;l'm'}$ requires an involved
procedure , described in Refs.~\cite{MF79,M82}. It is based on the observation
that $\bbox{\cal G}$---in its matrix representation---can be decomposed into a
block diagonal ($l=l'$) $\bbox{\cal G}_{D}$ and an off-diagonal $\bbox{\cal
G}_{OD}$ part ($l\neq l'$).  We could write then formally $\bbox{\cal G} =
\bbox{\cal G}_{D} + \bbox{\cal G}_{OD}$, which leads to, by taking the inner
product from left and right with $\bbox{\cal K}^{-1}$ and $\bbox{\cal
K}^{-1}_{D}$,respectively, and iterating indefinitely~:
\beq
\bbox{\cal K}^{-1} = \calKinv{D} -  \calKinv{D} \cdot\calGind{OD} \cdot
\calKinv{D}\, + \, \dots
\eeq 

We compute then first $(\calKinv{D})_{lm;lm'}$~:
\beq
(\calKinv{D})_{lm;lm'} = \eta_0 a \, (1 + W_{1}^{SP}) \, {1 \over \bessi{l +
\case{1}{2}} \, \bessk{l + \case{1}{2}}} \: \PLindinv{P}{lm;lm'}
\eeq
in which the last factor is the generalized inverse of $\PLind{P}{lm;lm'}$
defined by the relation~:
\beq
\sum_{m''=-l}^{l}\PLindinv{P}{lm;lm''}\cdot \PLind{P}{lm'';lm'} =
\openone\,\delta_{mm'} 
\eeq
and then one can calculate the non-diagonal part ({$l=l'+2$}) of
$\calKinv{lm;l'm'}$~:
\beq
\calKinv{lm;l'm'} = {\eta_0 a \, (1 + W_1^{SP}) \over \bessi{l'+\case{1}{2}}\,
\bessk{l+ \case{1}{2}}} \: \PLindinv{P}{lm;l'm'}
\eeq

Finally, we return  to (\ref{wbesselB}) to limit the expansion to factors
of order $k^2$ and using also the constraint $l-l' = \{0,2\}$ we get~:
\beqar
{\bf W}^{SP}(k\ll 1) & \simeq & -4\pi c^{SP} \left[ \calKinv{00;00}\,
{\text{j}}^2_0(ka) + 3\,\calKinv{10;10}\, {\text{j}}^2_1(ka)
\right. \nonumber\\ 
& - & \left. 2\sqrt{5}\,\bessph{2}\,\bessph{0} \,
\text{Re}\calKinv{20;00}\right] \label{wj0j2B} 
\eeqar

The actual values of the operators $\calKinv{lm;l'm'}$ are~:
\begin{mathletters}
\label{BKinvzero}
\beqar
\calKinv{00;00} & \; =\; & {3\over 2} \, a \eta_0 \, {1 + W_1^{SP} \over
\bessi{1/2} \, \bessk{1/2}} \: \openone \\
\calKinv{10;10} & \; =\; & {1\over 9} \, a \eta_0 \, {1 + W_1 \over \bessi{3/2}
\, \bessk{3/2}} \: \PLindinv{L}{10;10} 	\\
\calKinv{20;00} & \; =\; & {3 \over 8 \sqrt{30}}\, a \eta_0\, {1 + W_1^{SP}
\over \bessi{1/2} \, \bessk{5/2}} \: \PLindinv{L}{20;00} 
\eeqar
\end{mathletters}
with the $\PLindinv{L}{}$ tensors given by~:
\beqar
\PLindinv{L}{10;00} & \, = \, & 12 \,\openone - 2\, \hat{{\bf z}}\hat{{\bf z}}
\nonumber 	\\
\PLindinv{L}{20;00} & \, = \, & - 8 \sqrt{{3\over 2}}\, \openone +
24\sqrt{{3\over 2}}\, \hat{{\bf z}}\hat{{\bf z}} 
\eeqar

The last step in obtaining the self-energy of the suspension of fixed spheres
is to expand the spherical Bessel functions in (\ref{wj0j2B}) and, collecting
all factors, we get~:
\beqar
& {\bf W}& ^{SP}(\kvec) \, = \, - 6\pi c^{SP} a \eta_0 \, {1 + W_1^{SP} \over
\bessi{1/2} \bessk{1/2}}\, \openone \nonumber\\
& + & \, {3\over 2}\, \fsp\,  (1 + W_1^{SP})\, k^2 \left[ {1 \over
\bessi{1/2}\bessk{1/2}}\,\openone \right. \nonumber\\
& - &  {2\over 27}\, {1\over \bessi{3/2}
\bessk{3/2}}\, (12\,\openone - 2\,\hat{{\bf z}}\hat{{\bf z}}) 	\\
& + & \, \left. {1\over 10\sqrt{6}} \,{1 \over \bessi{1/2} \bessk{5/2}}\,\left(
-8\sqrt{\frac{3}{2}}\, \openone + 24\,\sqrt{\frac{3}{2}}\, \hat{{\bf
z}}\hat{{\bf z}} \right) \right] \nonumber 
\eeqar 
that reduces immediately to (\ref{wSP}), the equation we set to derive.

\section{Self-consistent calculation of ${\bf W}^{POL}(\kvec)$ for a
suspension of $\text{N}_P\/$ mobile polymer chains}

In order to compute the self-energy of a solution of polymers, we need to
make the transformations~:
\begin{itemize}
\item[a)] change the variable ${\bf R}_i$ to ${\bf R}(s)$ where $s$ is the
position of the $i$-th bead measured as the length of the arc along the chain,
that is $i = {s\over l}$, $l$ being the Kuhn length.
\item[b)] convert the sum $\sum_{i=1}^{n}(\cdots)$ into the integral ${1\over
l}\int_0^{l}\! ds\,(\cdots)$,and the Kronecker delta $\delta_{ij}$ to $l\,
\delta(s-s')$. 
\item[c)] the Fourier transform of a quantity $\la \bbox{A}(s,s')\ra$---the
average being taken over the distribution of the segments about the center of
mass of one polymer chain (assumed to be Gaussian)~:
\beqar
\la{\bf A}(s,s')\ra & = & \sum_{q,q' = -\infty}^{+\infty} {\bf
A}_{qq'}\exp\left( {2 i \pi q s\over L} - {2i\pi q's'\over L}\right)
\nonumber\\ 
{\bf A}_{qq'} & = & {1\over L^2} \int_{0}^{L}\!ds \int_{0}^{L}\! ds' \la{\bf
A}(s,s')\ra \\
& & \times \, \exp\left(-{2i\pi q s\over L} + {2 i \pi q's'\over
L}\right)\nonumber 
\eeqar
\end{itemize}

The generalized single chain inverse $\calKinv{\alpha}(\Sind{i},\Sind{j})$ of
the modified Oseen tensor ${\bbox{\cal G}}_{\alpha}(\Sind{i},\Sind{j})$ of the
polymer solution is defined by the relation~:
\beq
\sum_{j=1}^{n}\calKinv{\alpha}(\Sind{i},\Sind{j})\cdot{\bbox{\cal
G}}(\Sind{\alpha j},\Sind{\alpha k}) \, = \, \delta_{ik}\, \openone 
\eeq
Averaging over segments distribution and applying the previous transformations
we get~: 
\beq 
{1\over l^2}\int_{0}^{L}\! ds'\:
\calKinv{\alpha}(s,s')\cdot{\bbox{\cal G}}(s',s'') \, = \, \delta(s - s'')\,
\openone 
\eeq 
the Fourier transform of which being~: 
\beq
\sum_{q'=-\infty}^{+\infty}\calKinv{qq'}\cdot\calGind{q'q''} \; = \; {1\over
n^2} \delta_{qq''}\, \openone 
\eeq 
or, in operator notation ${\bbox{\cal K}}^{-1} = {1\over n^2} {\bbox{\cal
G}}^{-1}$.

Our starting point is Eq.\ (\ref{wdefPOL}) converted to an integral form~:
\beqar
& {\bf W} &^{POL}(\kvec) = - c^{POL}\, {1\over l^2}\int_0^{L}\! ds\int_0^{L}\!
ds'\: \la \exp[i \kvec\cdot ({\bf R}(s) \right. \nonumber\\
& & \;\;\; - \left. {\bf R}(s'))]\ra \left\{ \la{\bbox{\cal K}}^{-1}(s,s')\ra
\right. \label{wdefC}\\ 
& - & {1\over l^2} \, \la {\bf g}^{-1}_t\ra \cdot
\int_0^{L}\!\int_0^{L}\! dp\, dp'\: \left. \la{\bbox{\cal K}}^{-1}(s,p)\ra\cdot
\la {\bbox{\cal K}}^{-1}(p',s')\ra \right\}
\nonumber
\eeqar
where the averages are taken over the distribution of the segments about the
center of mass of the polymer chain. For a Gaussian probability distribution
function we have the following result~\cite{YAM}~:
\beq
\la\exp[i \kvec\cdot ({\bf R}(s) - {\bf R}(s'))]\ra = \exp\left[- {k^2 l |s -
s'| \over 6}\right]
\eeq
Inserting the above expression in (\ref{wdefC}), Fourier expanding the
$\la{\bbox{\cal K}}^{-1}\ra$ factors and using their definition gives~:
\begin{mathletters}
\beqar
& & {\bf W}^{POL}(\kvec) =  {\bf W}^{POL}_{a}(\kvec) + {\bf W}^{POL}_{b}(\kvec)
\nonumber\\ 
& & {\bf W}^{POL}_{a}(\kvec) =  - c^{POL} \sum_{q,q'=-\infty}^{+\infty}
\Iind{qq'} \, \calG{qq'}\, \openone  \\
& & {\bf W}^{POL}_{b}(\kvec) =  c^{POL} \la g^{-1}_t\ra\, {n^2 \over
l^2} \int_0^{L}\!\int_0^{L}\! dp\, dp'\: \sum_{q,q'}^{+\infty}\sum_{\omega
,\omega'}^{+\infty} \\
& & \; \Iind{qq'} \, \exp\left(- {2i\pi\omega p \over L} + {2i\pi\omega ' p'
\over L}\right) \, \calK{q\omega}\, \calK{\omega ' q'} \, \openone\nonumber
\eeqar
\end{mathletters}
with the notation~:
\beqar
\Iind{qq'} & = & \, {1\over L^2} \int_0^{L}\!\int_0^{L}\! ds\, ds'\:
\exp\left[- {k^2 l |s -s'| \over 6}\right] \nonumber\\
& & \,\times\, \exp\left[{2i\pi q s\over L} - {2i\pi q' s'\over L}\right]
\eeqar 
In the previous relations we implicitly assumed that all tensors are multiples
of the unit tensor, which will be shown to be true under certain
approximations.

The integrals over $p,p'$ in the ${\bf W}^{POL}_b$ expression are
straightforward, yielding~:
\beq
{\bf W}^{POL}_b(\kvec) = c^{POL} \, \la g^{-1}_t \ra\,  \sum_{qq'}\, \Iind{qq'}
\, \calG{q0}\,\calG{0q'}\, \openone
\eeq 
Now $\Iind{qq'}$ can be evaluated to be~:
\beqar
& & \Iind{qq'} = {2\over L^2} \left\{ \left({L \over 2\pi q}\right)^2 {k^2
R_g^2 \over 1 + \left({k^2 R_g^2 \over 2\pi q}\right)^2} \delta_{qq'}
\right. \nonumber\\
& & - \, \frac{\left[ \left({k^2 R_g^2 \over 2\pi}\right)^2 - qq'
\right]}{\left({2\pi\over L}\right)^2 q^2 q'^2 \left[1 + \left({k^2 R_g^2 \over
2\pi q}\right)^2 \right] \left[1 + \left({k^2 R_g^2 \over 2\pi q'}\right)^2
\right]} \label{IqqC}\\ 
& & \times\, \left. \left[ 1 - \exp(- k^2 R_g^2) \right] \right\}
\nonumber
\eeqar
in which $R_g = \sqrt{{n l^2 \over 6}}$ is the radius of gyration of any
polymer chain.  What remains to be calculated is the $\calGind{}^{-1}$
operator.

Approximating the transverse component of ${\bf W}^{POL}(\kvec)$ with
(\ref{wpoltrans}) and using the translational invariance of the suspension we
compute first the effective Oseen tensor~: 
\beqar
{\bbox{\cal G}}({\bf R}(s),{\bf R}(s')) & = & {\bbox{\cal G}}({\bf R}_s - {\bf
R}_s') \nonumber\\
& = & \, \int_{\kvec} {\bbox{\cal G}}(\kvec) \exp\left[-i \kvec\cdot({\bf R}_s
- {\bf R}_s' )\right] \nonumber\\
{\bbox{\cal G}}(\kvec) & = & {\unitkk \over \eta_0 k^2 -
W_{\perp}^{POL}(\kvec)}  \label{oseenkC}\\
& = & {\unitkk \over \eta_0 (1 + W_1^{POL})(k^2 + \xi
'^{-2}_{POL})} \nonumber\\ 
& & \xi '^{-2}_{POL} = {\xi^{-2} \over 1 + W_1^{POL}}\nonumber
\eeqar
where the Fourier integral is given by (\ref{Fourier}) and $\xi'_{POL}$ is the
effective screening length in the polymer solution.

Averaging over the distribution of the segments and integrating over the solid
angle, recalling that $W_{\perp}^{POL}(k)$ does not depend on the direction of
\kvec\ , we get~: 
\beqar
\la{\bbox{\cal G}}(s,s')\ra & = & {1\over 3 {\pi^2}}\int_0^{\infty}\! dk\: k^2
\, {1\over \eta_0 k^2 - W_{\perp}^{POL}(k)} \, \nonumber\\
& & \times \, \exp\left(- {k^2 l |s -s'| \over 6}\right)\, \openone 
\eeqar
Then, applying the Fourier transform defined at the beginning of the appendix,
one finds~: 
\beq 
\calGind{qq'} = {1\over 3 {\pi}^2} \int_0^{\infty}\! dk\: {k^2 \over \eta_0(1 +
W_1^{POL})(k^2 + {\xi '}^{-2}_{POL})} \, \text{I}_{qq'}(k)\, \openone
\label{GfourrierC}
\eeq
where we used $\int\! d\Omega_k\, (\unitkk) = {8\pi\over 3}\openone$ and
$\text{I}_{qq'}$  has been evaluated in (\ref{IqqC}).  

In order to make the calculation of ${\bbox{\cal G}}$ analytically tractable,
without coarsening the physical picture too much, we will make two
approximations~:
\begin{enumerate}
\item the Kirkwood-Riseman approximation,which amounts to discard the
off-diagonal terms in $\Iind{qq'}$.
\item the long chain limit, $n\gg 1$.
\end{enumerate}
Under these assumptions, both ${\bbox{\cal G}}$ and its inverse are diagonal
and their Fourier elements are simply related as~: 
\beq
(\bbox{{\cal G}}^{-1})_{qq'} = (\bbox{{\cal G}}^{-1})_q \delta_{qq'} =
(\bbox{{\cal G}}_q)^{-1}\, \delta_{qq'} 
\eeq 
For convenience we will denote these matrix elements by $\bbox{{\cal
G}}^{-1}_q$.

We will proceed further to evaluate ${\bf W}^{POL}(\kvec)$.  Noticing that
$\text{I}_q = \text{I}_{-q}$, one finds~:
\begin{mathletters}
\beqar
{\bf W}^{POL}_{a}(k) & = & - c^{POL} \left\{ \Ireal{0} \, \calG{0} + 2
\sum_{q=1}^{\infty} \Ireal{q} \, \calG{q}\right\}\, \openone 	\label{waqC}\\
{\bf W}^{POL}_{b}(k) & = & c^{POL} \, \Ireal{0} \, \calG{0}\, \openone 
\label{wbqC}
\eeqar
\end{mathletters}
in which we obtained $\la {\bf g}_t^{-1} \ra$ from~:
\beqar
& \la {\bf g}_t \ra & = \sum_{i,j=1}^{n}\la \calKinv{ij}\ra = {1\over
L^2}\int_0^{L}\!\!\int_0^{L}\! ds\, ds'\: \bbox{\cal G}^{-1}(s,s') =
\calGind{0}^{-1} \nonumber\\ 
& \la {\bf g}_t^{-1}\ra & = \bbox{\cal G}_0 
\eeqar

Working in the hydrodynamic limit $k\to 0$, we expand $\text{I}_q (k\ll 1)
\simeq {1\over \pi^2 q^2} k^2 R_g^2$ and converting the sum over $q$ to an
integral over the wavevector $\mu = {2\pi q\over L}$, we obtain the following
expression for ${\bf W}^{POL}(k)$ by adding the two terms
(\ref{waqC},\ref{wbqC})~:
\beq
{\bf W}^{POL}(k) = - {4\over \pi L} \, c^{POL} R_g^2 \, k^2\, \int_{{2\pi\over
L}}^{\infty}\! d\mu\: {1\over \mu ^2} \, \calG{\mu} \, \openone
\label{wmuC}
\eeq

To calculate the translational friction coefficient (and implicitly,the
diffusion coefficient), we have to consider one polymer chain
moving with some uniform velocity ${\bf u}_0$ inside an effective medium
replacing all others $\text{N}_{P}-1$ chains, medium characterized by the force
propagator ${\bbox{\cal G}}(\kvec)$ given by (\ref{oseenkC}).  The equation of
motion of the center of mass of the chain is~:
\beq
M {\ddot{{\bf R}}}^{0} = - \sum_{i=1}^{n} \bbox{\sigma}_i + {\bf f}_0
\eeq
$M$ is the mass of the chain and $\bbox{\sigma}_i$ is the force exerted by the
$i$-th bead of the polymer upon the fluid.  Neglecting the inertial term
(stationary problem) in the above equation of motion, the total external
average force upon the polymer chain is found to be~:
\beq
\la{\bf f}_0\ra = \la \sum_{i=1}^{n} \bbox{\sigma}_i \ra = \sum_{i,k}^{n} \la
\calKinv{ik}\ra \cdot {\bf u}_0
\eeq    
where the expression of $\bbox{\sigma}_i$ is a particular case of
(\ref{forcesAppA})---no spheres present and the other chains are {\sf absorbed}
in the effective fluid by replacing $\Kinv{ij}$ with $\calKinv{ij}$.  The
translational friction coefficient follows readily~:
\beq
\zeta_t \,\openone = \sum_{i,k}^{n} \la\calKinv{ik}\ra = \sum_{i,k}^{n}\la
\calK{ik}\ra \, \openone
\eeq
where first we had a configurational average as defined in (\ref{confav}) that
reduced to an average over the distribution of the segments
$\Sind{i}$, $\Sind{k}$ about the center of mass of the chain.  Introducing the
arc-length variables $s$ and $s'$ and applying the Fourier transform we
obtain for $\zeta_t$~:
\beq
\bbox{\zeta}_t  = \calG{0} \,\openone = \la g_t\ra\, \openone
\eeq
Next we insert (\ref{IqqC}) in (\ref{GfourrierC}) and  setting $q=q'=0$ we  
calculate ${\bbox{\cal G}}_0$~:
\beqar
{\bbox{\cal G}}_0 & = & \, {2\over 3 \pi^2}\, {1\over \eta_0 (1 + W_1^{POL})}\,
{1\over R_g \beta} \int_0^{\infty}\! dx\: {1\over x^2 + 1}\, \nonumber\\ 
& & \times \, \left\{1 - {1\over (\beta x)^2} [1 - \exp(- \beta^2 x^2)]\right\}
\, \openone
\eeqar
with $\beta$ a dimensionless variable~: $\beta  = R_g {\xi '}^{-1}_{POL}$. The
integral gives the exact result~:
\beqar
& & {\bbox{\cal G}}_0 = {1\over 3\pi}\, {1\over \eta_0 (1 + W_1^{POL})}\,
{1\over R_g} \, {\text{P}(\beta)\over \beta}\, \openone \nonumber\\ 
& & \text{P}(\beta) = 1 + {1\over \beta^2}\left[1 - \exp(\beta^2) + {2\over
\sqrt{\pi}} \exp(\beta^2) \Gamma(\case{3}{2},0,\beta^2)\right]
\label{gpAppC}\\ 
& & \Gamma (\case{3}{2},0,x) = \int_0^{x}\! dt\: t^{{1\over 2}}\exp(-t)
\nonumber  
\eeqar 
where the last expression is the incomplete Gamma function.

Thus the friction coefficient is given by~: 
\beq 
\zeta_t \,\openone = \calGind{0}^{-1} = 3\pi \eta_0 \, [1 + W_1^{POL}(\fpol)]
\, R_g \, {\beta \over \text{P}(\beta)} \,\openone
\label{frcoeffAppC}
\eeq

From (\ref{GfourrierC}), in the long-chain limit, $\bbox{\cal G}_{\mu}$
becomes~:
\beqar
\bbox{\cal G}_{\mu} & = & {1\over 3 \pi^2} \, {1\over \eta_o (1 + W_1^{POL})}\,
{2\over L^2} \int_0^{\infty}\! dp \: {p^2\over p^2 + {\xi'}^{-2}_{POL}}
\nonumber\\ 
& & \times \,{1\over \mu^2}\, {p^2 R_g^2 \over 1 + \left({p^2 R_g^2 \over \mu
L}\right)^2}\, \openone 
\eeqar 
where $\mu = {2\pi q\over L}$.  Working out the integral in terms of
dimensionless variables we get~: 
\beq 
\bbox{\cal G}_{\mu} = {1\over 6 \pi}\, {1\over \eta_0 (1 + W_1^{POL})}\,
{\sqrt{2}\over R_g \beta} \,{\sqrt{2} - \sqrt{{\mu L\over \beta^2}} +
\left({\mu L\over \beta^2}\right)^{3/2} \over 1 + \left({\mu L\over
\beta^2}\right)^2} \, \openone 
\eeq 
Also, $\calGind{\mu}^{-1}$ is immediately computed as~: $\calGind{\mu}^{-1} =
(\bbox{\cal G}_{\mu})^{-1}$.

Finally, changing the integration variable in (\ref{wmuC}) from $\mu$ to $x =
(\mu L/\beta^2)$ and defining the volume fraction of the polymers to be $\fpol
= {4\pi R_g^3\over 3} c^{POL}$, we calculate ${\bf W}^{POL}(\kvec)$~:
\beqar
{\bf W}^{POL}(\kvec) & = & -{36\over\pi} \,\eta_0\, (1 + W_1^{POL})\, \fpol \,
k^2 \, {1\over \beta} \nonumber\\
& & \times \, \int_{\sqrt{{2\pi\over \beta^2}}}^{\infty} \, dx\: {x^4 + 1\over
\sqrt{2}\,x^3(x^3 - x + \sqrt{2})} \: \openone
\eeqar
where the integration can be done exactly, leading eventually to the desired 
equation for the self-energy ${\bf W}^{POL}$ of a suspension of $\text{N}_P$
non-interacting, mobile polymer chains~:
\beqar
{\bf W}^{POL}(\kvec) & = & -{9\over\pi}\, \eta_0\, (1 + W_1^{POL})\,
\fpol\,k^2\, \text{Q}(\beta)\, \openone \label{sigpolAppC}\\ 
& &\text{Q}(\beta) = {1\over\beta} \, \ln \left(1 + {\beta \over \sqrt{\pi}}
\right) + {1\over\sqrt{\pi}} + {\beta\over 2\pi} \nonumber 
\eeqar 
with $\beta = {R_g \over {\xi '}_{POL}}$.

As a last note, we present here the series expansions and the asymptotic
behavior of the functions $\text{P}(x)$ and $\text{Q}(x)$ introduced in the
previous calculations~:
\beq
x\ll 1 \;\;\: \left\{
\begin{array}{rcl}
\text{P}(x) & \simeq & {4\over 3\sqrt{\pi}}x - {1\over 2}x^2 + {8\over
15\sqrt{\pi}} x^3 - \dots  \\
\\
\text{Q}(x) & \simeq & {2\over \sqrt{\pi}} + {1\over 3\pi\sqrt{\pi}}\,x^2 - {1
\over 4 \pi^2}\, x^3 + \dots 
\end{array} \right.
\eeq
\beq
x\gg 1 \;\;\;\; \left\{
\begin{array}{rcl}
\text{P}(x) & \simeq & 1 - {2\over \sqrt{\pi} x} + {1\over x^2} - {1\over
\sqrt{\pi} x^3} + \dots \\
\\
\text{Q}(x) & \simeq & {1\over 2 \pi}\, x + {1\over \sqrt{\pi}} + \dots
\end{array} \right.
\eeq

\end{multicols}

\twocolumn

\begin{figure}
\caption{The physical interpretation of the multiple scattering expression of
the microscopic velocity field ${\bf v}(\rvec)$.  O = observation point; {\bf
G} = the Oseen tensor of the pure fluid; ${\bf T}^{\text{OBJ}}$ =
single-object velocity propagator.}
\end{figure}

\begin{figure}
\caption{Diffusion coefficient $D$ function of $\fpol$ for the {\bf 1S/P} limit
case and for the {\bf S/P} system with $\fsp$ parameter; all $t$ values. From
top to bottom~: $\bbox{+}$~1 sphere limit; $\Diamond$~$\fsp = 0.01$;
$\triangle$~$\fsp = 0.11$; $\bigcirc$~$\fsp = 0.21$; $\Box$~$\fsp = 0.31$.}
\label{dSP1SP_P}
\end{figure}

\begin{figure}
\caption{Diffusion coefficient $D$ function of $\fsp$ for a probe sphere moving
among other fixed spheres in a suspension of polymers ({\bf S/P} system), with
$\fpol$ parameter; all $t$ values. From top to bottom~:
$\Diamond$~$\fpol=0.01$; $\triangle$~$\fpol =0.09$; $\bigcirc$~$\fpol = 0.16$;
$\Box$ $\fpol = 0.24$.}
\label{dSP_S}
\end{figure}

\begin{figure}
\caption{Total viscosity change $\veff$ function of $\fpol$ for the {\bf 1S/P}
limit case and for the {\bf S/P} system with $\fsp$ parameter; all $t$ values.
From bottom to top~: $\bbox{+}$~1 sphere limit; $\Diamond$~$\fsp = 0.01$;
$\triangle$~$\fsp = 0.14$; $\bigcirc$~$\fsp = 0.26$; $\Box$~$\fsp = 0.39$.}
\label{vT_SP1SP_P}
\end{figure}

\begin{figure}
\caption{Total viscosity change $\veff$ function of $\fsp$ for the {\bf S/P}
system, with $\fpol$ parameter; all $t$ values. From bottom to top~:
$\Diamond$~$\fpol = 0.01$; $\triangle$~$\fpol = 0.09$;
$\bigcirc$~$\fpol = 0.16$; $\Box$~$\fpol = 0.24$.}
\label{vT_SP_S}
\end{figure}

\begin{figure}
\caption{Relative viscosity change $\vrelspol$ function of $\fpol$ for the {\bf
S/P} system, with $\fsp$ parameter; all $t$ values. From bottom to top~:
$\Diamond$~$\fsp = 0.09$; $\triangle$~$\fsp = 0.24$; $\bigcirc$~$\fsp = 0.39$;
$\Box$~$\fsp = 0.46$.}
\label{six}
\end{figure}


\begin{figure}
\caption{Diffusion coefficient $D$ function of $\fsp$ for the {\bf 1P/S} limit
and for the {\bf P/S} system with $\fpol$ parameter; $t = 0.01$. From top curve
to bottom curve, Figs.~[7--10]~: $\bbox{+}$~1 chain limit; $\Diamond$~$\fpol =
0.01$; $\triangle$~$\fpol = 0.09$; $\bigcirc$~$\fpol = 0.16$; $\Box$~$\fpol =
0.24$.}
\label{dPS1PS_S_t001}
\end{figure}

\begin{figure}
\caption{Diffusion coefficient $D$ function of $\fsp$. \ {\bf 1P/S} and {\bf
P/S} systems, with $\fpol$ parameter; $t = 0.1$.}
\label{dPS1PS_S_t01}
\end{figure}

\begin{figure}
\caption{Diffusion coefficient $D$ function of $\fsp$. \ {\bf 1P/S} and {\bf
P/S} systems, with $\fpol$ parameter; $t = 1$.}
\label{dPS1PS_S_t1}
\end{figure}

\begin{figure}
\caption{Diffusion coefficient $D$ function of $\fsp$. \ {\bf 1P/S} and {\bf
P/S} systems, with $\fpol$ parameter; $t = 10$.}
\label{dPS1PS_S_t10}
\end{figure}

\begin{figure}
\caption{Differences in the diffusion coefficient of the polymer chain function
of $\fsp$ for four pairs of $t$ values ({\bf 1P/S} system). From top to
bottom,following the maximum of each curve~: $\bigcirc$~$D(t=0.01)-D(t=10)$;
$\Diamond$~$D(t=1) - D(t=10)$; $\triangle$~$D(t=0.01)-D(t=1)$;
$\Box$~$D(t=0.01) - D(t=0.1)$.}
\label{eleven}
\end{figure}


\begin{figure}
\caption{Diffusion coefficient $D$ function of $\fpol$ for one mobile polymer
chain inside a dispersion of other polymer chains and fixed spheres ({\bf P/S}
system), with $\fsp$ parameter; $t = 0.1$.  From top to bottom ,
Figs.~[12--14]~: $\Diamond$~$\fsp = 0.01$; $\triangle$~$\fsp = 0.14$;
$\bigcirc$~$\fsp = 0.26$; $\Box$~$\fsp = 0.39$.}
\label{dPS_P_t01}
\end{figure}

\begin{figure}
\caption{Diffusion coefficient $D$ function of $\fpol$. \ {\bf P/S} system,
with $\fsp$ parameter; $t = 1$.}
\label{dPS_P_t1}
\end{figure}

\begin{figure}
\caption{Diffusion coefficient $D$ function of $\fpol$. \ {\bf P/S} system,
with $\fsp$ parameter; $t = 10$.}
\label{dPS_P_t10}
\end{figure}


\begin{figure}
\caption{Total viscosity change $\veff$ function of $\fpol$. \ {\bf P/S}
system, with $\fsp$ parameter; $t = 0.01$. From bottom to top curve~:
$\Diamond$~$\fsp = 0.01$; $\triangle$~$\fsp = 0.14$; $\bigcirc$~$\fsp = 0.26$;
$\Box$~$\fsp = 0.39$.}
\label{vT_PS_P_t001}
\end{figure}

\begin{figure}
\caption{Total viscosity change $\veff$ function of $\fpol$. \ {\bf P/S}
system, with $\fsp$ parameter; $t = 10$. From bottom to top curve~:
$\Diamond$~$\fsp = 0.01$; $\triangle$~$\fsp = 0.14$; $\bigcirc$~$\fsp = 0.26$;
$\Box$~$\fsp = 0.39$.}
\label{vT_PS_P_t10}
\end{figure}

\begin{figure}
\caption{Total viscosity change $\veff$ function of $\fsp$. \ {\bf P/S} system,
with $\fpol$ parameter; $t = 0.01$.  From bottom to top curve, Figs.~[17--18]~:
$\Diamond$~$\fpol = 0.01$; $\triangle$~$\fpol = 0.09$; $\bigcirc$~$\fpol =
0.16$; $\Box$~$\fpol = 0.24$.}
\label{vT_PS_S_t001}
\end{figure}

\begin{figure}
\caption{Total viscosity change $\veff$ function of $\fsp$. \ {\bf P/S} system,
with $\fpol$ parameter; $t = 10$.}
\label{vT_PS_S_t10}
\end{figure}


\begin{figure}
\caption{The relative viscosity change $\vrelpolsp$ as a function of $\fpol$
for the {\bf P/S} system, with $\fsp$ parameter; $t = 1$ when symbols are
joined with lines and $t = 10$ when they are isolated. The corresponding $\fsp$
values are~: $\Diamond$ and $\Diamond$--$\Diamond$~$\fsp = 0.01$;
$\triangle$~$\fsp = 0.14$; $\bigcirc$~$\fsp = 0.26$; $\Box$ and
$\Box$--$\Box$~$\fsp = 0.39$.}
\label{vR_PS_P}
\end{figure}

\begin{figure}
\caption{The relative viscosity change $\vrelpolsp$ as a function of
$\fsp$. {\bf P/S} system, with $\fpol$ parameter; $t = 0.01$. From bottom to
top curve, $\triangle$~$\fpol = 0.09$; $\bigcirc$~$\fpol = 0.16$; $\Box$~$\fpol
= 0.24$}
\label{vR_PS_S_t001}
\end{figure}

\begin{figure}
\caption{The relative viscosity change $\vrelpolsp$ as a function of
$\fsp$. {\bf P/S} system, with $\fpol$ parameter; $t = 1$ when symbols are
joined with lines and $t = 10$ when they are isolated. The corresponding
$\fpol$ values are~: $\triangle$ and $\triangle$--$\triangle$~$\fpol = 0.09$;
$\bigcirc$ and $\bigcirc$--$\bigcirc$~$\fpol = 0.16$; $\Box$ and
$\Box$--$\Box$~$\fsp = 0.24$.}
\label{vR_PS_S_1t10}
\end{figure}

$${\epsfxsize=8.2 truecm \epsfbox{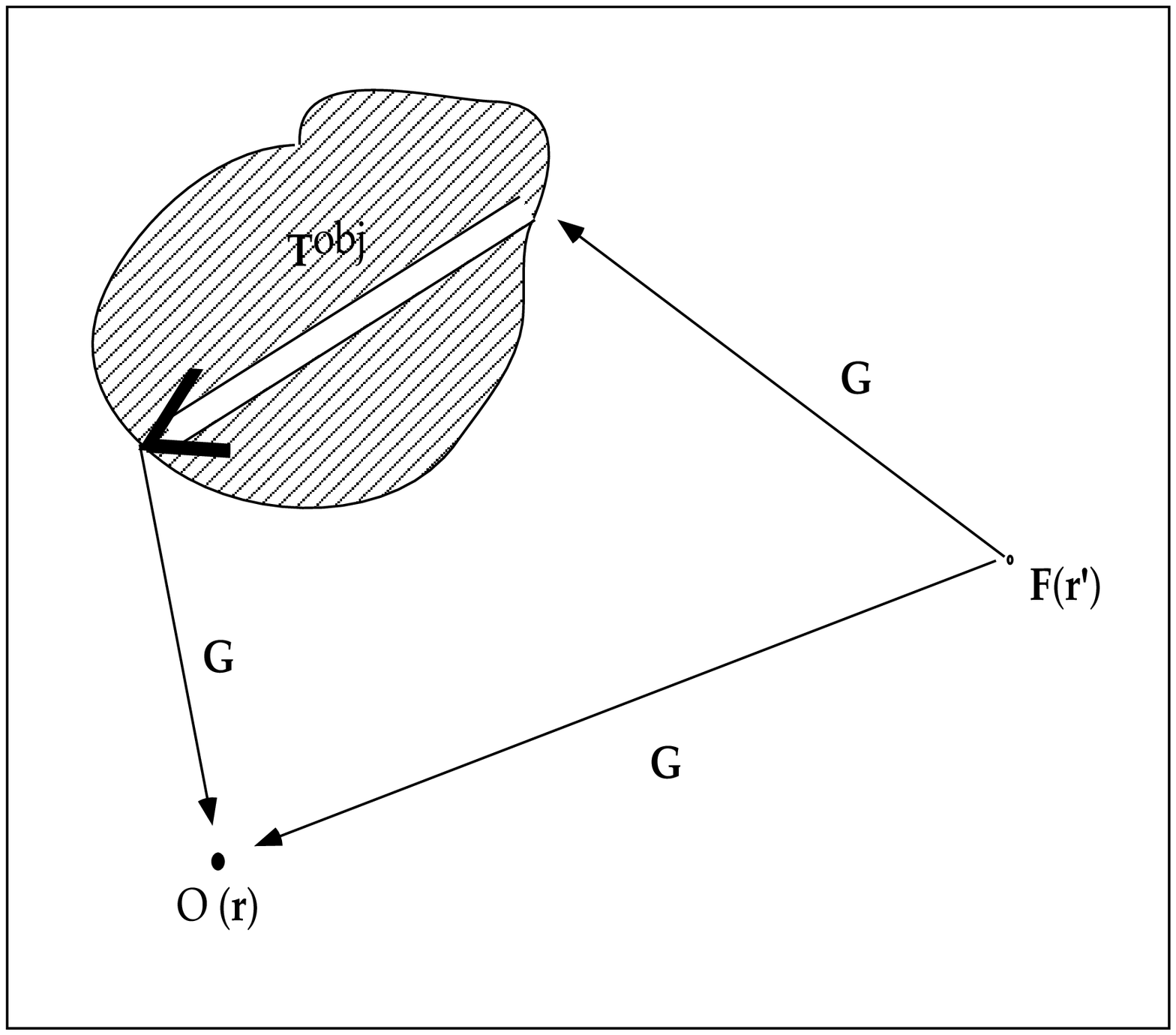}} $$
\begin{center} FIG.~1
\end{center}

\onecolumn
\widetext

$${\epsfxsize=8.2 truecm \epsfbox{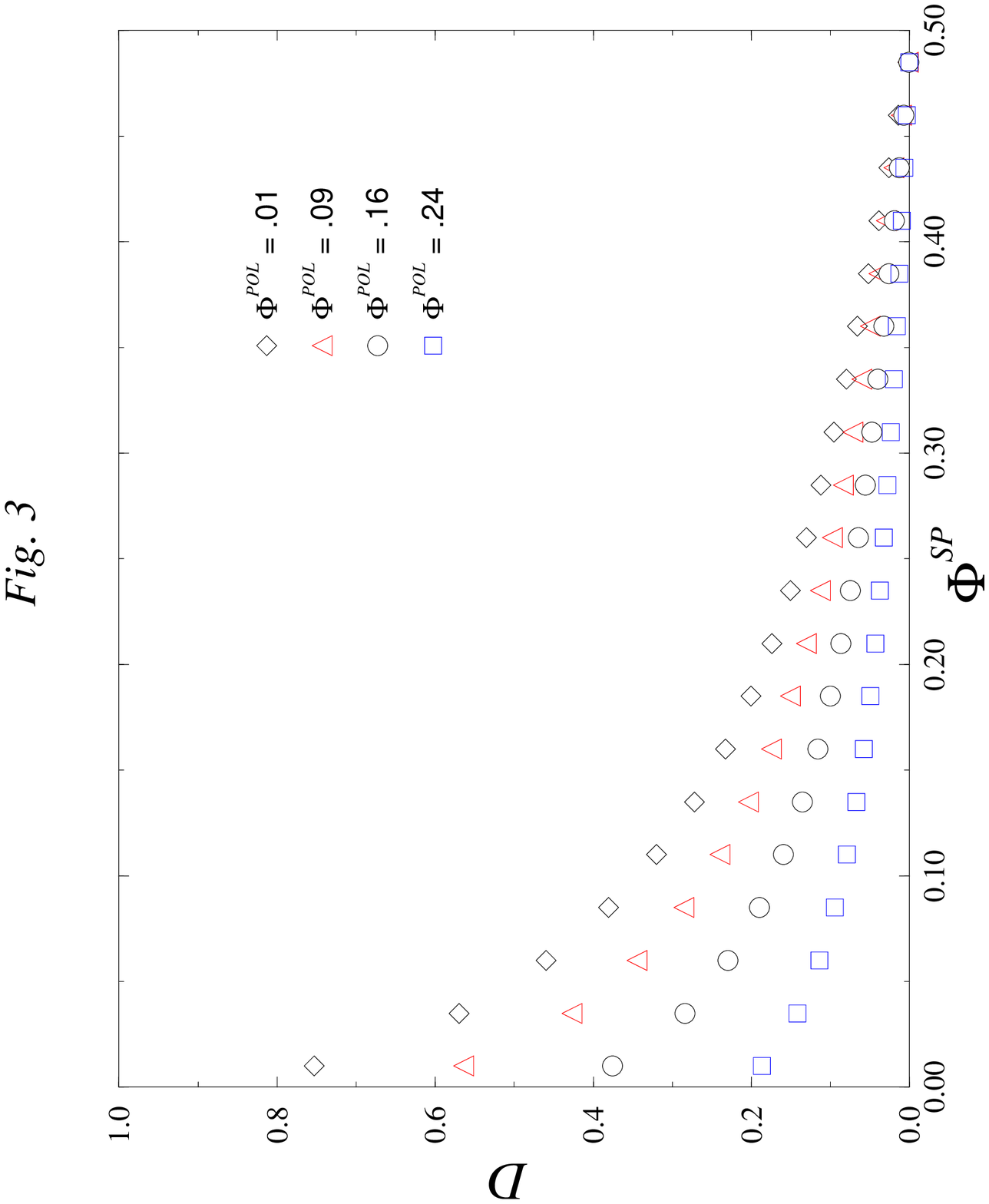}} \;\; {\epsfxsize=8.2 truecm
\epsfbox{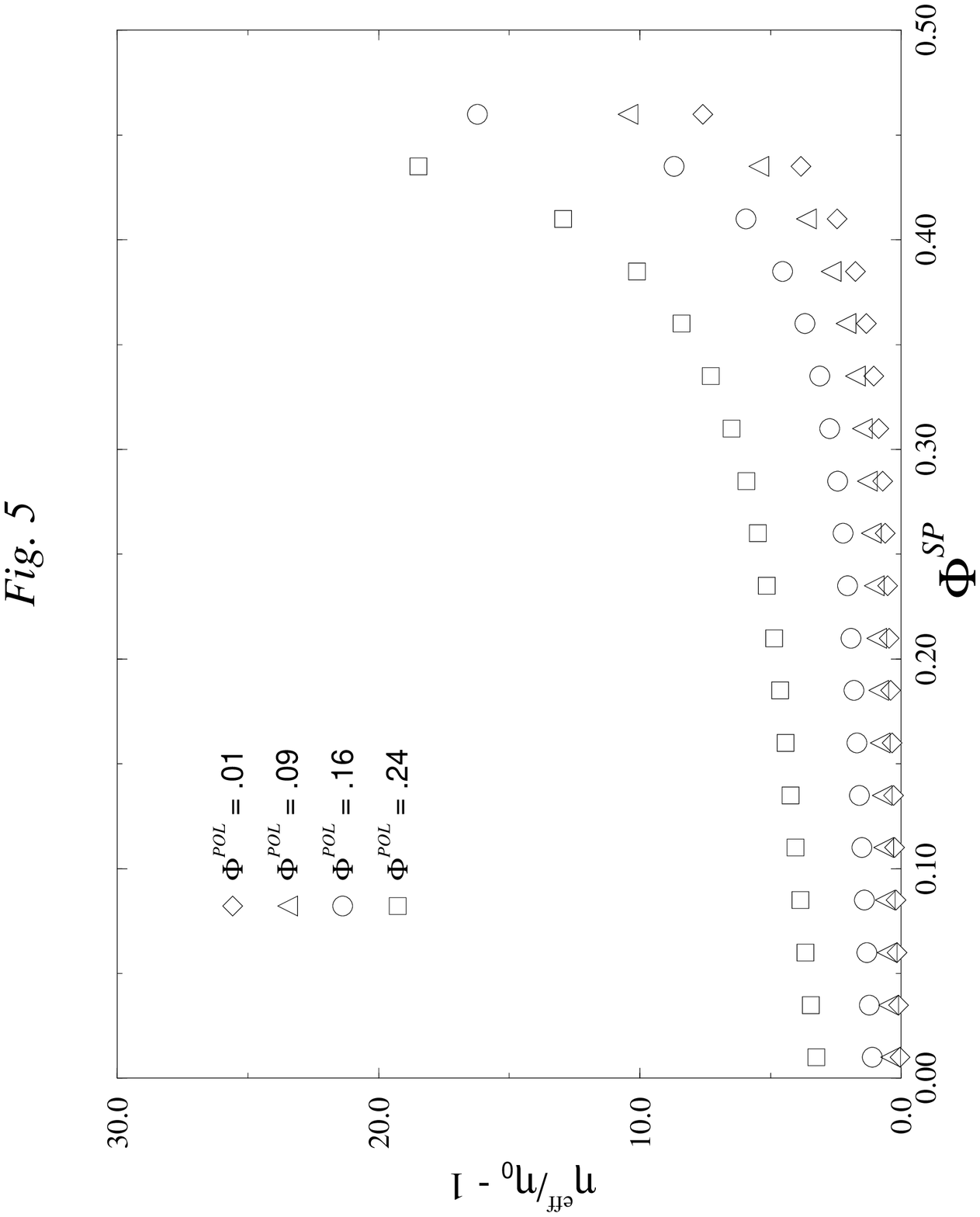}} $$ 
$${\epsfxsize=8.2 truecm \epsfbox{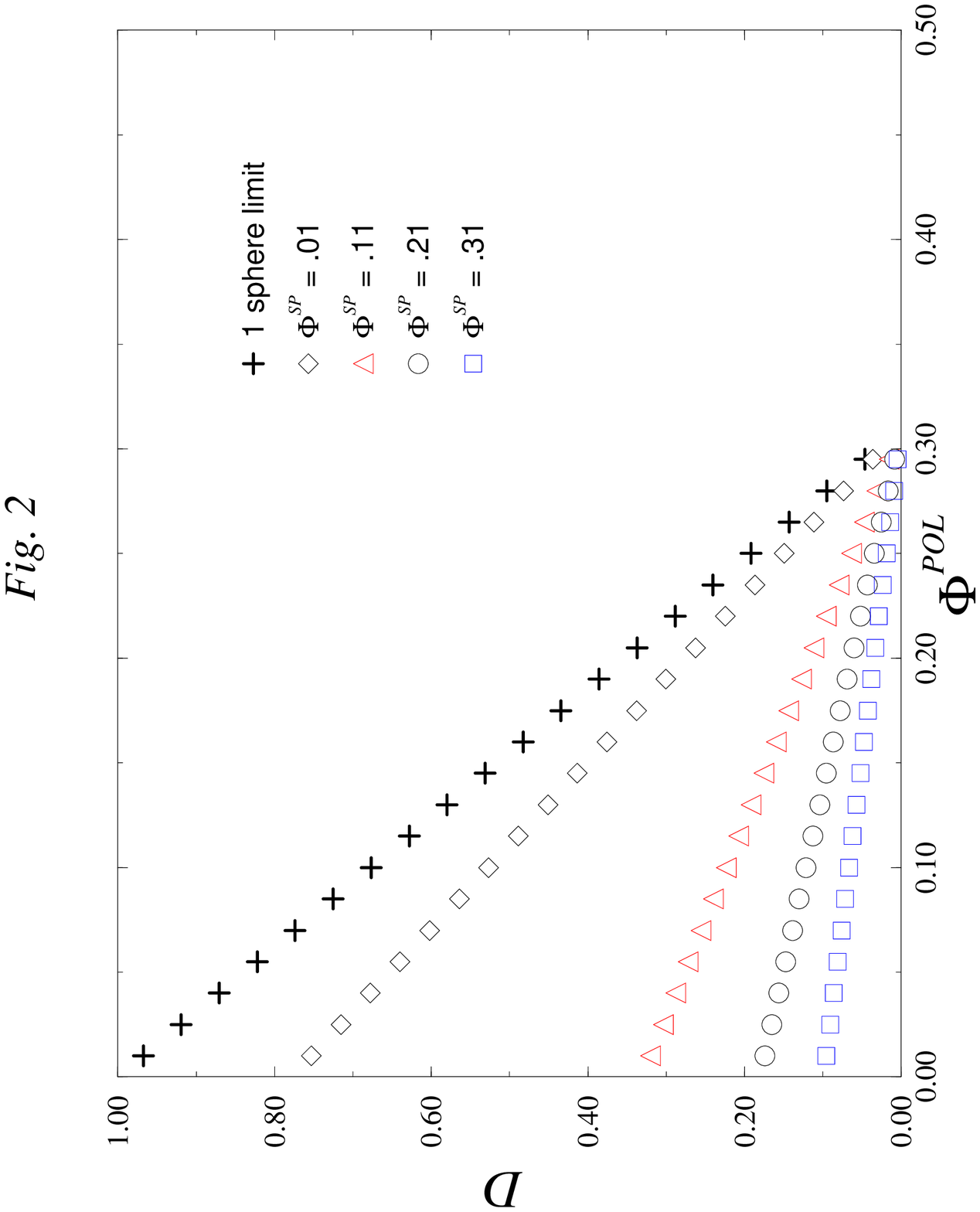}} \;\; {\epsfxsize=8.2 truecm
\epsfbox{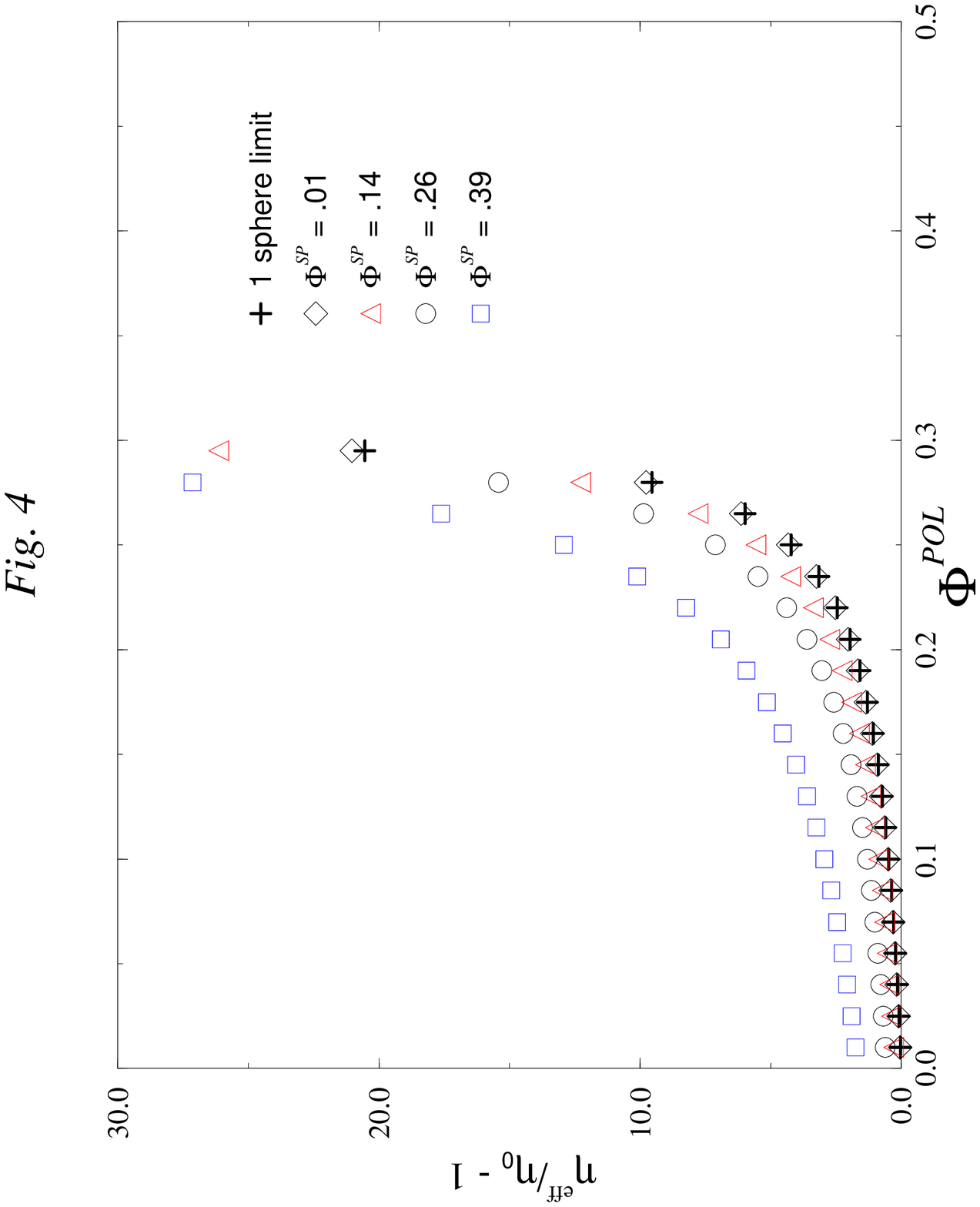}}$$ 

$$ {\epsfxsize=8.2 truecm \epsfbox{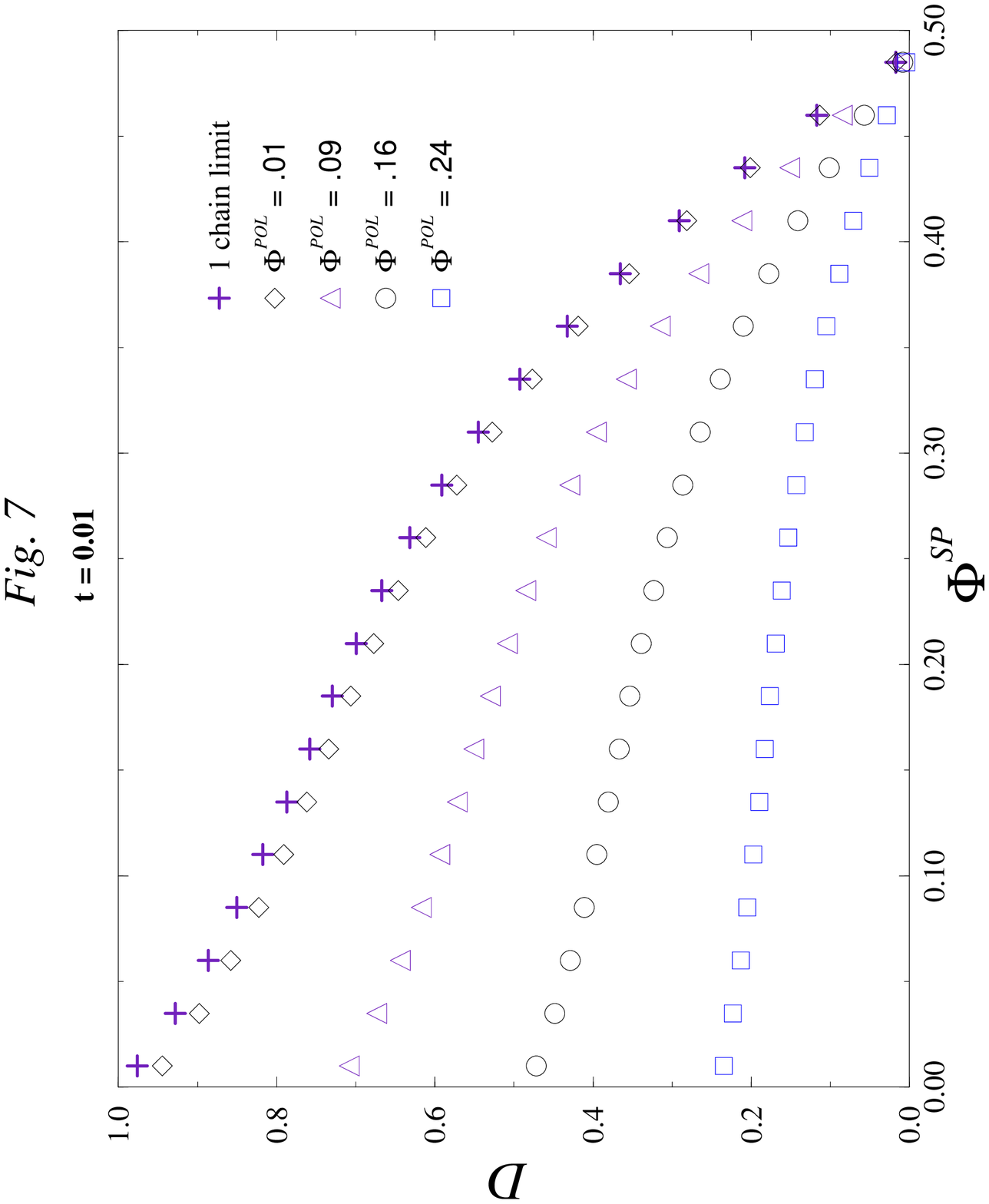}} \;\; {\epsfxsize=8.2 truecm
\epsfbox{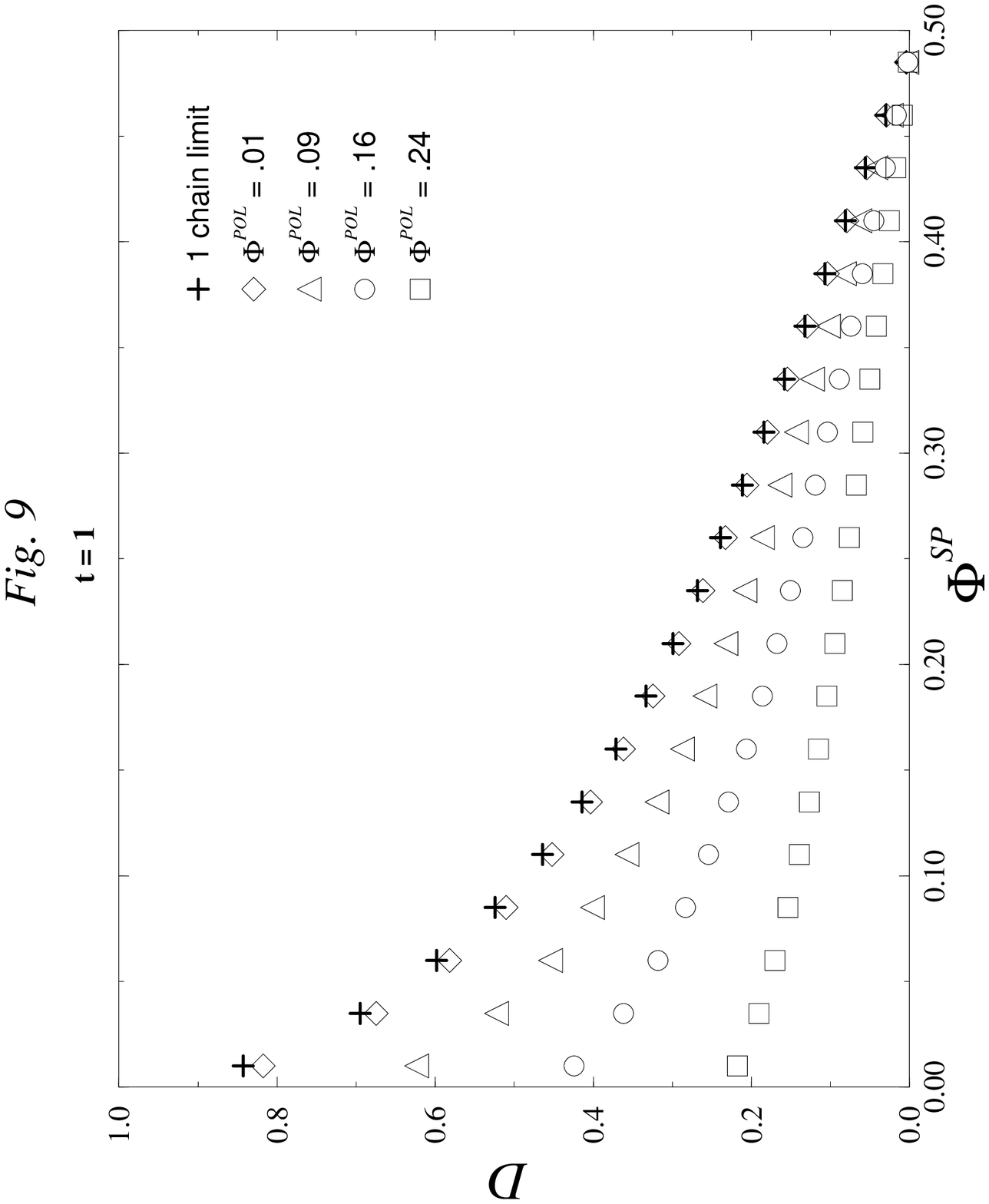}}$$ 
$$ {\epsfxsize=8.2 truecm \epsfbox{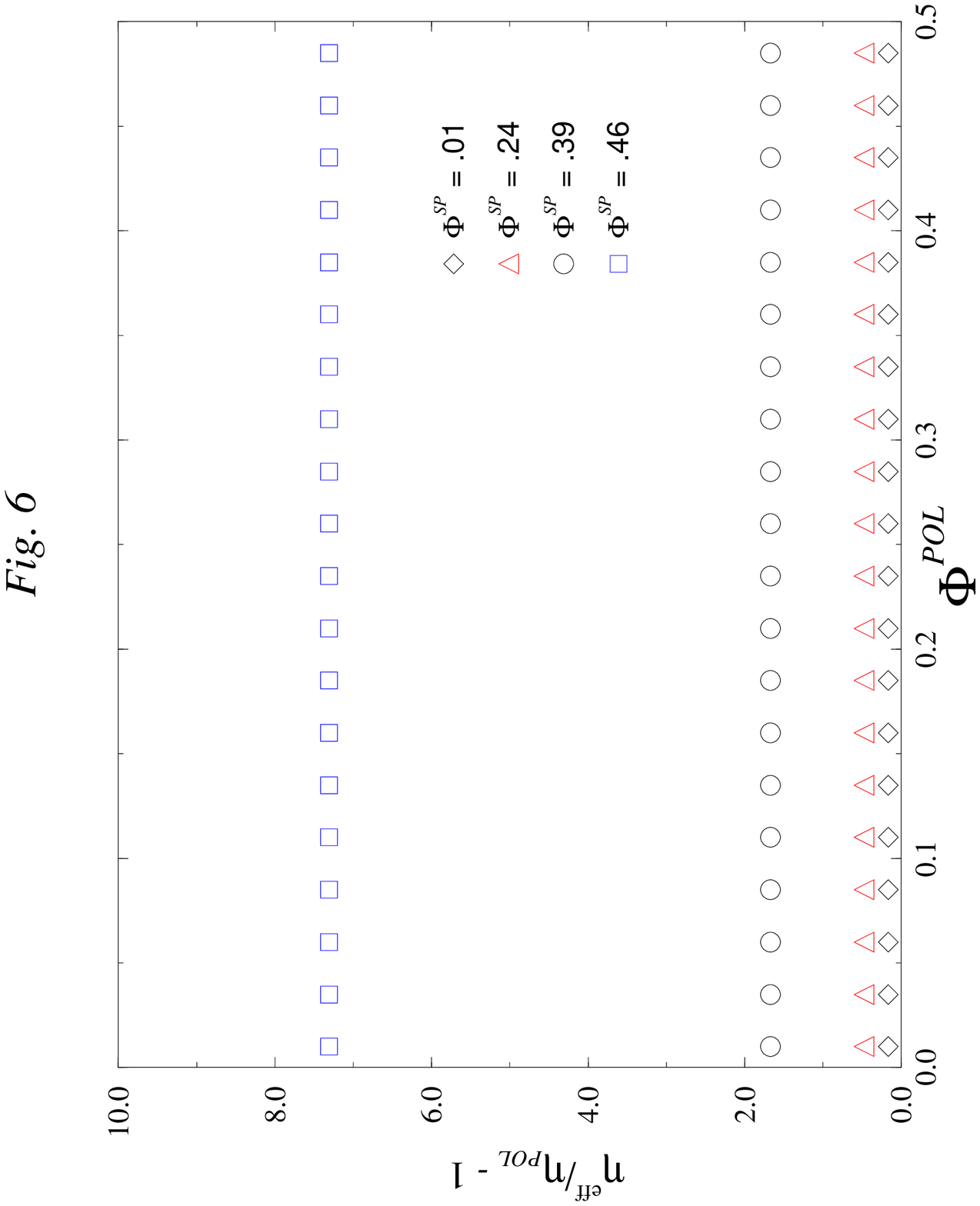}} \;\; {\epsfxsize=8.2 truecm
\epsfbox{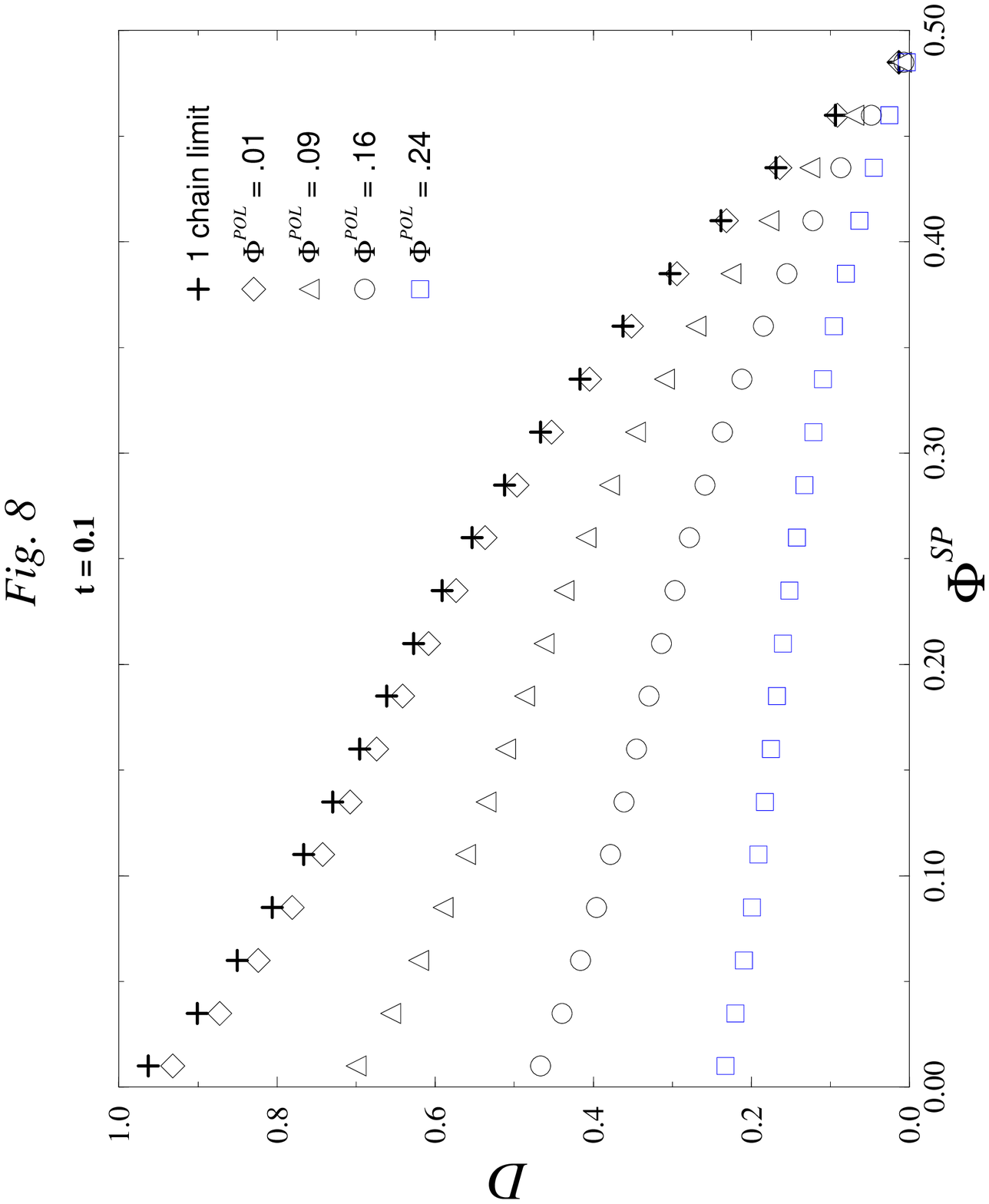}}$$ 

$$ {\epsfxsize=8.2 truecm \epsfbox{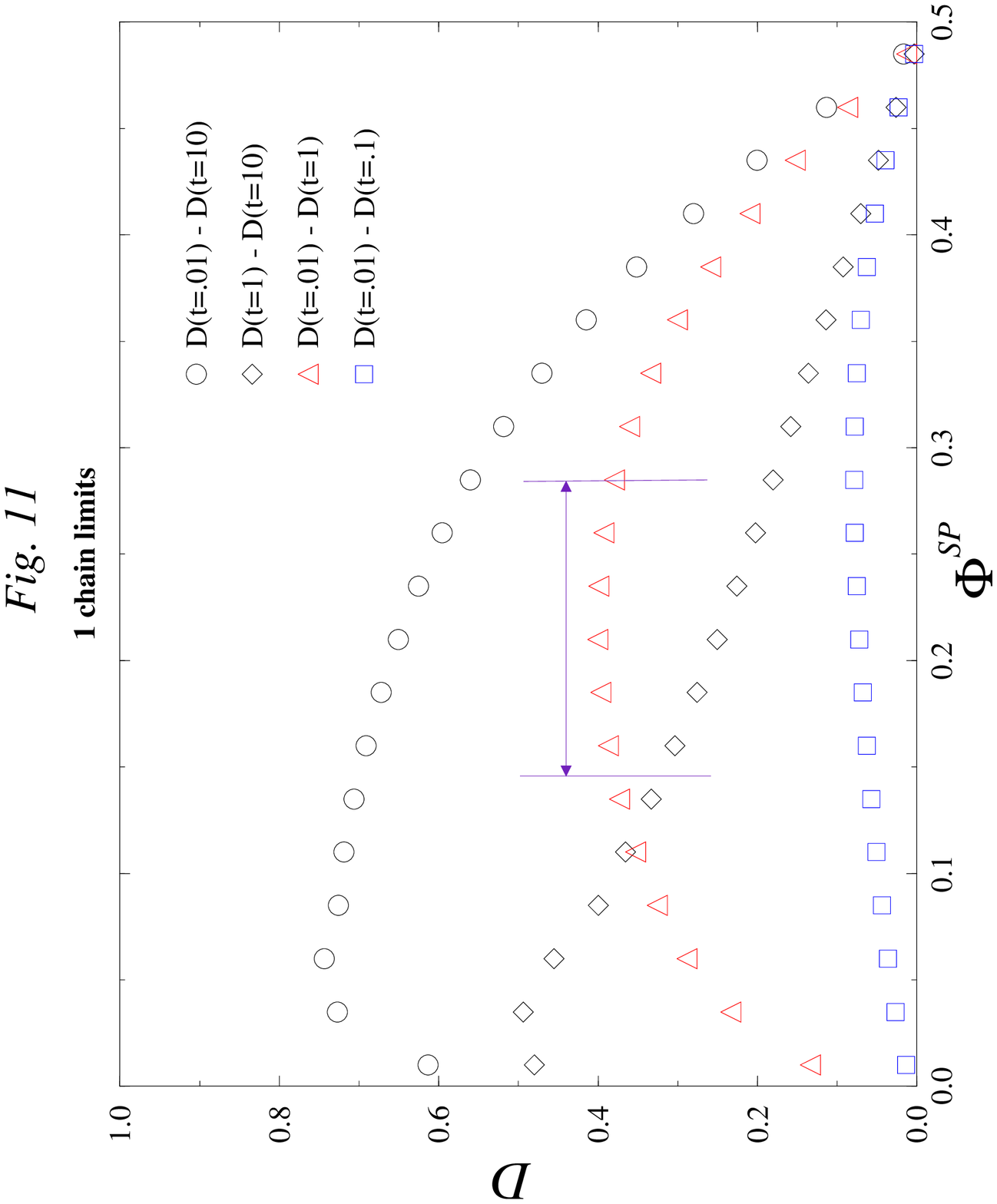}} \;\; {\epsfxsize=8.2
truecm \epsfbox{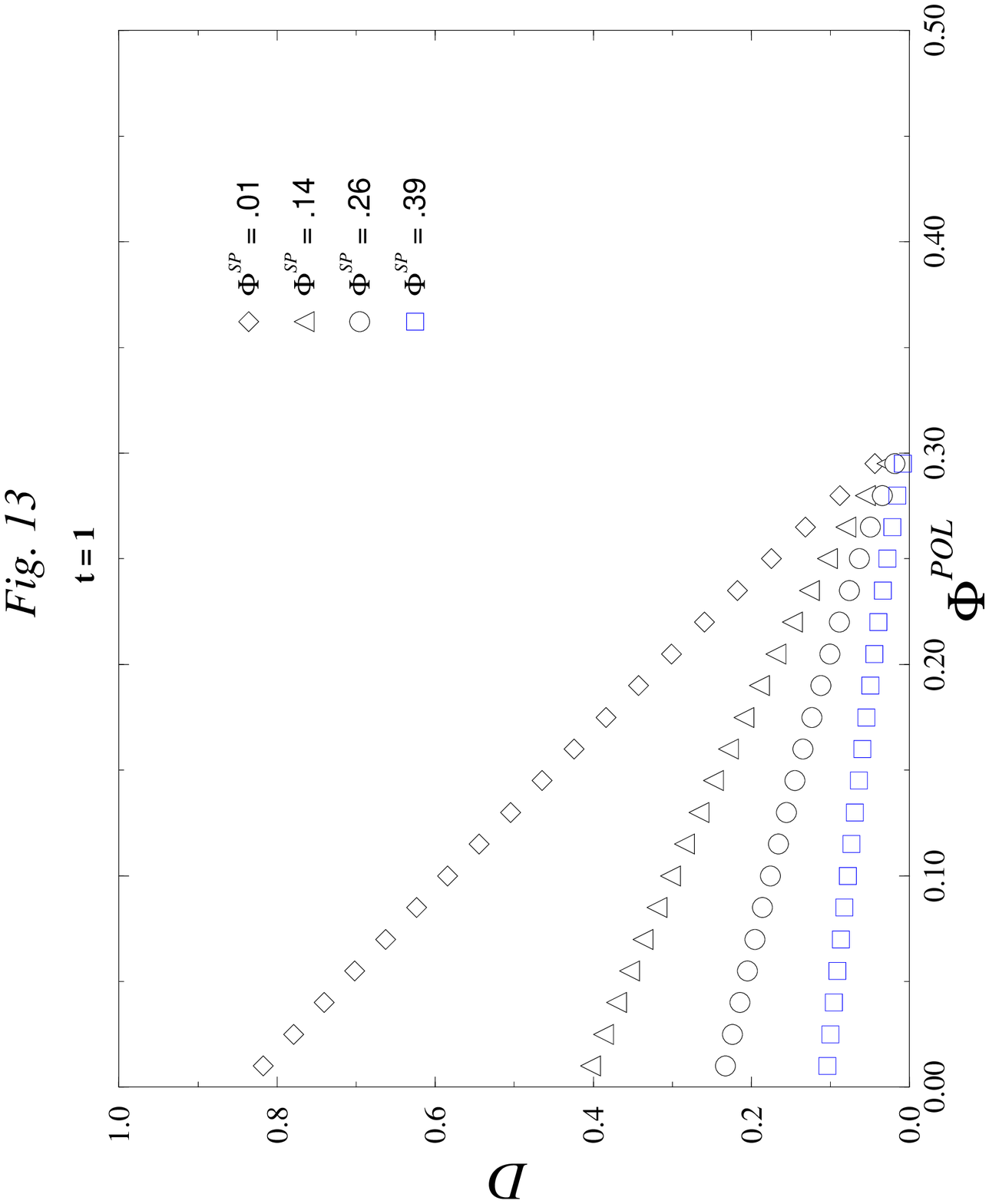}}$$
$$ {\epsfxsize=8.2 truecm \epsfbox{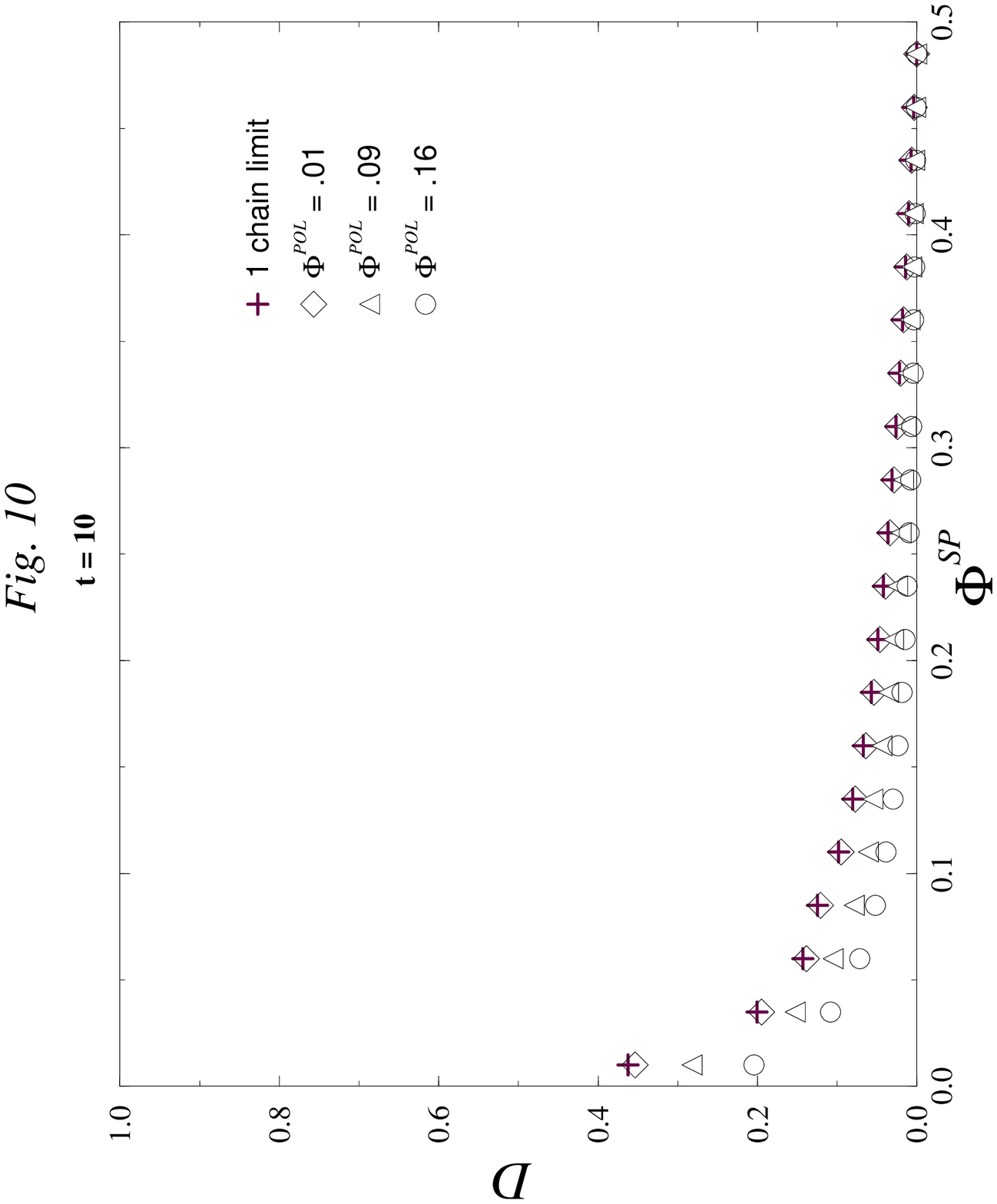}} \;\; {\epsfxsize=8.2
truecm \epsfbox{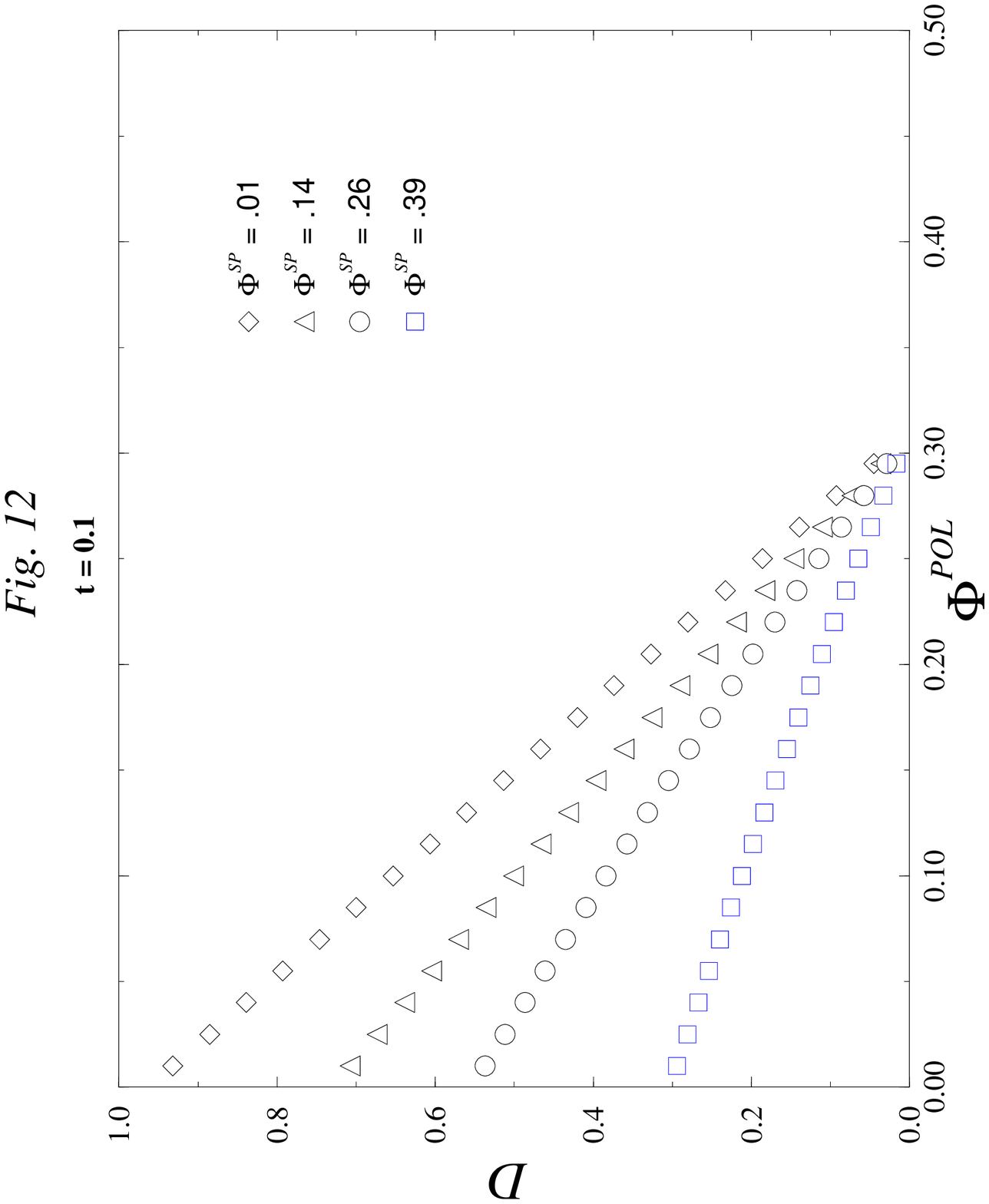}}$$

$$ {\epsfxsize=8.2 truecm \epsfbox{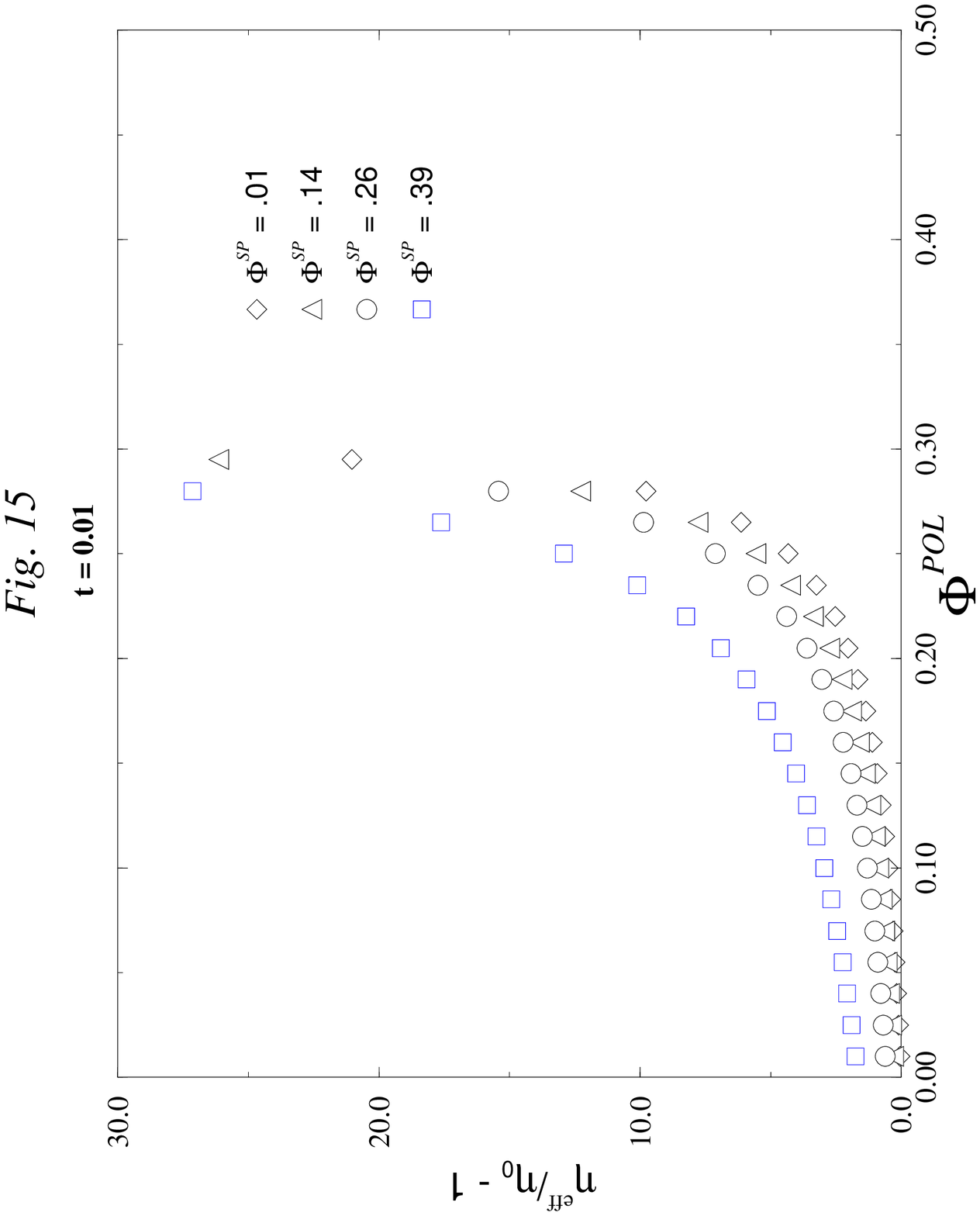}} \;\; {\epsfxsize=8.2
truecm \epsfbox{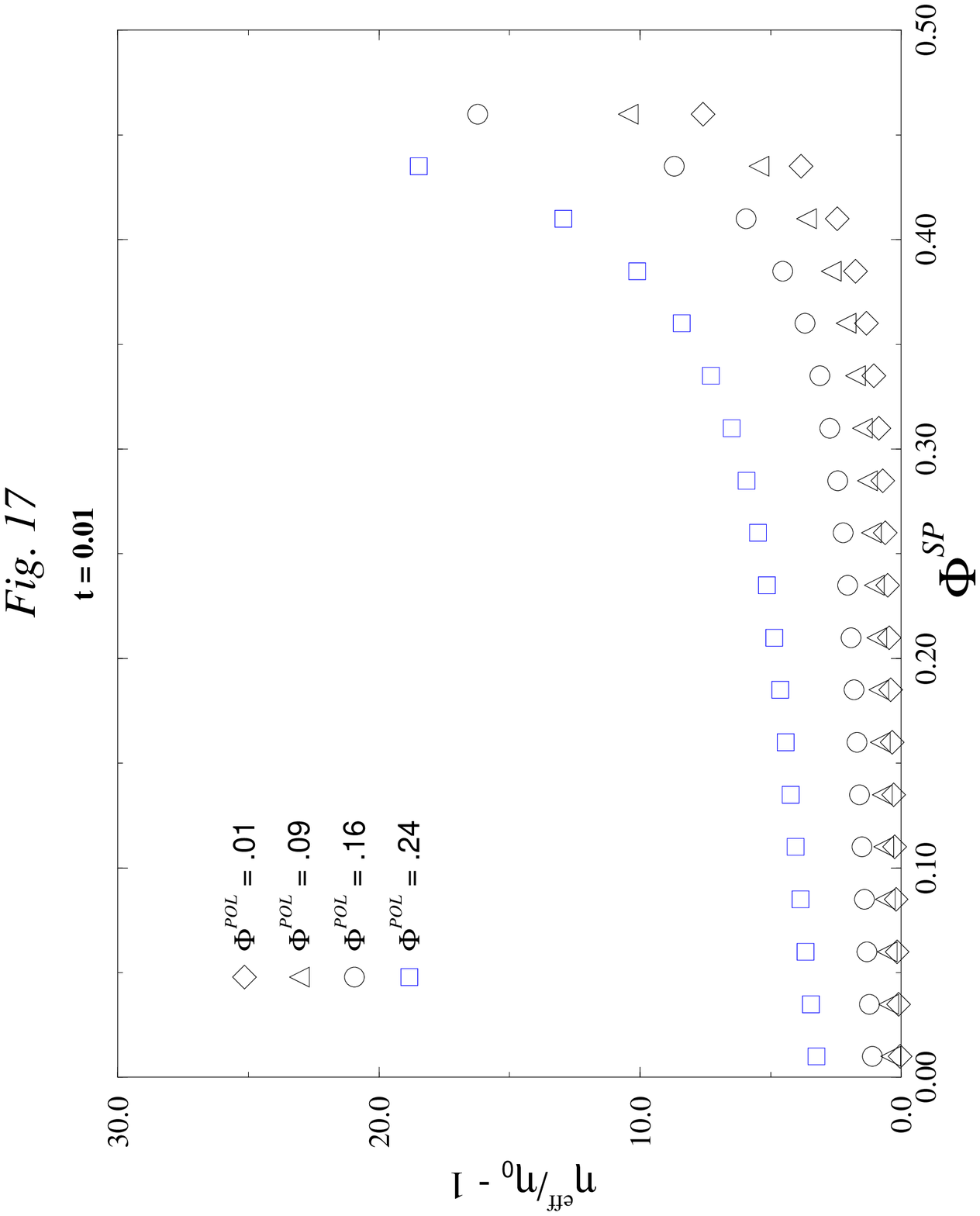}}$$
$$ {\epsfxsize=8.2 truecm \epsfbox{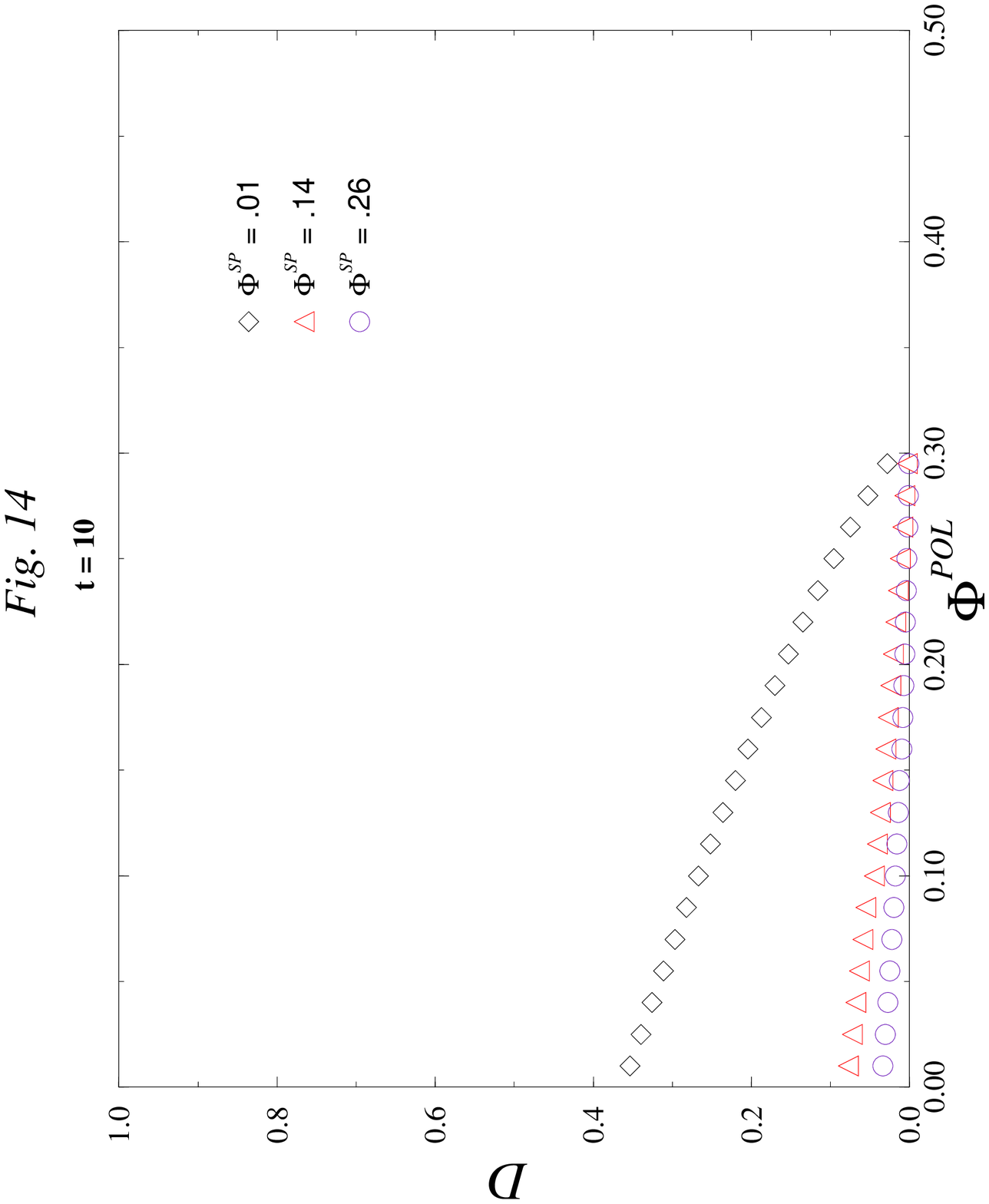}} \;\; {\epsfxsize=8.2
truecm \epsfbox{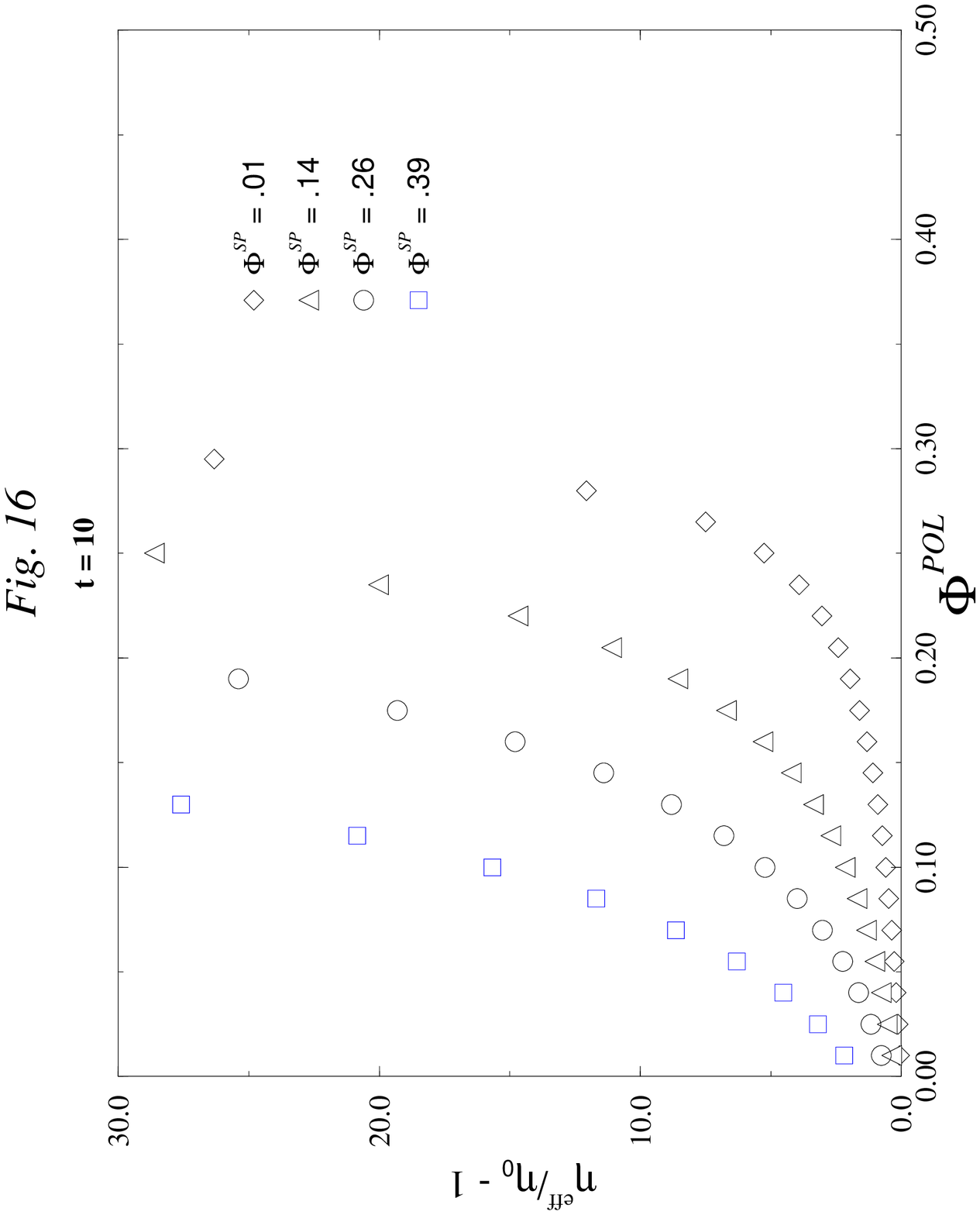}}$$

$$ {\epsfxsize=8.2 truecm \epsfbox{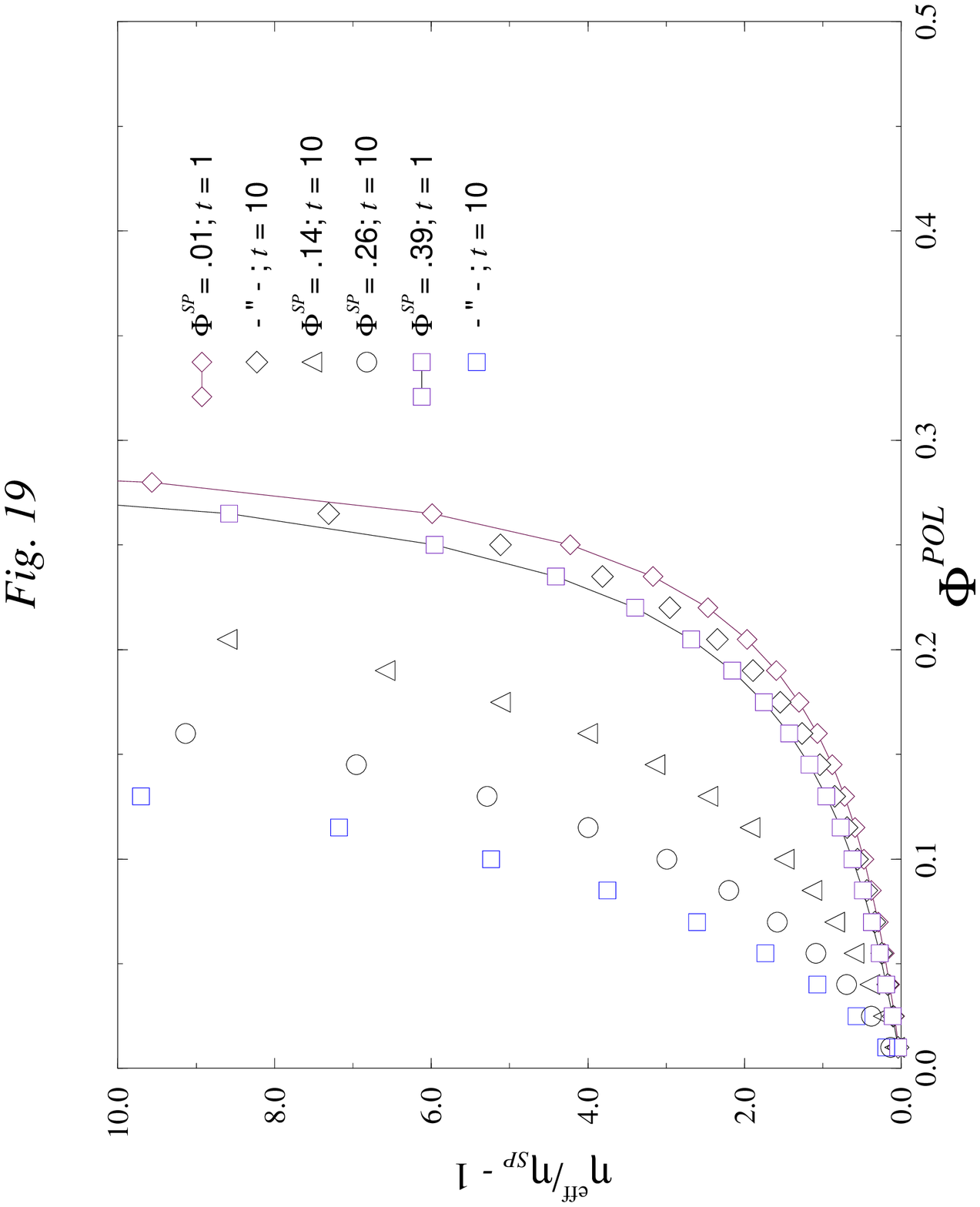}} \;\; {\epsfxsize=8.2
truecm \epsfbox{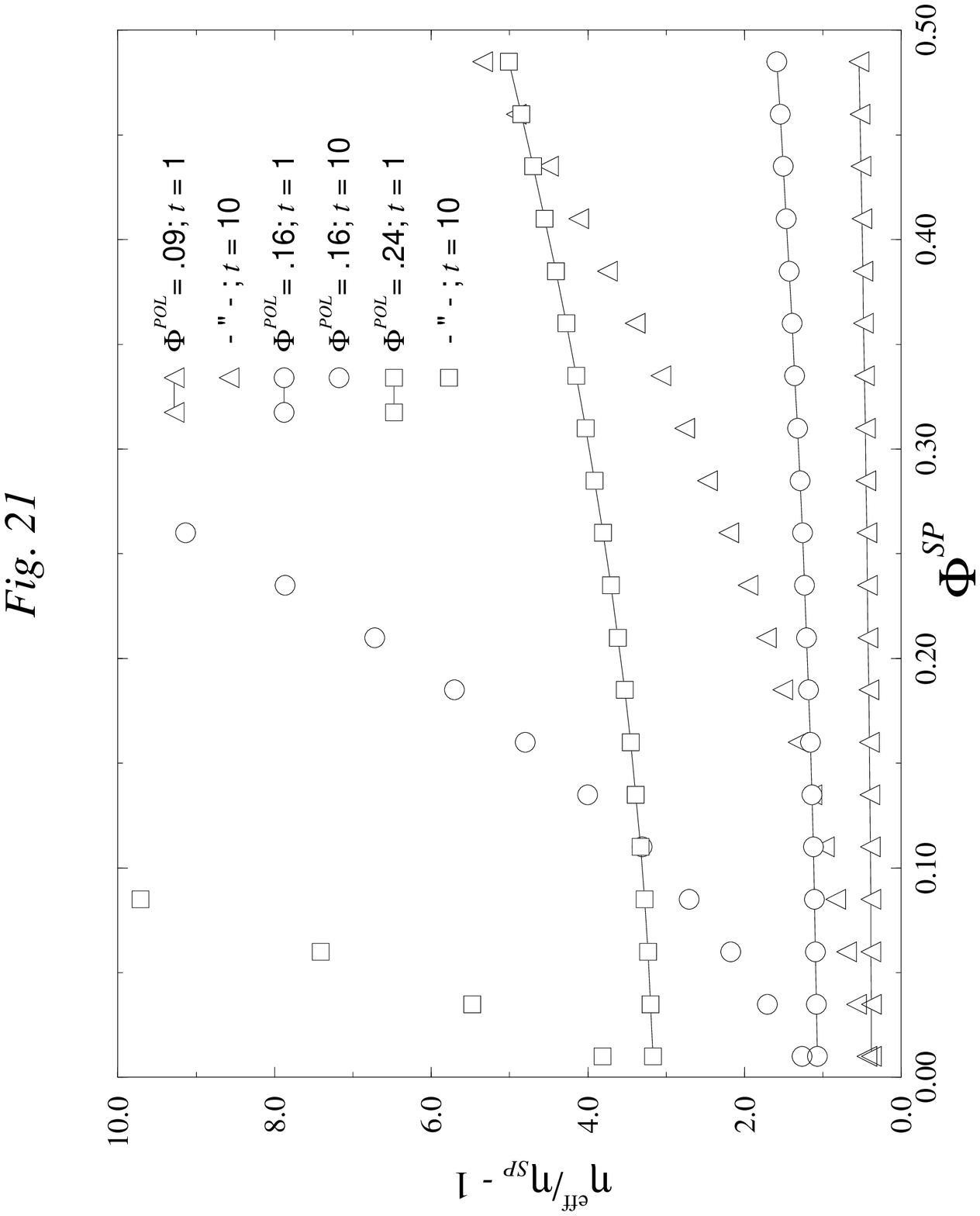}}$$
$$ {\epsfxsize=8.2 truecm \epsfbox{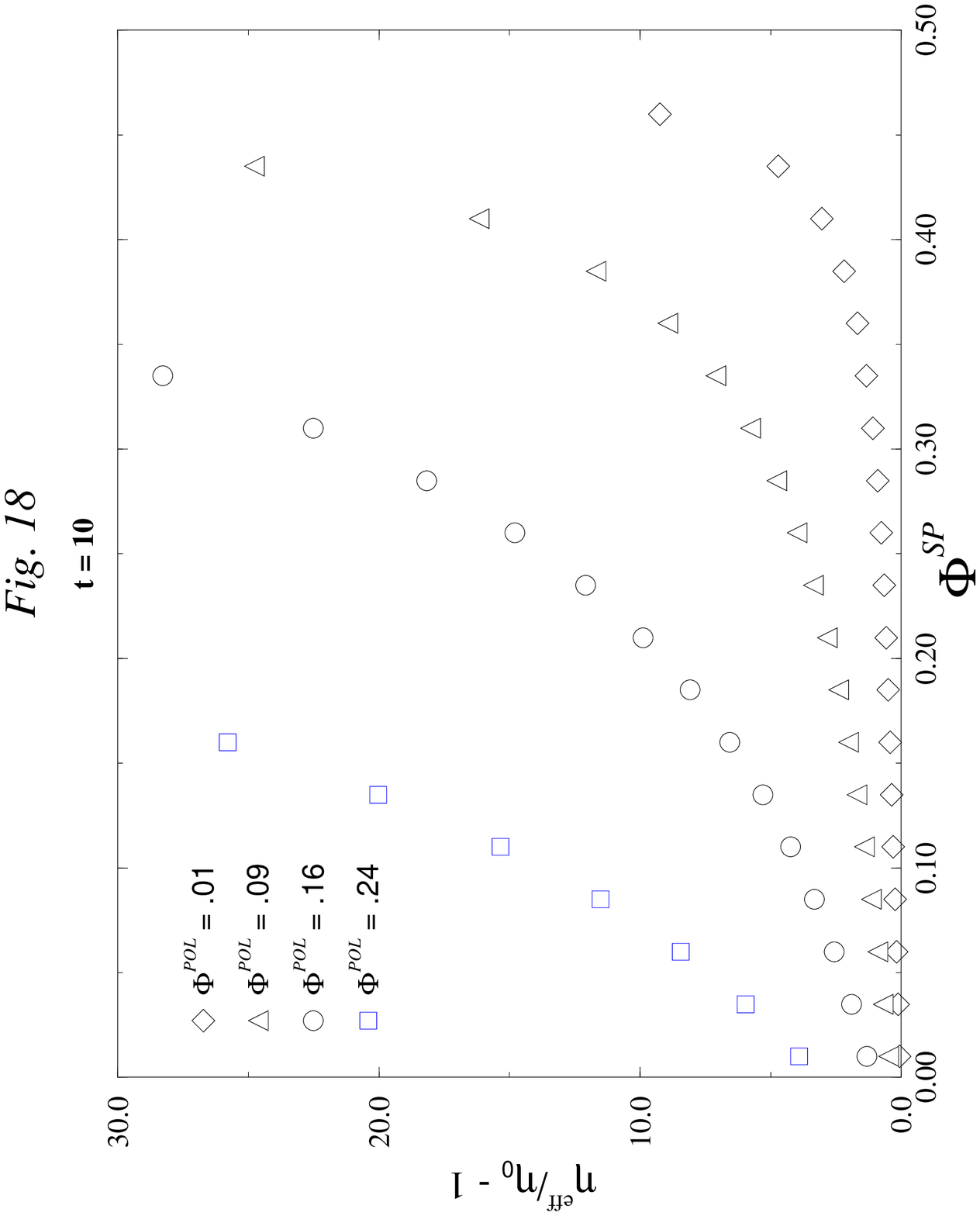}} \;\; {\epsfxsize=8.2
truecm \epsfbox{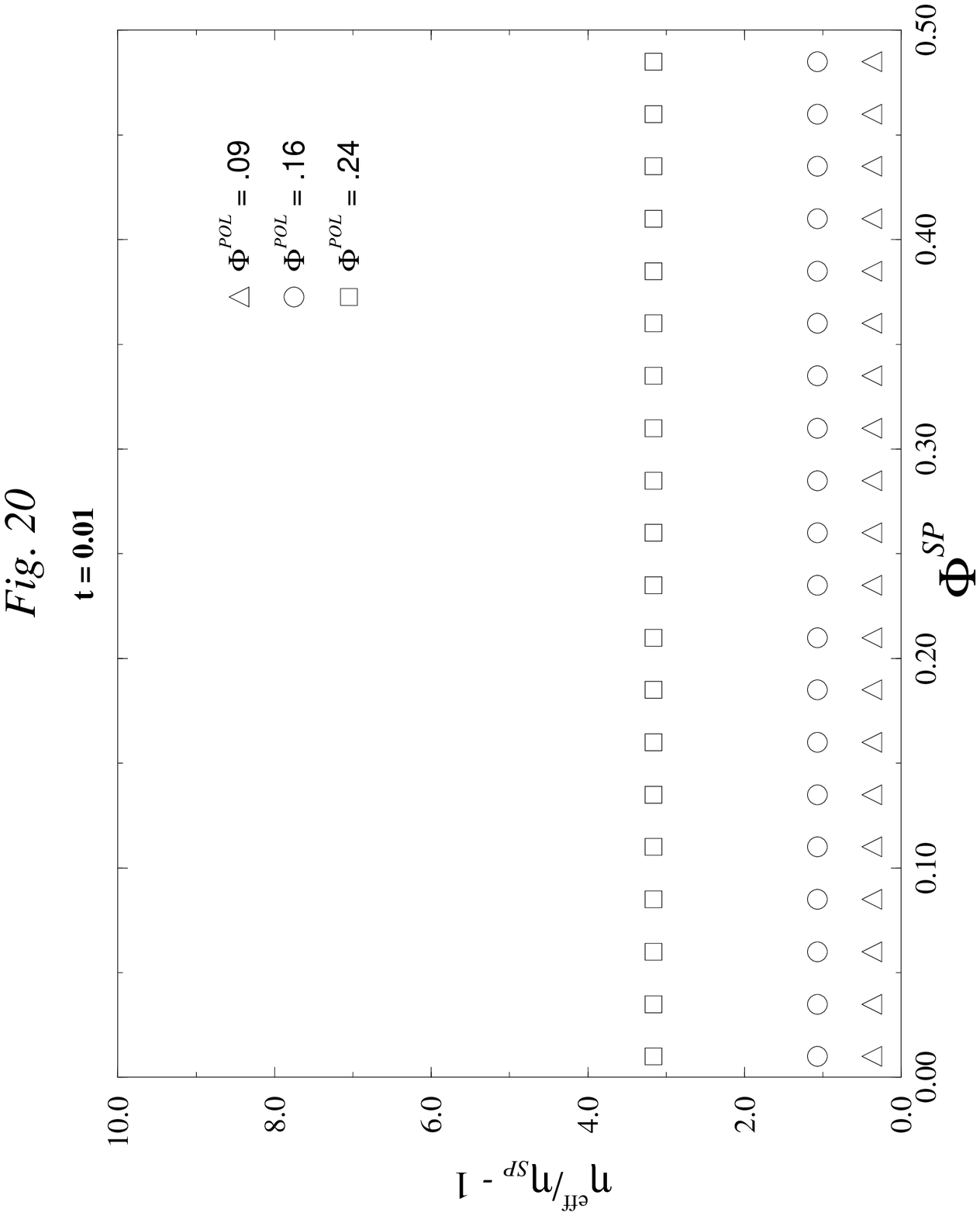}}$$

\end{document}